\documentclass[twocolumn]{aastex6}
\pdfoutput=1
\usepackage{natbib}
\usepackage{color}
\usepackage{hyperref} 

\def \bea {\begin{eqnarray}}
\def \ena {\end{eqnarray}}                  
\def \bee {\begin{equation}}
\def \ene {\end{equation}}
\def    \simlt  {\lower.5ex\hbox{$\; \buildrel < \over \sim \;$}}
\def    \simgt  {\lower.5ex\hbox{$\; \buildrel > \over \sim \;$}}

\newcommand     \mum    {\,\mu{\rm m}}  

\def	\cm		{\,{\rm {cm}}}

\def	\g		{\,{\rm g}}

\def	\K		{\,{\rm K}}

\def	\s		{\,{\rm s}}

\def    \yr  		{\,{\rm {yr}}}

\def	\H		{\rm H}

\def	\xhat		{\hat{\bf x}}
\def	\yhat		{\hat{\bf y}}
\def	\zhat		{\hat{\bf z}}

\def	\ahat		{\hat{\bf a}}
\def	\ehat		{\hat{\bf e}}

\def	\ehat		{\hat{\bf e}}

\def    \apjl  		{\rm {ApJL}}

\def    \Bv     	{\bf  B}

\def    \rv     	{{\bf  r}}

\def	\bJ		{{\bf J}}

\def   	\bQ  		{{\bf Q}}

\def	\br		{{\rv}}
\def	\bv		{{\bf v}}

\def	\gas		{\rm {gas}}

\def	\eff		{\rm {eff}}

\font\mib=cmmib10

\def\bGamma{\hbox{\mib\char"00}}

\shorttitle{Mechanical alignment of irregular grains}
\shortauthors{Hoang, Cho, and Lazarian}

\begin{document}
\title{Alignment of Irregular Grains by Mechanical Torques}

\author{Thiem Hoang}
\affil{Korea Astronomy and Space Science Institute, Daejeon 34055, Korea, email: thiemhoang@kasi.re.kr}
\affil{Korea University of Science and Technology, Daejeon, 34113, Korea}
\author{Jungyeon Cho}
\affil{Department of Astronomy and Space Science, Chungnam National University, Daejeon, Korea}
\author{A. Lazarian}
\affil{Astronomy Department, University of Wisconsin, Madison, WI 53706, USA}

\begin{abstract}
We study the alignment of irregular dust grains by mechanical torques due to the drift of grains through the ambient gas. We first calculate mechanical alignment torques (MATs) resulting from specular reflection of gas atoms for seven irregular shapes: one shape of mirror symmetry, three highly irregular shapes (HIS), and three weakly irregular shapes (WIS). We find that the grain with mirror symmetry experiences negligible MATs due to its mirror-symmetry geometry. Three highly irregular shapes can produce strong MATs which exhibit some generic properties as radiative torques (RATs), while three weakly irregular shapes produce less efficient MATs. We then study grain alignment by MATs for the different angles between the drift velocity and the ambient magnetic field, for paramagnetic and superparamagnetic grains assuming efficient internal relaxation. We find that for HIS grains, MATs can align subsonically drifting grains in the same way as RATs, with low-J and high-J attractors. For supersonic drift, MATs can align grains with low-J and high-J attractors, analogous to RAT alignment by anisotropic radiation. We also show that the joint action of MATs and magnetic torques in grains with iron inclusions can lead to perfect MAT alignment. Our results point out the potential importance of MAT alignment for HIS grains predicted by the analytical model of Lazarian \& Hoang (2007b), although more theoretical and observational studies are required due to uncertainty in the shape of interstellar grains. We outline astrophysical environments where MAT alignment is potentially important.

\end{abstract}


\section{Introduction}\label{sec:intro}
Interstellar dust is an intrinsic component of the interstellar medium and plays an important role in many astrophysical processes, including gas heating and cooling, and formation of planets and stars. Observations of starlight polarization reveal that dust grains are non-spherical and aligned with the interstellar magnetic field (\citealt{Hall:1949p5890}; \citealt{Hiltner:1949p5856}). The alignment of interstellar grains allows us to trace the topology and to measure the strength of interstellar magnetic field through starlight polarization (\citealt{1951PhRv...81..890D}; \citealt{1953ApJ...118..113C}) and polarized thermal dust emission (\citealt{Hildebrand:1988p2566}), in various astrophysical environments. Moreover, polarized thermal emission from aligned grains is a significant Galactic foreground contaminating cosmic microwave background (CMB) radiation, which is considered the most critical challenge to the quest of CMB B-modes as demonstrated by the joint BICEP2/Keck and Planck data analysis (\citealt{Ade:2015ee}).

After more than 65 years since the discovery of starlight polarization (\citealt{Hall:1949p5890}; \citealt{Hiltner:1949p5856}), the problem of grain alignment of dust grains might be solved eventually (see latest reviews by \citealt{Andersson:2015bq} and \citealt{LAH15}). The radiative torque (RAT) alignment (\citealt{1976Ap&SS..43..291D}) is established as a dominant mechanism (\citealt{Andersson:2015bq}; \citealt{LAH15}). This mechanism relies on anisotropic radiation source to spin-up (\citealt{1996ApJ...470..551D}; hereafter DW96) and align irregular grains (\citealt{1997ApJ...480..633D}, hereafter DW97; \citealt{2007MNRAS.378..910L}; \citealt{Hoang:2008gb}), and thus it is valuable for not very dense environments (\citealt{2005ApJ...631..361C}; \citealt{2008ApJ...674..304W}; \citealt{2014MNRAS.438..680H}). In very dense regions, such as the shielded regions of protoplanetary disks, the alignment of classical grains is lost due to low radiation intensity (\citealt{2007ApJ...669.1085C}; \citealt{2014MNRAS.438..680H}).\footnote{Very large grains can still be aligned by radiative torques induced by long-wavelength photons \citep{2017ApJ...839...56T}.} Powerful observation capabilities in the era of submm/mm polarization (e.g., ALMA, JCMT) allow us to look into extreme dense regions where little radiation but active gas dynamics are present, such as protoplanetary disks. Therefore, it is of great importance to reexamine the alignment by non-radiative mechanisms, such as the mechanical one.

Mechanical alignment was pioneered by \cite{{Gold:1952p5848},{1952MNRAS.112..215G}} in which random collisions with gas atoms stochastically increase the grain rotational energy, leading to the final alignment state with the long grain axis parallel to the gas flow. It was then elaborated by other researchers (e.g. \citealt{Purcell:1969p3641}; \citealt{1971ApJ...167...31P}; \citealt{1976Ap&SS..43..291D}; \citealt{1994MNRAS.268..713L}; \citealt{1995ApJ...453..238R}). While the original mechanism could deal with thermally rotating grains only, two modifications of the mechanism that were introduced in \cite{1995MNRAS.277.1235L} and elaborated later in \cite{Lazarian:1996p6083} enabled the alignment of grains rotating at much higher rates. The latter were introduced to the field by \cite{1979ApJ...231..404P} where the limitations on the size for suprathermally rotating grains were discussed).
      
The main shortcoming of the classical mechanical mechanism was that it required supersonic gas-dust drift to get any appreciable degree of alignment (see \citealt{Purcell:1969p3641}). Although subsequent studies indicated that such grain drift can be produced by ambipolar diffusion in star-forming clouds (\citealt{1990BAAS...22..862R}; \citealt{1995ApJ...453..238R}) or interactions of charged grains with magnetohydrodynamic (MHD) turbulence (\citealt{1994MNRAS.268..713L}; \citealt{2002ApJ...566L.105L}; \citealt{Yan:2004ko}; \citealt{Hoang:2012cx}), the degree of alignment achievable for the Mach number of the order of unity (typical conditions of the ISM) is insufficient to explain observations (see estimates in \citealt{1997ApJ...483..296L}). 

An alternative process based on the interaction of a gaseous flow with a {\it helical} grain was considered in \cite{2007ApJ...669L..77L} (henceforth LH07b). The authors extended their analytical model of RAT alignment to include the anisotropic gas flow instead of a photon flux. LH07b found that the gas flow can produce strong regular mechanical torques when interacting with the helical grain. Those torques can align grains up to degrees from 30\% to 100\% even for {\it subsonic} gaseous flows (subsonic grain drift). In comparison, elaborate calculations in \cite{1997ApJ...483..296L} show that the alignment in \cite{1952MNRAS.112..215G} original process does not exceed $20\%$ for realistically flattened grains. Another mechanical alignment process, namely cross-section mechanism (\citealt{1995MNRAS.277.1235L}; \citealt{1997ApJ...483..296L};) does not produce a high degree of alignment either. 

The conclusions in LH07b have recently been supported by \cite{2016MNRAS.457.1958D} who numerically computed MATs for 13 shapes built from Gaussian random spheres. {In the absence of paramagnetic relaxation, they found that MAT alignment in the subsonic regime varies with the grain shapes (i.e., more efficient alignment for shapes with stronger MATs).} Nevertheless, their results for MAT alignment of supersonic grains are consistent with the theoretical predictions in LH07b. {To understand why MATs substantially vary with the grain shape (cf. radiative torques), in this paper, we will first conduct an analytical estimate of MATs for an irregular shape comprising many facets of random orientation and compare with the MAT of a helical grain from LH07b. Then we will calculate the torques for several irregular shapes that exhibit much different degree of irregularity (see Section \ref{sec:shape}). Moreover, since we are interested in the analog of RATs and MATs, we will compute MATs of two irregular shapes chosen by DW97 that are found to produce strong RATs. Furthermore, to achieve a better understanding of MAT alignment for realistic conditions, we will study MAT alignment for the different drifting direction of grains with the ambient gas, in the presence of magnetic relaxation, for both ordinary paramagnetic grains and grains with iron inclusions. Note that the effect of magnetic relaxation was ignored in \cite{2016MNRAS.457.1958D}.}

We note that the incorporation of iron clusters into big grains plays an important role in the RAT alignment paradigm, which can produce universal high-J attractors (\citealt{Lazarian:2008fw}). Interestingly, a recent experiment by \cite{2017SciA....3E1992K} reveals that pure iron grains are rare due to low sticking probability, suggesting that iron is more likely present in dust grains as inclusions/compounds. \cite{2016ApJ...831..159H} carried out extensive simulations on the alignment of RATs for grains with iron inclusions and found that grains can perfectly be aligned with a moderate fraction (above $10\%$) of iron abundance in the form of nanoparticles. Thus, a similar effect should be applicable for MAT alignment.

The structure of the paper is as follows. In Section \ref{sec:AMO_MAT} we present analytical estimates of MATs from AMO and a simple irregular grain, finding a cancellation effect of MATs. Section \ref{sec:shape} describes the irregular shapes used to compute mechanical torques. In Section \ref{sec:MAT} we present the model setup for calculations and the results of MATs. Section \ref{sec:align} is devoted to studying in detail the alignment of grains by MATs for paramagnetic and superparamagnetic grains. Discussion and summary are shown in Section \ref{sec:dis} and \ref{sec:sum}, respectively.

\section{Analytical estimates of mechanical torques}\label{sec:AMO_MAT}
\subsection{MATs from an Analytical Model}
{\citealt{2007MNRAS.378..910L} (hereafter LH07a) first introduced a helical grain model built from an oblate spheroid attached to the massless mirror inclined with respect to the principal plane of the grain. They obtained an analytical description of RATs and made testable predictions of RAT alignment. Such a model is referred to as Analytical MOdel (AMO).

LH07b used the LH07a's model but considered that atoms rather than photons impinge on the model grain. The corresponding calculation of torques for the supersonic case were identical to the calculation of the photon-grain interaction, but the calculations were also performed for the subsonic case where the effect of random thermal velocities of atoms is important. In both cases, substantial torques were reported. {LH07b noticed that MATs for a realistic grain shape would experience reduction due to the {\it reflection efficiency factor} $E$ and {\it helicity reduction factor} $D$. Nevertheless, the physics of the factor $D$ is not yet quantified.} }

Following Equation (7) of LH07b, the MAT from AMO induced by a helical grain drifting through the gas of density $n_{\H}$ is given by: 
\bea
\Gamma_{\rm AMO}&=&\frac{m_{\H}n_{\H}v_{\rm th}^{2}Al_{1}}{2} Q_{\rm AMO}(s_{d})\nonumber\\
&\sim& 2\times 10^{-27}a_{-5}^{3}Q_{\rm AMO}(s_{d})\nonumber\\
&&\times\left(\frac{n_{\H}}{30 \cm^{-3}} \right) \left( \frac{T_{\gas}}{100 \K} \right) \g\cm^{2}\s^{-2},~~~\label{eq:Gam_AMO}
\ena
where $A$ is the area of the mirror and $l_{1}$ is the length from the center of mass to the mirror, $a_{-5}=a_{\rm eff}/10^{-5}\cm$ with $a_{\rm eff}=(3Al_{1}/4\pi)^{1/3}$ being the effective grain size,\footnote{The effective grain radius, $a_{\rm eff}$, is defined as the radius of equivalent sphere of the same volume as the grain.} and $Q_{\rm AMO}$ is the MAT efficiency (see Eq. 6 in LH07b). For the above estimate, the thermal velocity $v_{\rm th}=(2kT_{\gas}/m_{\H})^{1/2}$ is evaluated for gas temperature $T_{\gas}=100\K$ (see Table \ref{tab:MAT_RAT}), and $Q_{\rm AMO}$ is roughly unity at $v_{d}=v_{th}$ and increases with $s_{d}$.

{The AMO appears to exhibit a highest degree of helicity because it has only one twisted facet exposed to the photon or gas flow. Any attempt adding one or more additional facets to the spheroid result in the decrease in grain helicity.\footnote{Here, irregularity is analogous to helicity.} Therefore, we expect the MAT from AMO is strongest.}

\subsection{MATs from a simple irregular shape}\label{sec:AMO}
Now, let us estimate the torque for a more realistic shape. We assume that the grain surface consists of $N_{\rm facet}$ facets with different orientations. When a stream of gas particles hit the grain surface, each facet acts as a mirror and provides a random contribution to the total torque.

The mechanical torque due to specular reflection by a facet is approximately given by
\bea
\delta \Gamma_{\rm MAT}\sim \left(n_{\H} \frac{4\pi a_{\eff}^2}{N_{\rm facet}} v_{d}\right) \gamma_r \left( m_{\H} v_{d} a_{\rm eff}\right),
\ena
where $\gamma_r$ is the reflection coefficient, and the grain surface area is $4\pi a_{\eff}^{2}$. 

The net torque from $N_{\rm facet}$ facets can be calculated using the random walk formula:
\bea
\Gamma_{\rm MAT} &\sim & \delta \Gamma_{\rm MAT} \sqrt{\gamma_e N_{\rm facet}} \nonumber \\
&\sim& 4 \pi \gamma_r \sqrt{\gamma_e} n_{\H} m_{\H}  v_{\gas}^2  \frac{a_{\eff}^3}{\sqrt{N_{\rm facet}}}, 
\ena
where $\gamma_e$ denotes the fraction of the grain surface area that is substantially
exposed to the stream of particles. 

By plugging the typical physical parameters into the above equation, one obtains
\bea
\Gamma_{\rm MAT}&\sim& \frac{2\times 10^{-28}a_{-5}^{3}}{\sqrt{N_{\rm facet}}}\left(\frac{n_{\H}}{30 \cm^{-3}} \right) \nonumber\\
&&\times\left( \frac{T_{\gas}}{100 \K} \right)\left( \frac{v_{d}}{v_{\rm th}} \right)^2~\g\cm^{2}\s^{-2},\label{eq:GamMAT}
\ena
where $\gamma_e=1/6$ is adopted. 

{From Equations (\ref{eq:GamMAT}) and (\ref{eq:Gam_AMO}), we see that an arbitrary grain shape of $N_{\rm facet}$ facets  has MATs reduced by a factor $\sqrt{N_{\rm facet}}$ from the AMO. For instance, a spheroidal or spherical shape has $N_{\rm facet}\gg 1$, leading to negligible MATs. The reduction of MATs from the AMO arises from averaging individual torques of random facets, which we term {\it cancellation effect}. In the next section, we will compute MATs of grains made of finite facets.}

\section{Selective Irregular Grain Shapes}\label{sec:shape}
\subsection{Highly and Weakly Irregular Shapes}
An irregular grain shape is constructed by assembling $N_{\rm block}$ cubic blocks of unit volume. Our strategy to create an irregular shape is first to assembly a number of cubic blocks on a plane, so-called {\it principal plane}. Then, we add a few blocks above and below the principal plane. Shape 1 is built from $5\times 3$ blocks in the principal plane, 4 blocks above and 4 blocks below the principal plane. Shape 2 is made by a principal plane of $5\times 5$ blocks and 16 blocks above and below the plane. Shapes 3-5 are slightly different in which there are three blocks on top of the principal plane. The principal plane of shape 3 is made by a layer of $5\times5$ blocks, shape 4 by two layers of $4\times 4$, and shape 5 is built from two layers of $5\times 5$. We also consider two shapes (denoted by shapes 6 and 7) built from 13 and 11 cubic blocks, similar to the shapes used for RAT calculations by DW96 and DW97. The grain shape is described by coordinates of the blocks and three principal axes $\ahat_{1},\ahat_{2}, \ahat_{3}$ (see Tables 1-6).

\begin{figure*}
\includegraphics[width=0.3\textwidth]{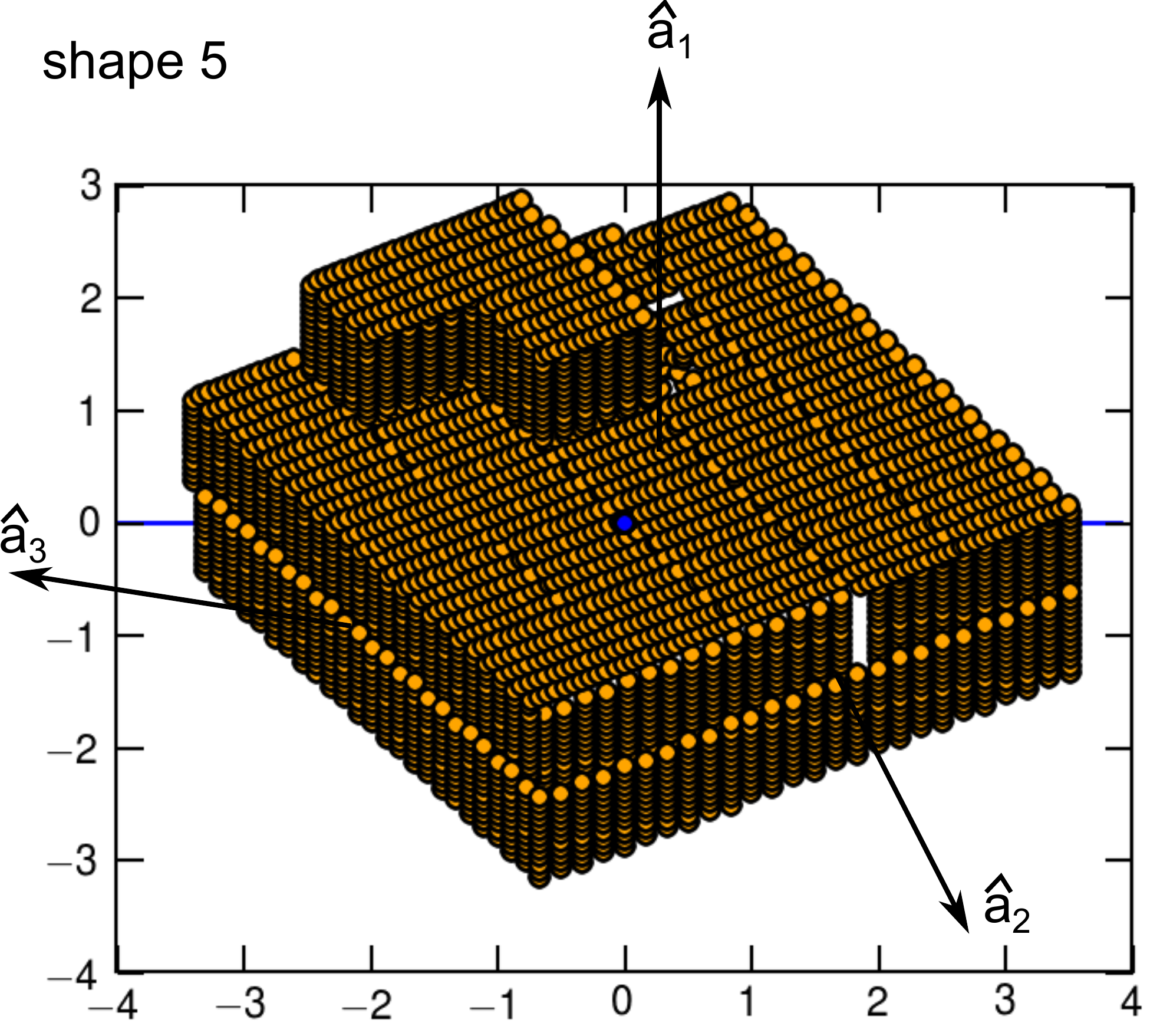}
\includegraphics[width=0.3\textwidth]{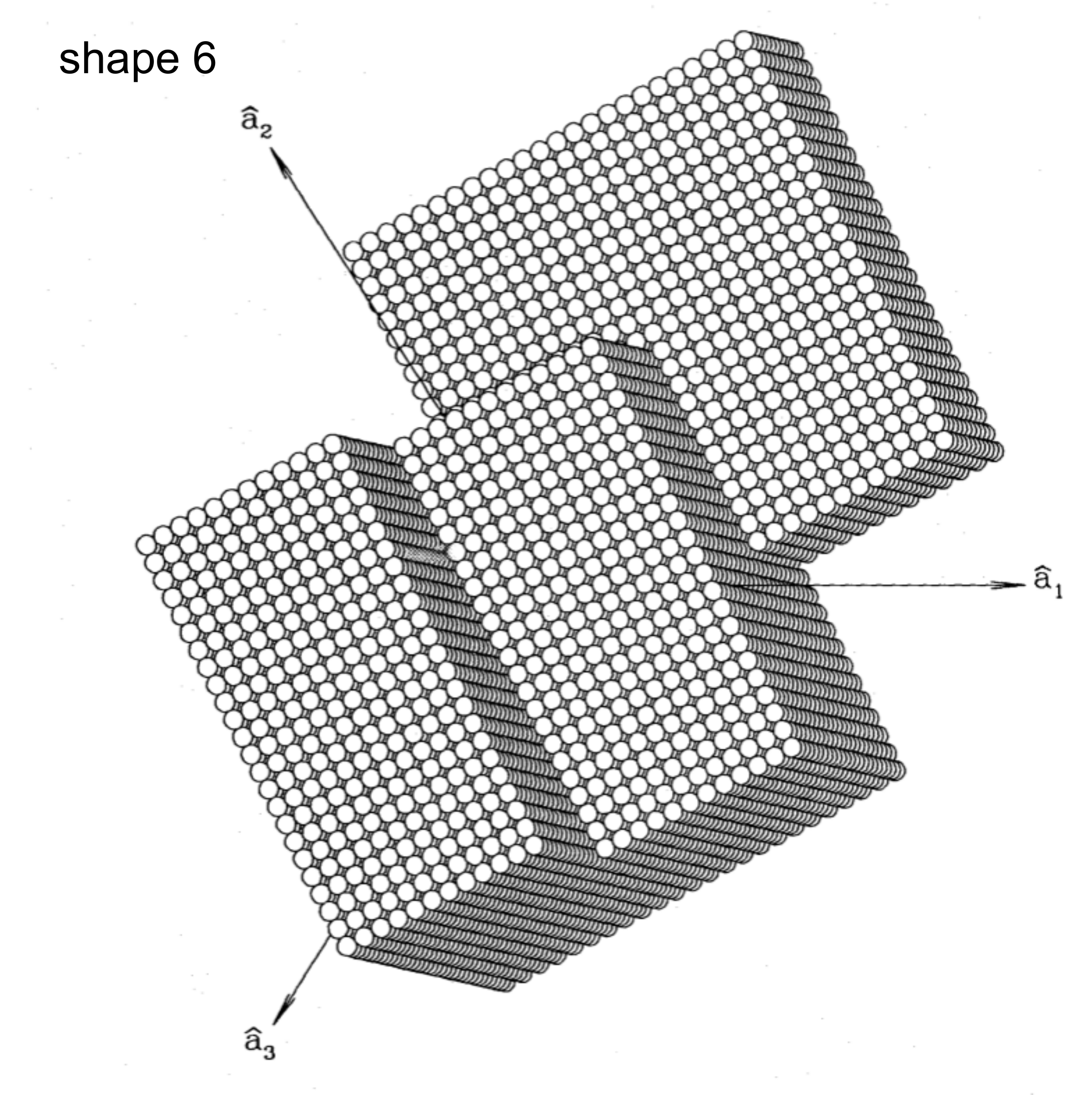}
\includegraphics[width=0.3\textwidth]{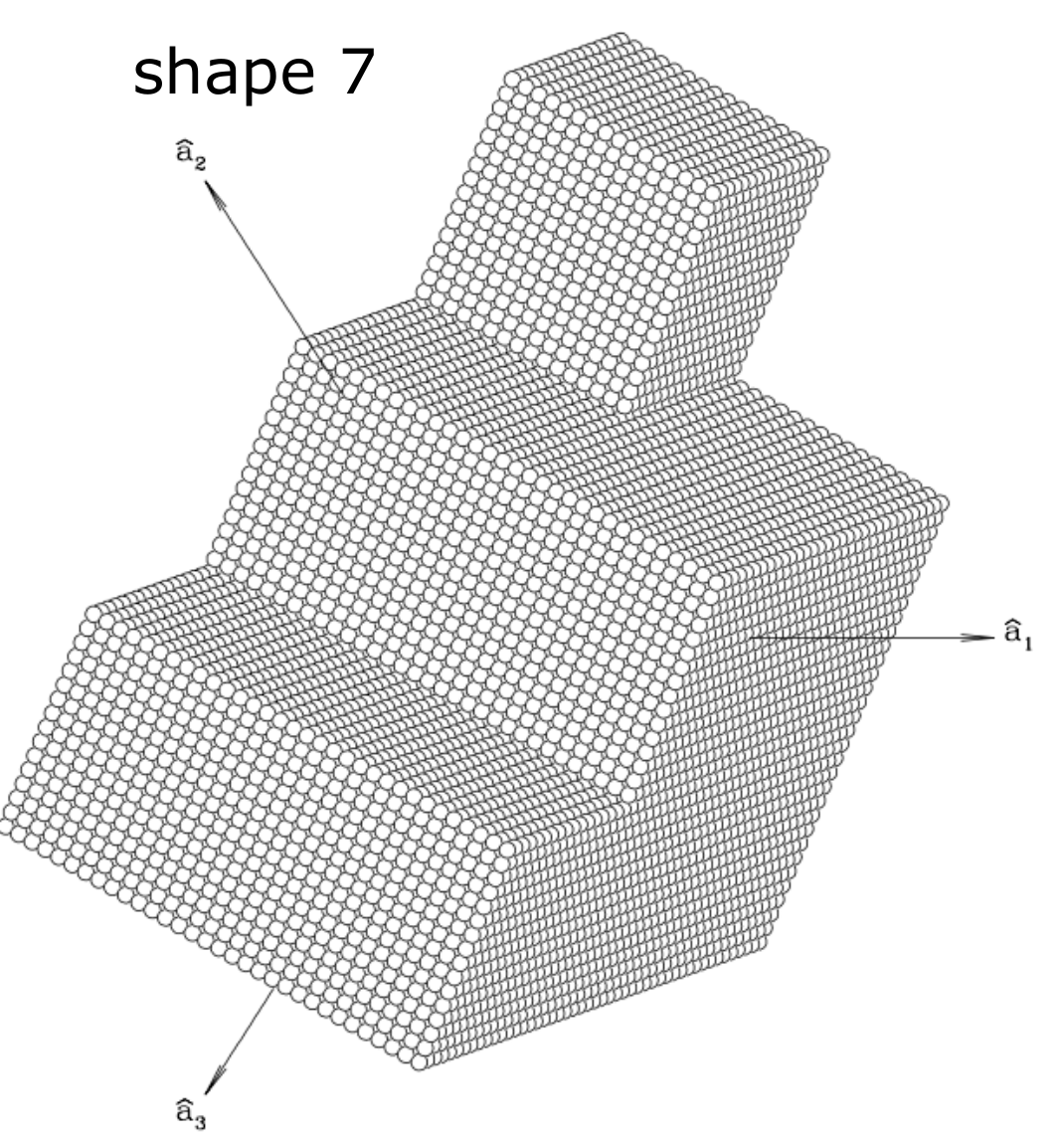}
\includegraphics[width=0.3\textwidth]{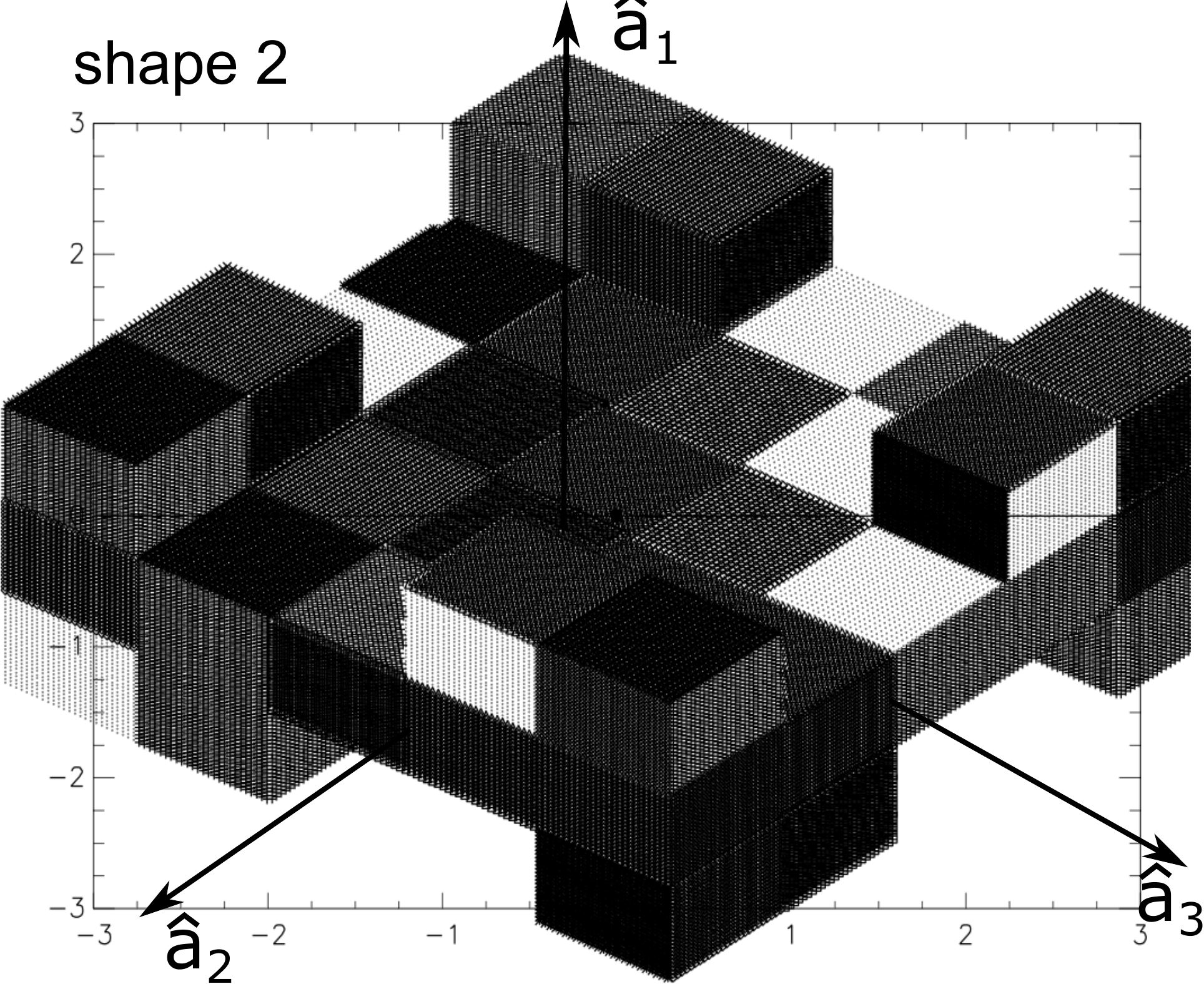}
\includegraphics[width=0.3\textwidth]{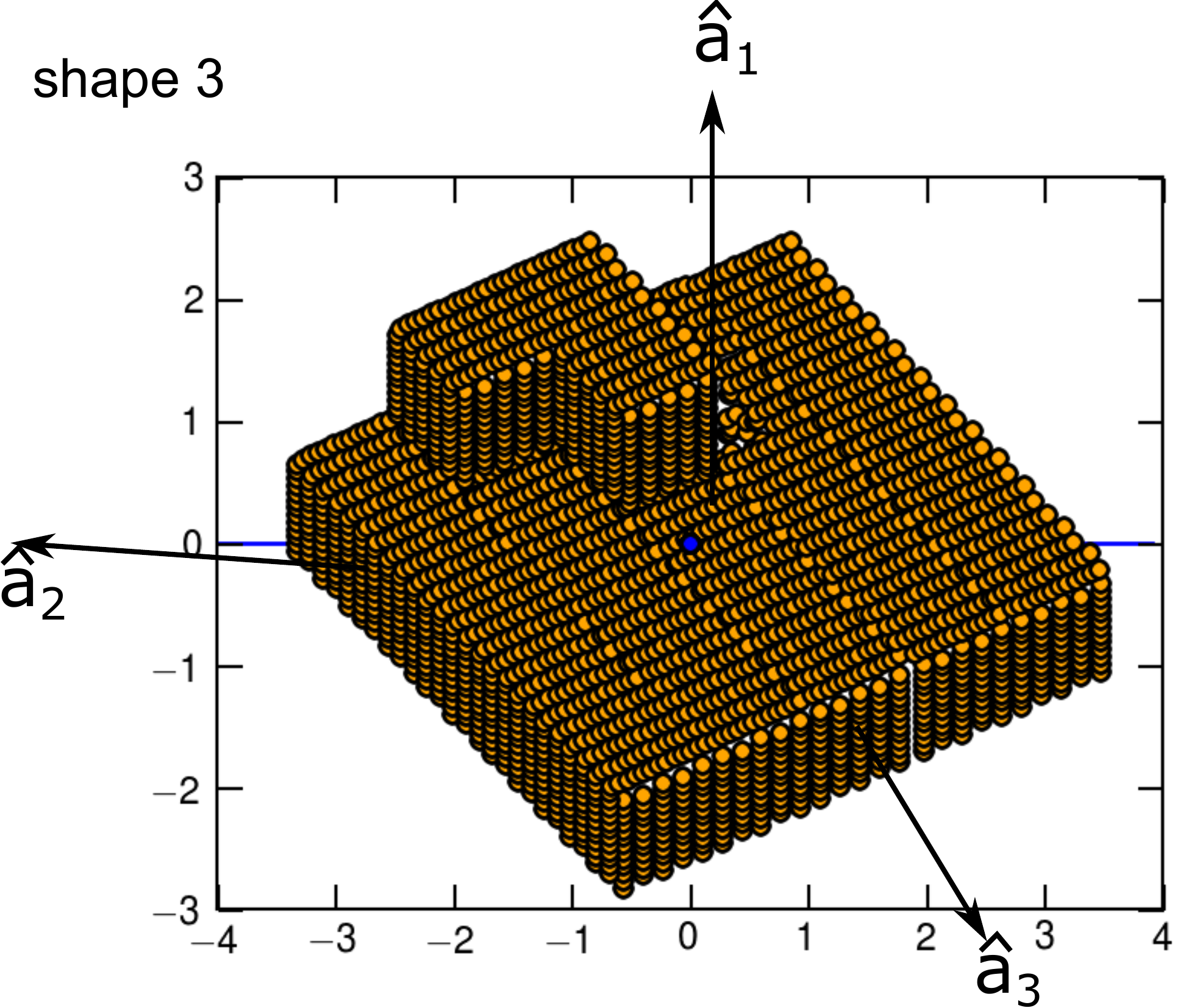}
\includegraphics[width=0.3\textwidth]{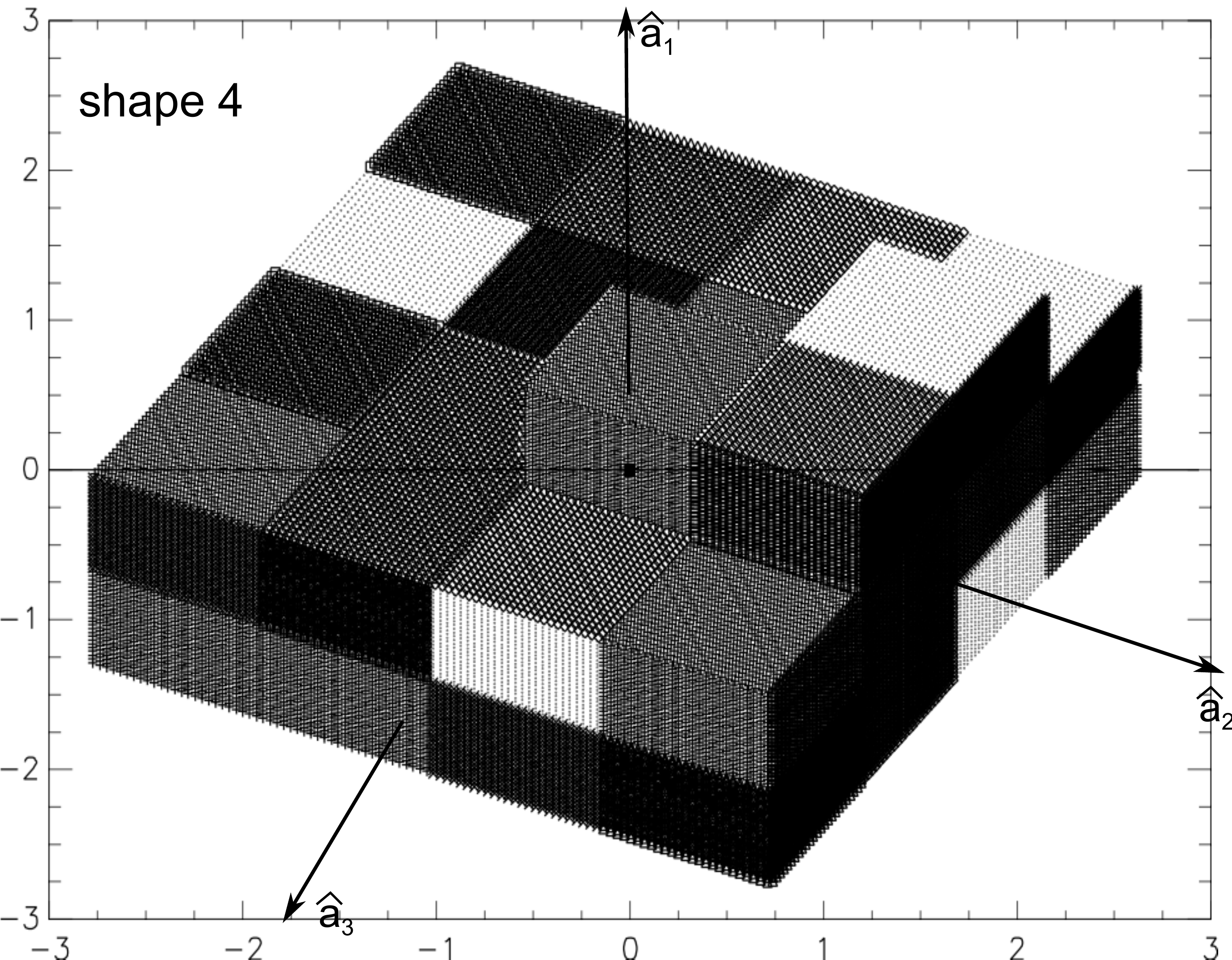}
\includegraphics[width=0.3\textwidth]{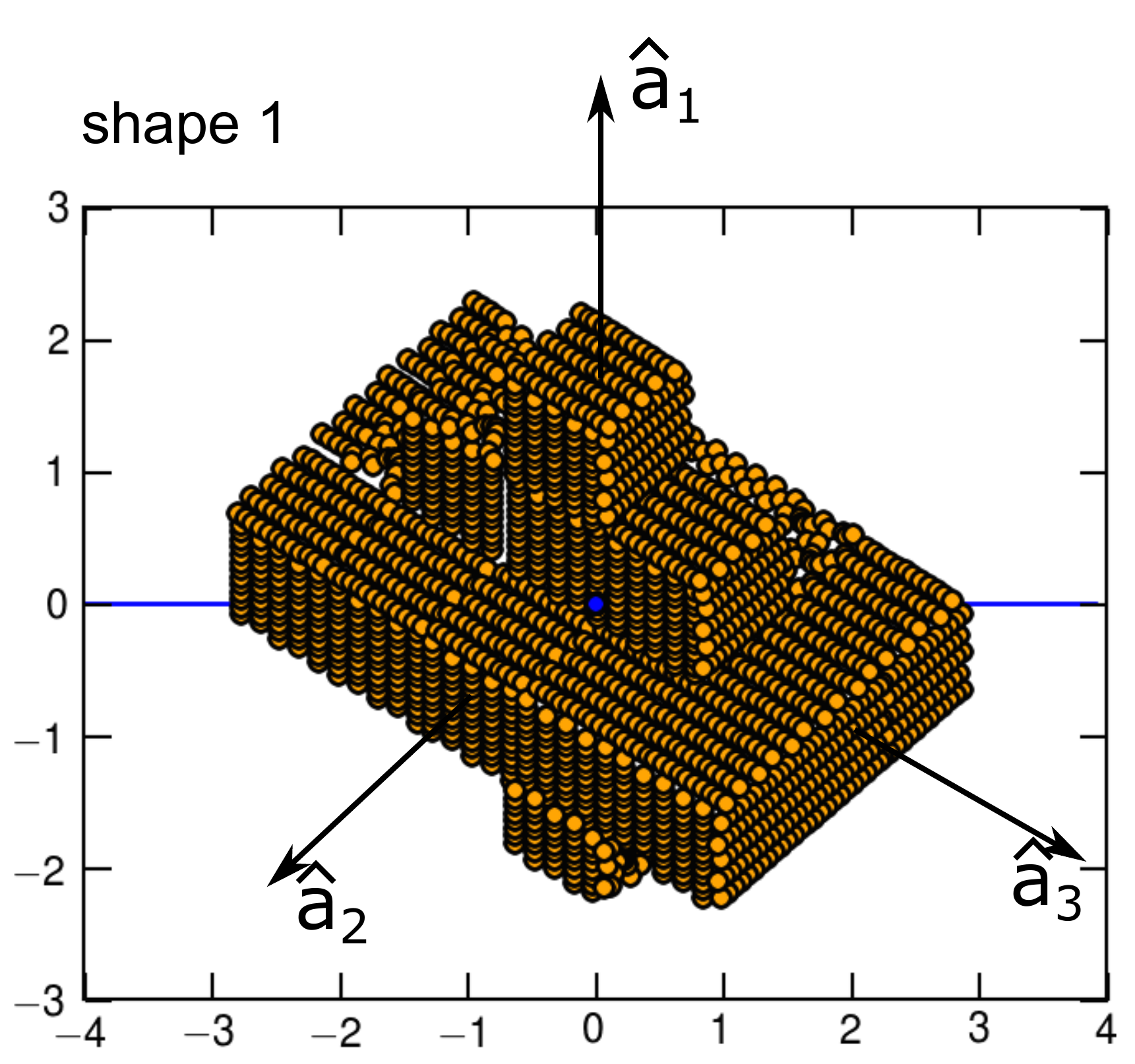}
\caption{Visualization of seven irregular shapes constructed for our study of MAT alignment. Shapes 1-5 consist of a principal plane $\ahat_{2}\ahat_{3}$ and several blocks placed on the top/bottom of the plane. Shape 1 exhibits mirror symmetry.}
\label{fig:shape}
\end{figure*}

Figure \ref{fig:shape} shows the visualization of our selected irregular shapes. As seen, shape 1 exhibits mirror symmetry with respect to the principal axes $\ahat_{j}$. Shapes 3-5 are {\it weakly irregular shapes} (hereafter WIS), i.e., have a low degree of irregularities arising from the blocks on top of the principal plane, while shapes 2, 6, and 7 are {\it highly irregular shapes} (hereafter HIS), i.e., have a high degree of irregularities.

The coordinates of constituent blocks for shapes 1-7 are shown in Tables \ref{tab:sh1-2}-\ref{tab:dw96}. Table \ref{tab:sh1-sh5} shows the coordinates of the principal axes $\ahat_{j}$ of seven shapes. For shapes 1 and 2,  the principal axes are directed along the cube unit vectors. Shapes 3-5 have the principal axes almost aligned with the cube unit vectors. For this particular choice, when the grain drift is along $\ahat_{1}$ axis, the torque for the rotation along $\ahat_{1}$ is expected to be negligible because incident atoms mostly hit a single symmetric surface.

\begin{table}
\centering
\caption{Coordinates of blocks for shapes 1 and 2}\label{tab:sh1-2}
\begin{tabular}{lllllll} \hline\hline\\
& \multicolumn{3}{c}{{shape 2}} & \multicolumn{3}{c}{{shape 1}}\\
$j$ & x$_{j}$ & y$_{j}$ & z$_{j}$ & x$_{j}$ & y$_{j}$ & z$_{j}$\cr
\hline\cr

           1 &  1.0 &       1.0 &       1.0  & 1.0 &       1.0 &       1.0 \cr  
           2 &  2.0 &       1.0 &       1.0  & 2.0 &       1.0 &       1.0 \cr    
           3 &  3.0 &       1.0 &       1.0  & 3.0 &       1.0 &       1.0 \cr    
           4  & 4.0 &       1.0 &       1.0  & 1.0 &       2.0 &       1.0 \cr    
           5 &  5.0 &       1.0 &       1.0  & 2.0 &       2.0 &       1.0 \cr    
           6 &  1.0 &       2.0 &       1.0  & 3.0 &       2.0 &       1.0 \cr    
           7 &  2.0 &       2.0 &       1.0  & 1.0 &       3.0 &       1.0 \cr    
           8 &  3.0 &       2.0 &       1.0  &  2.0 &       3.0 &       1.0 \cr  
           9 &  4.0 &       2.0 &       1.0  & 3.0 &       3.0 &       1.0 \cr    
          10 &  5.0 &       2.0 &       1.0  & 1.0 &       4.0 &       1.0 \cr    
          11 &  1.0 &       3.0 &       1.0  &  2.0 &       4.0 &       1.0 \cr   
          12 &  2.0 &       3.0 &       1.0  & 3.0 &       4.0 &       1.0 \cr    
          13 &  3.0 &       3.0 &       1.0  &  1.0 &       5.0 &       1.0 \cr    
          14 &  4.0 &       3.0 &       1.0  & 2.0 &       5.0 &       1.0 \cr    
          15 &  5.0 &       3.0 &       1.0 & 3.0 &       5.0 &       1.0 \cr    
          16 &  1.0 &       4.0 &       1.0 &  2.0 &       2.0 &      2.0 \cr   
          17 &  2.0 &       4.0 &       1.0 &  2.0 &       3.0 &       2.0 \cr    
          18 &  3.0 &       4.0 &       1.0 &  2.0 &       4.0 &       2.0 \cr    
          19 &  4.0 &       4.0 &       1.0 &  2.0 &       2.0 &      0.0 \cr    
          20 &  5.0 &       4.0 &       1.0 &  2.0 &       3.0 &      0.0 \cr    
          21 &  1.0 &       5.0 &       1.0 & 2.0 &       4.0 &      0.0 \cr    
          22 &  2.0 &       5.0 &       1.0 & 2.0 &       3.0 &      3.0 \cr    
          23 &  3.0 &       5.0 &       1.0 & 2.0 &       3.0 &      -1.0 \cr    
          24 &  4.0 &       5.0 &       1.0 \cr   
          25 &  5.0 &       5.0 &       1.0 \cr  
            
          26 &  1.0 &       1.0 &       2.0 \cr    
          27 &  1.0 &       1.0 &       0.0 \cr   
          28 &  2.0 &       1.0 &       0.0 \cr    
          29 &  4.0 &       1.0 &       2.0 \cr    
          30 &  5.0 &       1.0 &       2.0 \cr    
          31 &  5.0 &       1.0 &       0.0 \cr 
             
          32 &  1.0 &       2.0 &       2.0 \cr    
          33 &  5.0 &       2.0 &       0.0 \cr    
          34 &  1.0 &       4.0 &       0.0 \cr    
          35 &  5.0 &       4.0 &       2.0 \cr    
          36 &  1.0 &       5.0 &       2.0 \cr    
          37 &  1.0 &       5.0 &       0.0 \cr    
          38 &  2.0 &       5.0 &       2.0 \cr    
          39 &  4.0 &       5.0 &       0.0 \cr    
          40 &  5.0 &       5.0 &       2.0 \cr    
          41 &  5.0 &       5.0 &       0.0 \cr    
\hline\hline\\
\end{tabular}
\end{table}

\begin{table}
\centering
\caption{Coordinates of blocks for shapes 3 and 4}\label{tab:sh3-4}
\begin{tabular}{lllllll} \hline\hline\\
& \multicolumn{3}{c}{{shape 4}} & \multicolumn{3}{c}{{shape 3}}\\
$j$ & x$_{j}$ & y$_{j}$ & z$_{j}$ & x$_{j}$ & y$_{j}$ & z$_{j}$\cr
\hline\cr
           1  & 1.0   &    1.0    &   1.0 &  1.0 &       1.0 &       1.0 \cr    
           2  & 2.0   &    1.0    &   1.0 &  2.0 &       1.0 &       1.0 \cr    
           3  & 3.0   &    1.0    &   1.0  &  3.0 &       1.0 &       1.0 \cr    
           4  & 4.0   &    1.0    &   1.0  &  4.0 &       1.0 &       1.0 \cr    
           5  & 1.0   &    2.0    &   1.0  &  5.0 &       1.0 &       1.0 \cr   
           6  & 2.0   &    2.0    &   1.0  & 1.0 &       2.0 &       1.0 \cr    
           7  & 3.0   &    2.0    &   1.0  & 2.0 &       2.0 &       1.0 \cr    
           8  & 4.0   &    2.0    &   1.0  & 3.0 &       2.0 &       1.0 \cr    
           9  & 1.0   &    3.0   &    1.0  & 4.0 &       2.0 &       1.0 \cr    
          10  & 2.0	  &    3.0	&    1.0  & 5.0 &       2.0 &       1.0 \cr    
          11  & 3.0	  &    3.0	&    1.0  & 1.0 &       3.0 &       1.0 \cr    
          12  & 4.0	  &     3.0 &        1.0  &  2.0 &       3.0 &       1.0 \cr    
          13  & 1.0	  &      4.0 &       1.0 &  3.0 &       3.0 &       1.0 \cr    
          14  & 2.0	  &      4.0 &       1.0 &  4.0 &       3.0 &       1.0 \cr    
          15  & 3.0	  &      4.0 &       1.0 &  5.0 &       3.0 &       1.0 \cr    
          16  & 4.0   &      4.0 &       1.0 &   1.0 &       4.0 &       1.0 \cr    
          17  & 1.0   &      1.0 &       0.0 &  2.0 &       4.0 &       1.0 \cr    
          18  & 2.0   &      1.0 &       0.0 &  3.0 &       4.0 &       1.0 \cr    
          19  & 3.0   &      1.0 &       0.0 &  4.0 &       4.0 &       1.0 \cr    
          20  & 4.0   &      1.0 &       0.0 &  5.0 &       4.0 &       1.0 \cr    
          21  & 1.0   &      2.0 &       0.0 &  1.0 &       5.0 &       1.0 \cr    
          22  & 2.0   &      2.0 &       0.0 &  2.0 &       5.0 &       1.0 \cr    
          23  & 3.0   &      2.0 &       0.0 &  3.0 &       5.0 &       1.0 \cr    
          24  & 4.0   &      2.0 &       0.0  &  4.0 &       5.0 &       1.0 \cr    
          25  & 1.0   &      3.0 &       0.0 &  5.0 &       5.0 &       1.0 \cr    
          26  & 2.0   &      3.0 &       0.0 &  1.0 &       2.0 &       2.0 \cr    
          27  & 3.0   &      3.0 &       0.0 &  1.0 &       3.0 &       2.0 \cr    
          28  & 4.0   &      3.0 &       0.0  &  2.0 &       3.0 &       2.0 \cr    
          29  &  1.0  &      4.0 &       0.0 \cr    
          30  &  2.0  &      4.0 &       0.0 \cr    
          31  &  3.0  &      4.0 &       0.0 \cr    
          32  &  4.0  &      4.0 &       0.0 \cr    
          33  &  1.0  &      2.0 &       2.0 \cr    
          34  &  1.0  &      3.0 &       2.0 \cr    
          35  &  2.0  &      3.0 &       2.0 \cr    

\hline\hline\\
\end{tabular}
\end{table}
 
\begin{table}
\centering
\caption{Coordinates of blocks for shape 5}\label{tab:sh5}
\begin{tabular}{llllllll} \hline\hline\\
$j$ & x$_{j}$ & y$_{j}$ & z$_{j}$  &$j$ & x$_{j}$ & y$_{j}$ & z$_{j}$\cr
\hline\cr

 		   1  &  1.0 &       1.0 &       1.0 &    28   & 3.0 &       1.0 &       0.0 \cr    
 
           2  & 2.0 &       1.0 &       1.0 &     29   & 4.0 &       1.0 &       0.0 \cr    

           3  & 3.0 &       1.0 &       1.0 &     30   & 5.0 &       1.0 &       0.0 \cr    
     
           4  &  4.0 &       1.0 &       1.0 & 31   & 1.0 &       2.0 &       0.0 \cr    
           5  &  5.0 &       1.0 &       1.0 & 32   & 2.0 &       2.0 &       0.0 \cr    
           6   & 1.0 &       2.0 &       1.0 &  33   & 3.0 &       2.0 &       0.0 \cr  
           7   & 2.0 &       2.0 &       1.0 &  34   & 4.0 &       2.0 &       0.0 \cr     
           8   & 3.0 &       2.0 &       1.0 &  35   & 5.0 &       2.0 &       0.0 \cr  
           9   & 4.0 &       2.0 &       1.0 &  36   & 1.0 &       3.0 &       0.0 \cr  
          10  &  5.0 &       2.0 &       1.0 &  37   & 2.0 &       3.0 &       0.0 \cr   
          11  &  1.0 &       3.0 &       1.0 &  38   & 3.0 &       3.0 &       0.0 \cr    
          12  &  2.0 &       3.0 &       1.0 &   39   & 4.0 &       3.0 &       0.0 \cr 
          13  &  3.0 &       3.0 &       1.0 &  40   & 5.0 &       3.0 &       0.0 \cr      
          14  &  4.0 &       3.0 &       1.0 &  41   & 1.0 &       4.0 &       0.0 \cr   
          15  &  5.0 &       3.0 &       1.0 &   42   & 2.0 &       4.0 &       0.0 \cr 
          16   & 1.0 &       4.0 &       1.0 &   43   & 3.0 &       4.0 &       0.0 \cr 
          17   & 2.0 &       4.0 &       1.0 &   44   & 4.0 &       4.0 &       0.0 \cr 
          18   & 3.0 &       4.0 &       1.0 &   45   & 5.0 &       4.0 &       0.0 \cr 
          19   & 4.0 &       4.0 &       1.0 &   46   & 1.0 &       5.0 &       0.0 \cr    
          20   & 5.0 &       4.0 &       1.0 &   47   & 2.0 &       5.0 &       0.0 \cr     
          21   & 1.0 &       5.0 &       1.0 &    48   & 3.0 &       5.0 &       0.0 \cr 
          22   & 2.0 &       5.0 &       1.0 &   49   & 4.0 &       5.0 &       0.0 \cr   
          23   & 3.0 &       5.0 &       1.0 &   50   & 5.0 &       5.0 &       0.0 \cr
          24   & 4.0 &       5.0 &       1.0 &   51   & 1.0 &       2.0 &       2.0 \cr 
          25   & 5.0 &       5.0 &       1.0 &   52   & 1.0 &       3.0 &       2.0 \cr  
          26   & 1.0 &       1.0 &       0.0 &   53   & 2.0 &       3.0 &       2.0 \cr  
          27   & 2.0 &       1.0 &       0.0 &  \cr
   
\hline\hline\\
\end{tabular}
\end{table}

\begin{table}
\centering
\caption{Coordinates of blocks for shapes 6 and 7}\label{tab:dw96}
\begin{tabular}{lllllll} \hline\hline\\
& \multicolumn{3}{c}{{shape 6}} & \multicolumn{3}{c}{{shape 7}}\\
$j$ & x$_{j}$ & y$_{j}$ & z$_{j}$ & x$_{j}$ & y$_{j}$ & z$_{j}$\cr
\hline\cr
1 & 0 & 1 & 0 & 0 & 0 & 0 \cr
2 & 0 & 1 & 1 & 1 & 0 & 0\cr
3 & 0 & 2 & 0 & 0 & 1 & 0\cr
4 & 0 & 2 & 1 &  1 & 1 &0\cr
5 & 1 & 1 & 0 & 0 & 0 & 1\cr
6 & 1 & 1 & 1 & 1 & 0 & 1\cr
7 & 1 & 2 & 0 & 0 & 1 & 1\cr
8 & 1 & 2 & 1 & 1 & 1 & 1 \cr
9 & 0 & 0 & 1 & 2 & 0 & 0 \cr
10 &0 & 0 & 2& 2 & 1 & 0 \cr
11 &0 & 1 & 2& 0 & 0 & 2 \cr
12 &2 & 1 & 0 \cr
13 &2 & 2 & 0 \cr
\hline\hline\\
\end{tabular}
\end{table}

\begin{table}
\centering
\caption{Coordinates of the principal axes for the different shapes}\label{tab:sh1-sh5}
\begin{tabular}{l l l l} \hline\hline\\
Shape & $\ahat_{1}$ & $\ahat_{2}$ & $\ahat_{3}$ \cr
\hline\cr

1 & (0,0,1) & (1,0,0) & (0,1,0)\cr
2 & (0,0,1) & (1,0,0) & (0,1,0)\cr
3 & (0.08,0.02,0.99) & (0.23,-0.97,-0.01) & (0.97,0.23,-0.08)\cr
4 & (0.15,-0.02,0.98) &(0.07,0.99,0.01) & (-0.99,0.07,0.15)\cr
5 & (0.08,0.015,0.99) &(0.23,-0.97,0.004) & (0.97,0.23,-0.079)\cr
6 & (0.45,0.43,0.78) & (0.67,-0.74,0.03) & (0.59,0.50,-0.62)\cr
7 & (0.22,0.83,0.49) & (0.57,-0.52,0.63) & (0.79,0.13,-0.59)\cr

\hline\hline\\
\end{tabular}
\end{table}

\begin{table}
\centering
\caption{Coefficients $\alpha_{j}$ for the different shapes}\label{tab:alpha}
\begin{tabular}{l l l l} \hline\hline\\
Shape & $\alpha_{1}$ & $\alpha_{2}$ & $\alpha_{3}$ \cr
\hline\cr
1 & 1.810 &1.670  & 0.972\cr
2 & 2.809 & 1.663 & 1.663\cr
3 & 2.871 & 1.658 & 1.447\cr
4 & 1.591 & 1.117 & 1.040\cr
5 & 1.902 & 1.187 & 1.109\cr
6 & 1.745 & 1.610 & 0.876\cr
7 & 1.561 & 1.464 & 0.889\cr

\hline\hline\\
\end{tabular}
\end{table}

The equivalent sphere radius of the grain in the code unit is given by
\bea
R_{\rm code} =\left(\frac{3N_{\rm block}}{4\pi}\right)^{1/3},
\ena
which can be converted to the physical unit $a_{\rm eff}$ via a scaling parameter $f_{cp} = a_{\rm eff}/R_{\rm code}$.

The inertia moments along the principal axes are defined in terms of the equivalent sphere as follows:
\bea
I_{j}=\alpha_{j}\frac{8\pi \rho a_{\rm eff}^{5}}{15},\label{eq:Ia}
\ena
where $\rho$ is the mass density, $\alpha_{j}$ with $j=1-3$ are coefficients with $\alpha_{j}=1$ for spheres. Table \ref{tab:alpha} shows the coefficients $\alpha_{j}$ for the chosen irregular shapes.

\section{Mechanical Torques: Numerical Method and Results}\label{sec:MAT}
\subsection{Model Setup and Coordinate systems}
Let define a lab system $\ehat_{1},\ehat_{2},\ehat_{3}$ in which $\ehat_{1}$ is directed along the drift velocity $\bv_{d}$, $\ehat_{2}\perp \ehat_{1}$, and $\ehat_{3}=\ehat_{1}\times \ehat_{2}$ (see Fig. \ref{fig:labRF}). The orientation of the grain in the lab frame is completely determined by the orientation of $\ahat_{1}$ and the rotation of the grain axes $\ahat_{2}$ around the $\ahat_{1}$ axis. The orientation of $\ahat_{1}$ is defined by two angles $\Theta$ between $\ahat_{1}$ and $\bv_{d}$ and the precession $\Phi$. The rotation of the grain around $\ahat_{1}$ axis is determined by the angle $\beta$.

The impact position of an atom on a block is determined by the radius vector $\br$ centered at the center of mass (CM) and its normal vector ${\bf N}$. Let $\xhat,\yhat,\zhat$ be unit vectors along the three normal vectors of the cubic block. Each surface of the block is then divided into a grid of $N_{1}\times N_{1}$ cells, with the regular resolution $dx=dy=1/(N_{1}-1)$. The cell center of the $m$th block is determined by $\br_{c;m}$ for $m=1-N_{block}$. Thereby, the position of the cell $ij$ of the surface $m$ is given by 
\bea
\br_{ij;m} = \br_{c;m} + id{\bf x} + jd{\bf y},\label{eq:rijm}
\ena
for $i,j$ from $0-N_{1}-1$, and ${\bf dx}=[dx,0,0], d{\bf y}=[0,dy,0]$. 

The block's center radial vector is described by
\bea
\br_{c;m} = X_{m} \pm \frac{1}{2}\xhat + \pm \frac{1}{2}\yhat + \frac{1}{2}\zhat.\label{eq:rcm}
\ena

\subsection{Numerical Method}

\subsubsection{Single scattering}

\begin{figure*}
\centering
\includegraphics[width=0.4\textwidth]{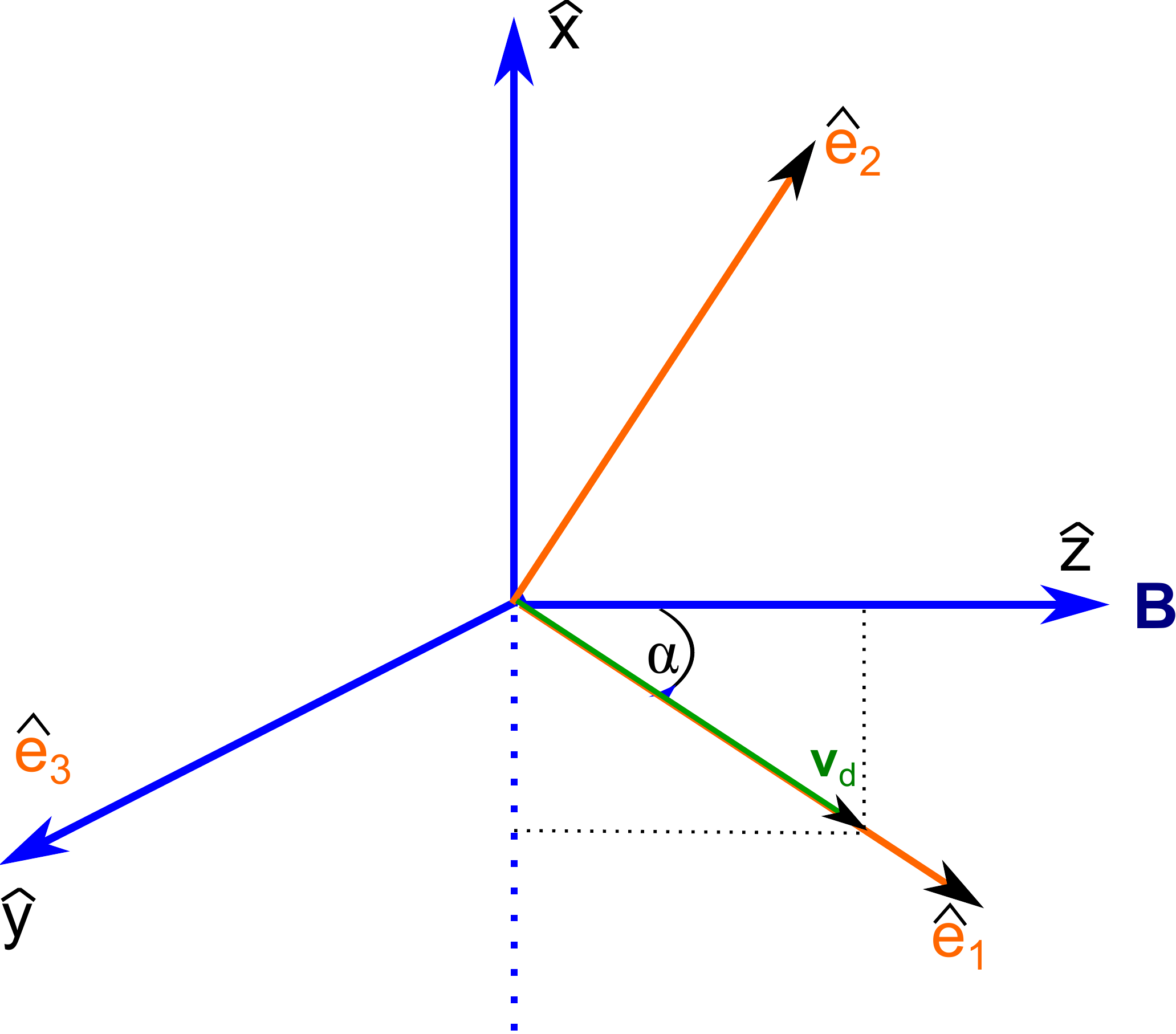}
\includegraphics[width=0.4\textwidth]{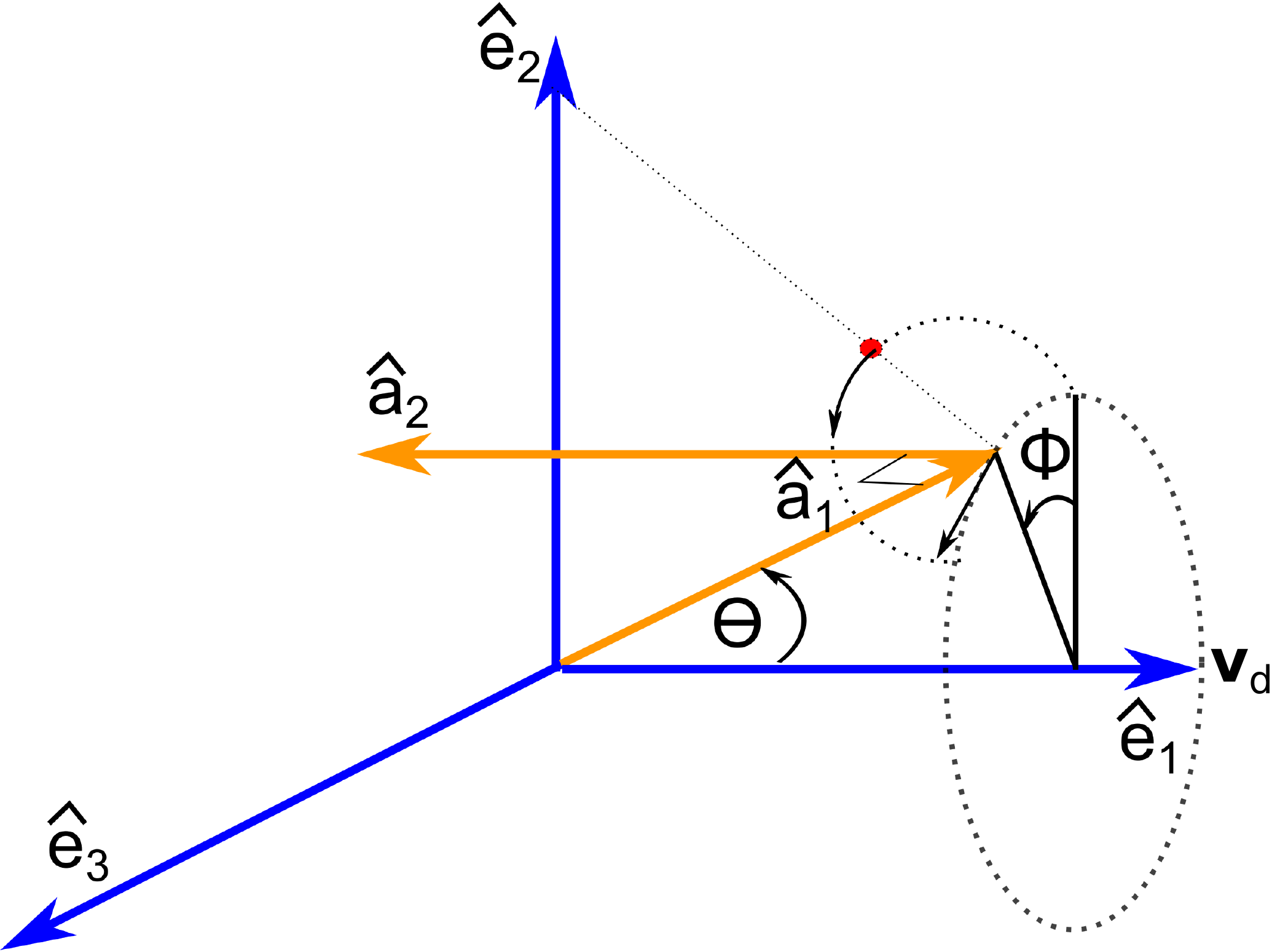}
\caption{Coordinate systems used for calculations. Left panel: The direction of the grain drift in the fixed frame of reference $\ehat_{1}\ehat_{2}\ehat_{3}$ defined by the magnetic field $\Bv$ and $\ehat_{1}\| \bv_{d}$. The grain drift direction $\bv_{d}$ lies in the plane $\xhat\zhat$ and makes an angle $\psi$ with $\Bv$. Right panel: Orientation of grain principal axes $\ahat_{1},\ahat_{2},\ahat_{3}$ in the lab frame of reference.}
\label{fig:labRF}
\end{figure*}

For calculations, we assume the perfect reflection of gas atoms by the grain surface. The torque produced by the single scattering (i.e., reflection) of atoms on an element of surface area $dA$ of cell $ij$ on block $m$ is given by
\begin{equation}
d{\bf \Gamma}_{ij;m}=[ \br_{ij;m}\times \Delta {\bf P}] dA f({\bf s}-{\bf s}_{d}) d^{3} s,\label{ap1}
\end{equation}
where ${\bf s}={\bf v}/v_{\rm th}$, ${\bf s}_{d}=\bv_{d}/v_{\rm th}$, and 
\begin{equation}
f({\bf s}-{\bf s}_{d})=n_{\H} v_{\rm th}({\bf s}-{\bf s}_{d}).{\bf N}_{ij;m} e^{-s^{2}},\rm { ~for~} ({\bf s}-{\bf s}_{d}).{\bf N}_{ij;m}<0,  \label{ap2} 
\end{equation}
is the flux of incoming atoms that can collide with the grain surface element determined by the normal unit vector ${\bf N}_{ij;m}$. 

The angular momentum element $\Delta {\bf P}$ is antiparallel to the normal vector and given by
\begin{equation}
\Delta {\bf P}=-2 m_{\H} {\bf N}_{ij;m}|{\bf v}-{\bf v}_{d}|\cos\gamma,\label{ap3}
\end{equation}
where $\cos\gamma=-({\bf v}-{\bf v}_{d}).{\bf N}_{ij;m}/|{\bf v}-{\bf v}_{d}|$. In the following, we disregard the minor effect of thermal collisions and set $s=0$.

To find the total torque, Equation (\ref{ap1}) is summed over all $ij$ and $m$ that are exposed to the gas flow:
\bea
\bGamma_{\rm MAT} = \sum_{ij,m} d\bGamma_{ij;m},
\ena
where the interior cells of the grain are excluded.

The torque efficiency ${\bf Q}$ is defined as
\bea
\bGamma_{\rm MAT} = \pi a_{\eff}^{2} (n_{\H}v_{\rm th})(m_{\H}v_{\rm th}a_{\eff}){\bf Q},
\ena
where ${\bf Q}$ has the torque components $Q_{ai}$ with $i=1-3$ and depends on $s_{d}$ and the grain shape.

For convenience, the torque efficiency $\bQ$ is first calculated in the coordinate system $\xhat\yhat\zhat$. Then, we evaluate the torque in the $\ahat_{1}\ahat_{2}\ahat_{3}$ frame as $
Q_{a{i}}={\bf Q}.\ahat_{i}$ for $i=1,2,3$. Finally, we obtain the torque components $Q_{ei}$ in the lab system $\ehat_{1}\ehat_{2}\ehat_{3}$ as follows:
\bea
Q_{ei} = {\bf Q}.\ehat_{i}=\sum_{j=1}^{3}Q_{a{j}}\ahat_{j}.\ehat_{i}, {\rm for~} i=1,2,3,
\ena
where the coordinate transformation from $\ahat_{j}$ to $\ehat_{i}$ is described in \cite{Hoang:2008gb}.

\subsubsection{Multiple scattering}
After reflection, gas atoms may continue to hit other facets of the grain, leading to multiple scattering. To calculate the torque by multiple scattering, we trace the trajectory of atoms after each reflection and calculate the torque contribution when the atom hits each surface.

\subsection{Results and Torque Properties}
For our calculations, we consider $N_{1}=N_{2}=50$ patches. Calculations are performed for 37 angles of $\Theta$ ($N_{\Theta}=37$) and 36 angles of $\beta$ ($N_{\beta}=36$), having evenly divided grids. The torque efficiency is averaged over $\beta$ due to fast rotation of the grain around $\ahat_{1}$. Table \ref{tab:para} shows parameters adopted for our study.

\begin{table}
\centering
\caption{Model parameters used for MAT calculations}\label{tab:para}
\begin{tabular}{l l l} \hline\hline\\
Parameter & Meaning & Value \cr
\hline\cr
$n_{\H} (\cm^{-3})$ & Gas density & 20\cr
$T_{\gas}(\K)$ &Gas temperature & 50 \cr
$v_{d}(\cm/\s)$ & Grain drift velocity & 10$^{5}$\cr
$a_{\rm eff}(\mum)$ & Effective grain size & 0.1 \cr
\hline\hline\\
\end{tabular}
\end{table}

\begin{figure*}
\includegraphics[width=0.33\textwidth]{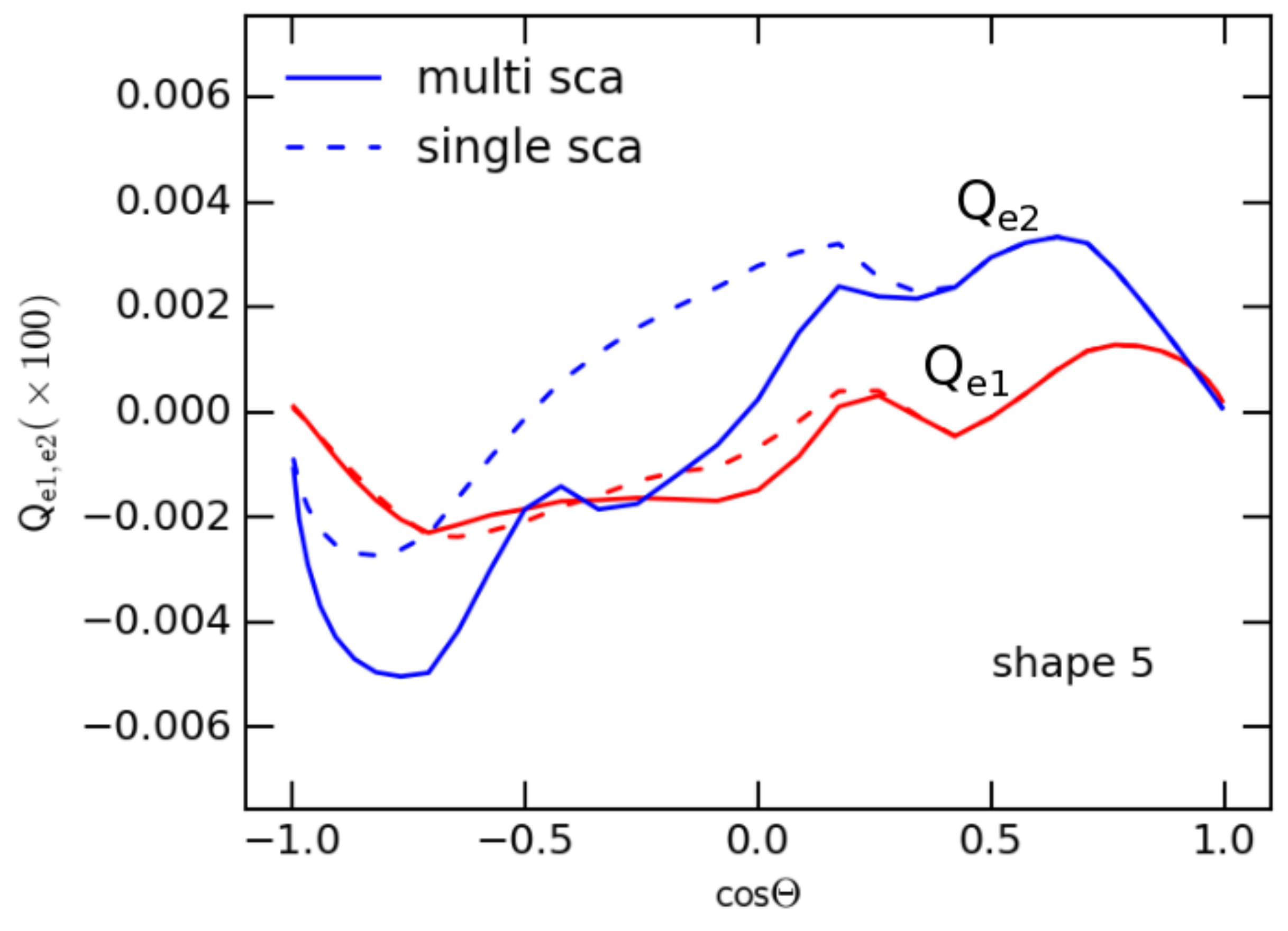}
\includegraphics[width=0.33\textwidth]{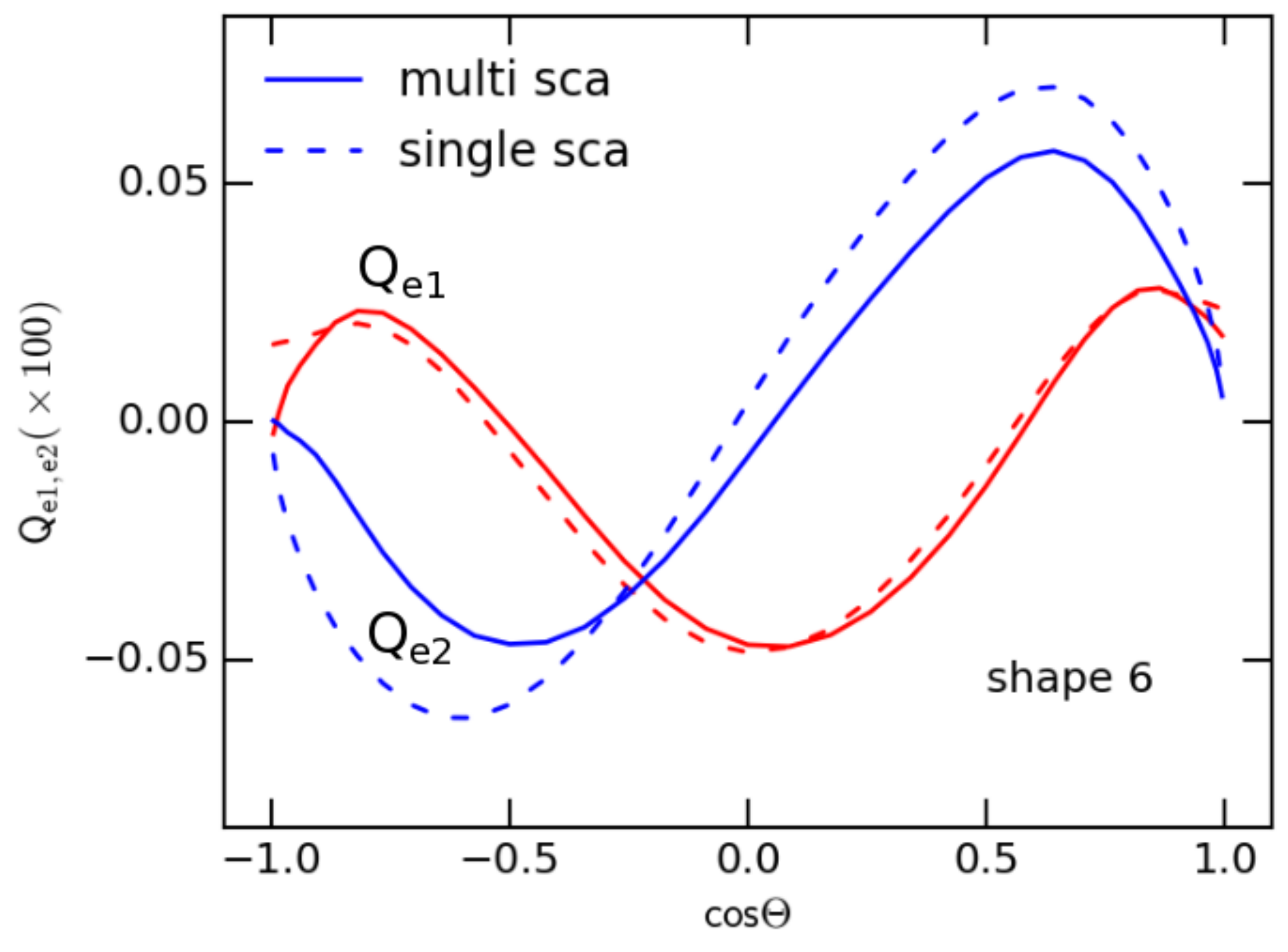}
\includegraphics[width=0.33\textwidth]{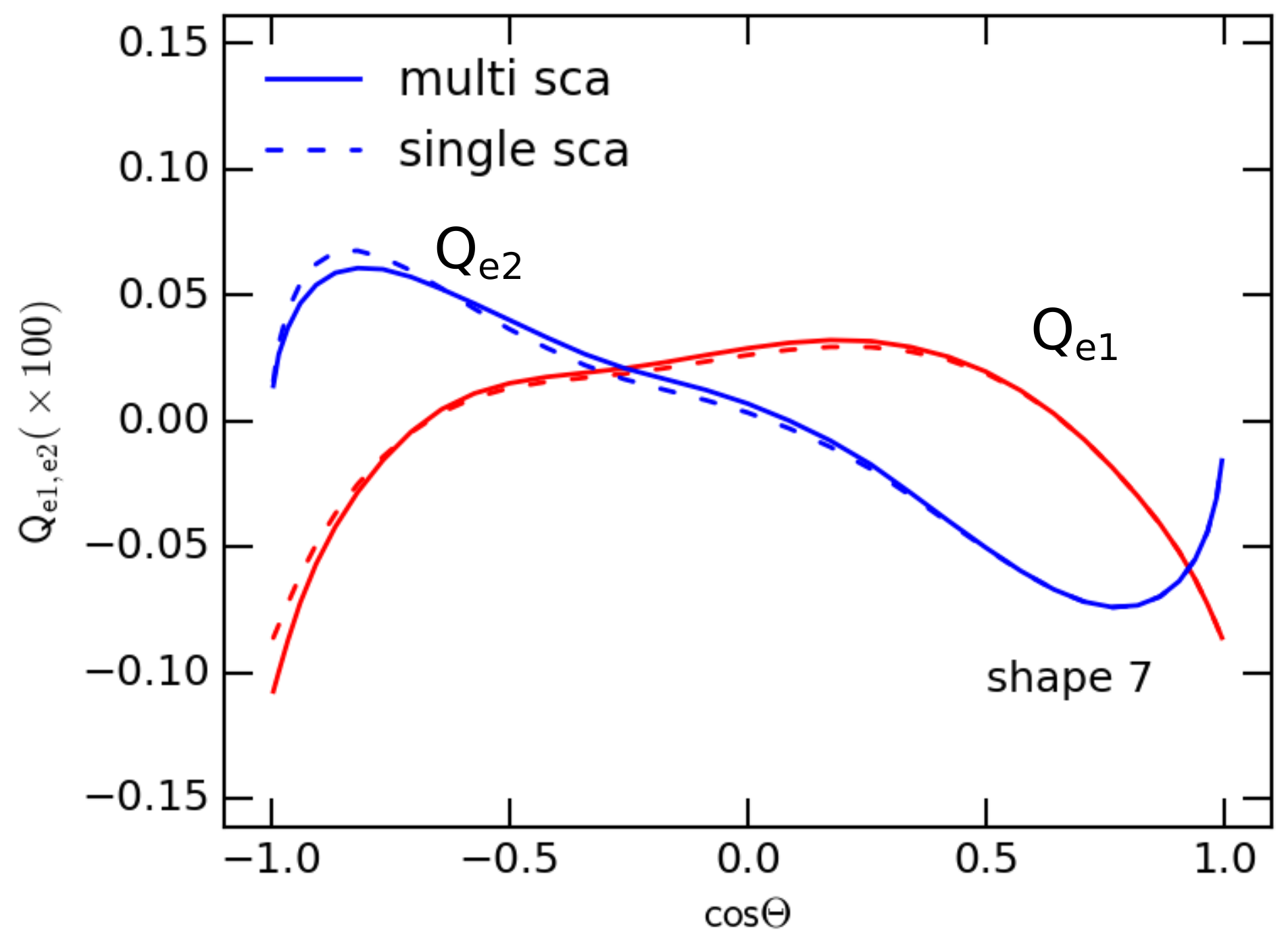}
\includegraphics[width=0.33\textwidth]{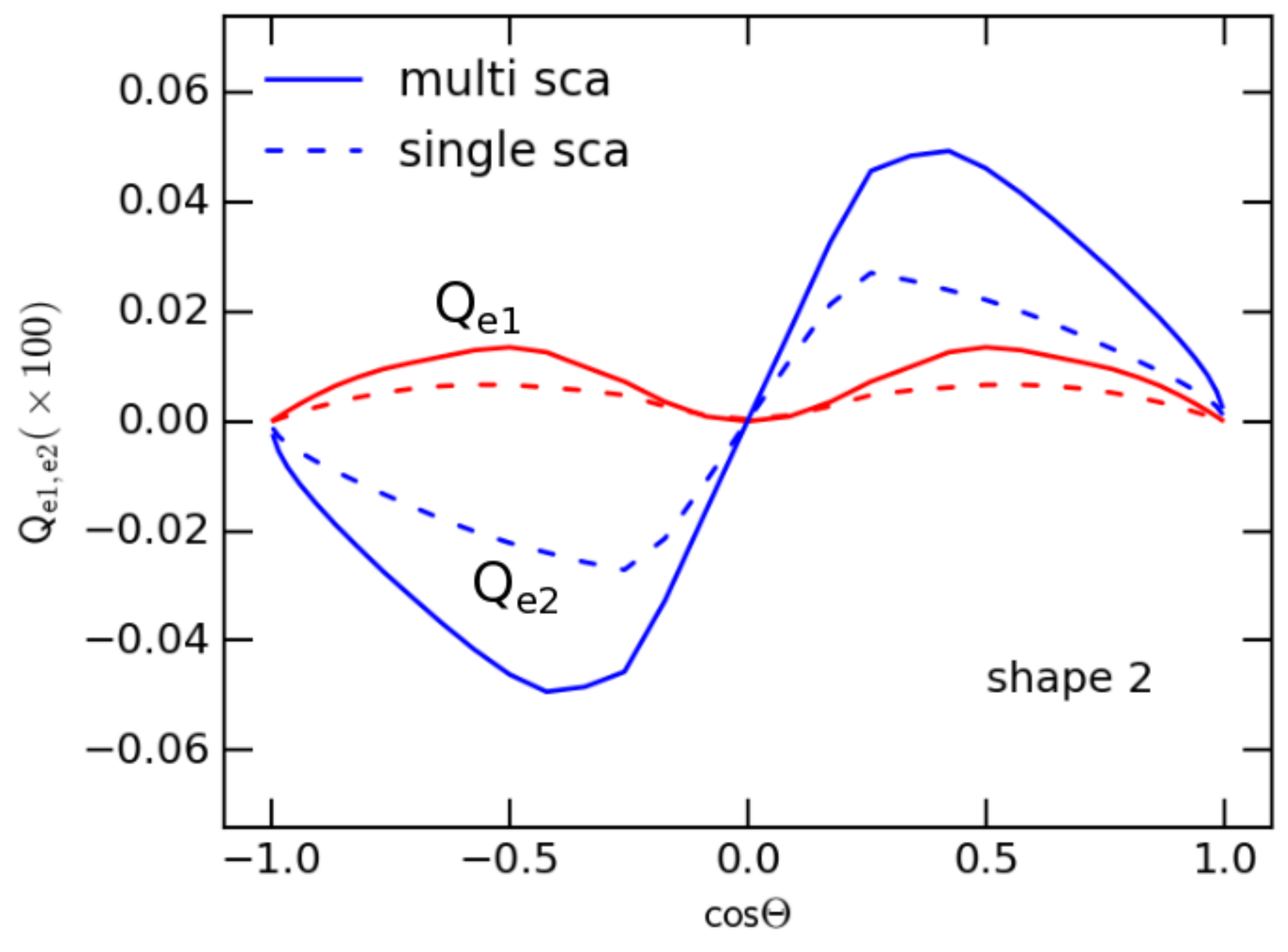}
\includegraphics[width=0.33\textwidth]{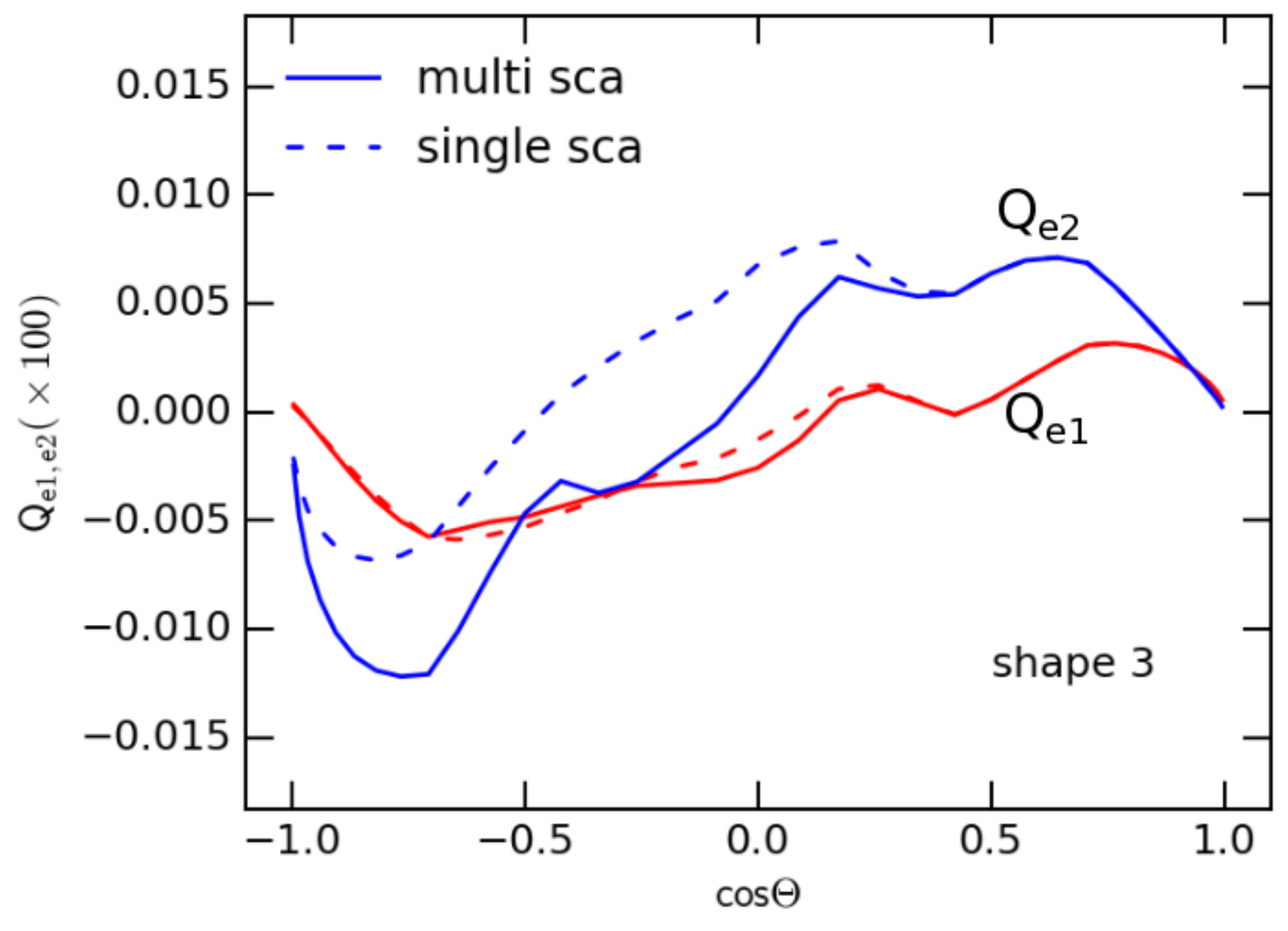}
\includegraphics[width=0.33\textwidth]{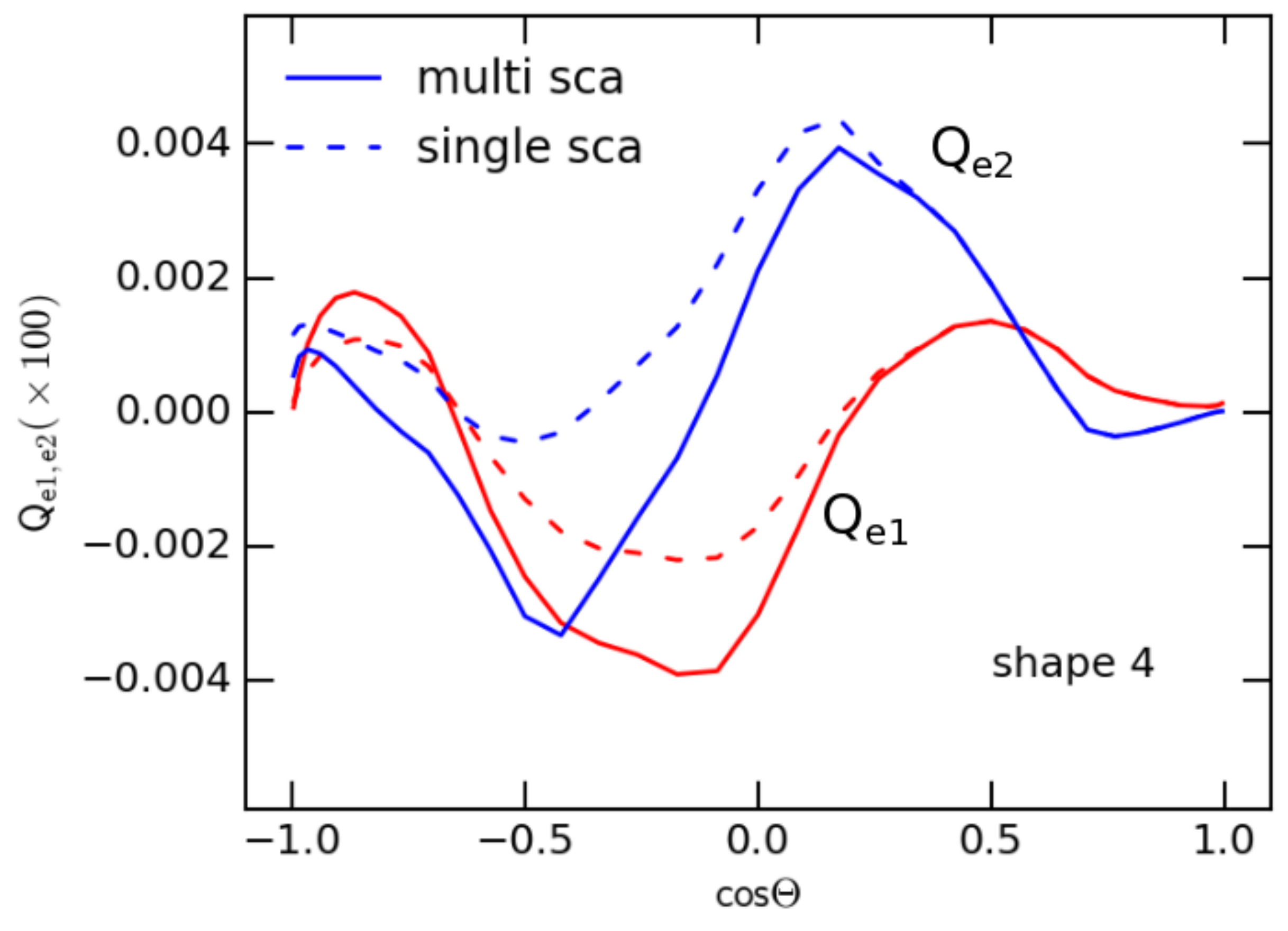}
\includegraphics[width=0.33\textwidth]{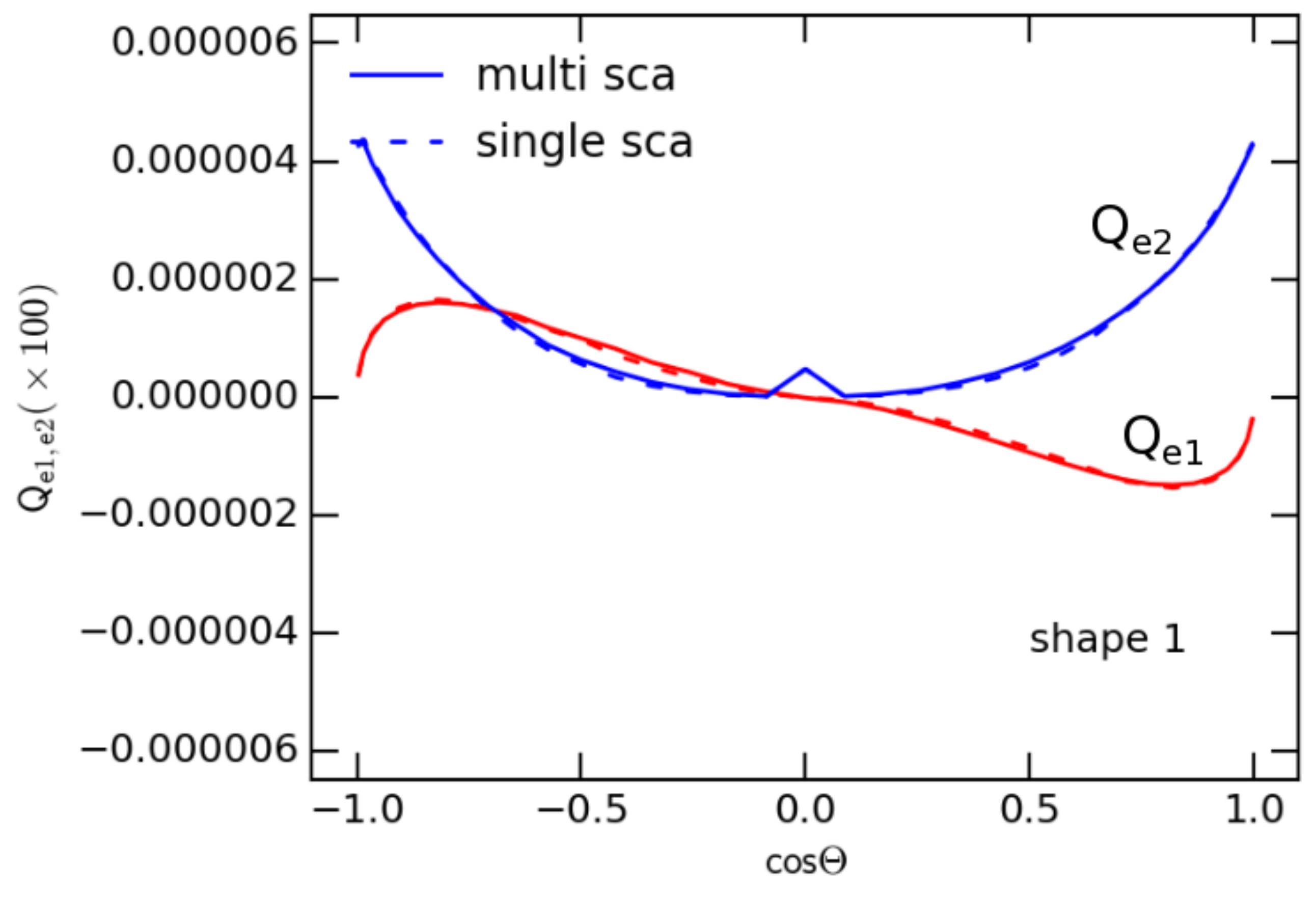}
\caption{MAT components for the different shapes for the cases of single scattering (dashed lines) and multiple scattering (solid lines). Shape 1 has much lower MAT efficiency than other shapes due to its mirror-symmetry geometry, shapes 2, 6, and 7 have strong MATs with generic properties, and shapes 3-5 have weaker MATs. Multiple scattering slightly modifies the MAT.}
\label{fig:MAT}
\end{figure*}

Figure \ref{fig:MAT} shows the values of $Q_{e1}$ and $Q_{e2}$ for the seven different irregular shapes. Since the component $Q_{e3}$ acts mainly to induce precession around the drift direction (LH07a), it is not shown here. The multiple scattering has a noticeable effect on the MAT, but the torques are essentially similar. From Figure \ref{fig:MAT}, we find the following properties of MAT.

First, shape 1 has negligible torques, about $10^{4}$ order of magnitude smaller than the other shapes. This naturally arises from the mirror symmetry of the grain shape combined with fast rotation around the axis of maximum moment of inertia.

Second, shapes 2, 6, and 7 (HIS) experience strong MATs. Moreover, MATs exhibit some basic properties as seen with RATs, having symmetric $Q_{e1}$ and zeros points of $Q_{e2}$ at $\cos\Theta=\pm 1$. Note that shape 6 has right helicity, while 2 and shape 7 have negative helicity, as seen in their RATs (see LH07a). 

Third, shapes 3-5 (WIS) have comparable torques, but an order of magnitude lower than shapes 2, 6 and 7. Such a comparable MAT efficiency is expected from the fact that the geometry of shapes 3-5 is only slightly different from each other (see Figure \ref{fig:shape}). Moreover, since these shapes contain only three blocks on top of the principal plane, the fraction of the gas flow that hits the flat principal plane is substantial, which results in the averaging out of the individual torques.

Finally, the MAT efficiency of shapes 2-5 is negligibly small at $\cos\Theta=1$, i.e., when the direction of grain drift is parallel to the the major axis $\ahat_{1}$ (see Figure \ref{fig:MAT} (panels (c)-(f)). This is due to the fact that for these chosen shapes, the principal axes $\ahat_{j}$ are nearly along the basic unit vectors of the block (see Figure \ref{fig:shape}). Thus, when $\bv_{d}$ is along $\ahat_{1}$, gas atoms bombard a single symmetric surface of the grain at the right angle, producing negligible torques. 

In summary, it is demonstrated that the degree of grain surface irregularity is an important factor of MAT efficiency. 

\section{Grain alignment by mechanical torques}\label{sec:align}
\subsection{Dynamical timescales}
Sticking collisions of gas atoms to the dust grain surface followed by evaporation of thermalized molecules results in the loss of grain angular momentum. The rate of gas damping is given by
\bea
\bGamma_{\gas}=-\frac{\bJ}{\tau_{\gas}},
\ena
where the $\tau_{\gas}$ is the characteristic damping timescale for the rotation along the axis of major inertia $\ahat_{1}$ given by
\bea
\tau_{\gas}&=&\frac{\pi \alpha_{1}\rho a_{\rm eff}}{3\delta n_{\H}(2\pi m_{\H}kT_{\gas})^{1/2}},\nonumber\\
&\simeq & 8.74\times 10^{4}\frac{\alpha_{1}}{\delta}\hat{\rho}a_{-5}\hat{T}^{1/2}\left(\frac{3000\cm^{-3}\K}{n_{\H}T_{\gas}} \right)\yr,\label{eq:tgas}
\ena
where $a_{-5}=a_{\eff}/10^{-5}\cm$, $\hat{\rho}=\rho/3\g\cm^{-3}$, $n_{\H}$ and $\hat{T}=T_{\gas}/100\K$ are gas density and temperature, and $\delta$ is a parameter comparable to $\alpha_{1}$ (DW97). For oblate spheroids with semimajor and minor axes $a$ and $b$, we have $\alpha_{1}=s^{-2/3}$ with $s=b/a<1$.

Rotating paramagnetic grains experience paramagnetic relaxation, leading to the gradual alignment of grains with the magnetic field (\citealt{1951ApJ...114..206D}). The characteristic time of such a paramagnetic relaxation is given by 
\bea
\tau_{m} &=& \frac{I_{1}}{K(\omega)VB^{2}}=\frac{2\rho \alpha_{1}a_{\rm eff}^{2}}{5K(\omega)B^{2}},\nonumber\\
&\simeq & 1.5\times 10^{6}\alpha_{1}\hat{\rho}a_{-5}^{2}\hat{B}^{-2}\hat{K}^{-1} \yr,\label{eq:tau_DG_sup}
\ena
where $V=4\pi a_{\rm eff}^{3}/3$ is the grain volume, $\hat{B}=B/5\mu$G is the normalized magnetic field strength, and $\hat{K}=K(\omega)/10^{-13}\s$ and $K(\omega)=\chi_{2}(\omega)/\omega$ with $\chi_{2}(\omega)$ is the imaginary part of complex magnetic susceptibility of the grain material (see \citealt{2016ApJ...831..159H}).

\subsection{Equation of steady motion}
For convenience, we assume the magnetic field to be in the plane $\ehat_{1}\ehat_{2}$ and makes an angle $\psi$ with $\ehat_{1}$ (also the grain drift direction $\bv_{d}$; see Figure \ref{fig:labRF}). 

To capture the essence of MAT alignment, we assume that the axis of maximum inertia moment $\ahat_{1}$ is coupled to the angular momentum (DW97; LH07a). The effect of thermal fluctuations within the grain that induces the fluctuations of $\ahat_{1}$ with $\bJ$ \citep{1997ApJ...484..230L} is disregarded. We also disregard stochastic effect by gas random collisions (\citealt{Hoang:2008gb}) and consider the steady rotation dynamics.

MAT alignment of grains is studied by following the evolution of the grain angular momentum subject to MATs, gas damping torque, and magnetic torque. The equation of motion is then described by
\bea
\frac{d\bJ}{dt} = {\bGamma}_{\rm MAT} + \bGamma_{\gas} + \bGamma_{m},
\ena
where $\bGamma_{m}$ is the magnetic damping torque due to paramagnetic relaxation. Here we also disregard the damping due to infrared emission which is subdominant for large grains \citep{Hoang:2010jy}. 

\begin{figure}
\includegraphics[width=0.45\textwidth]{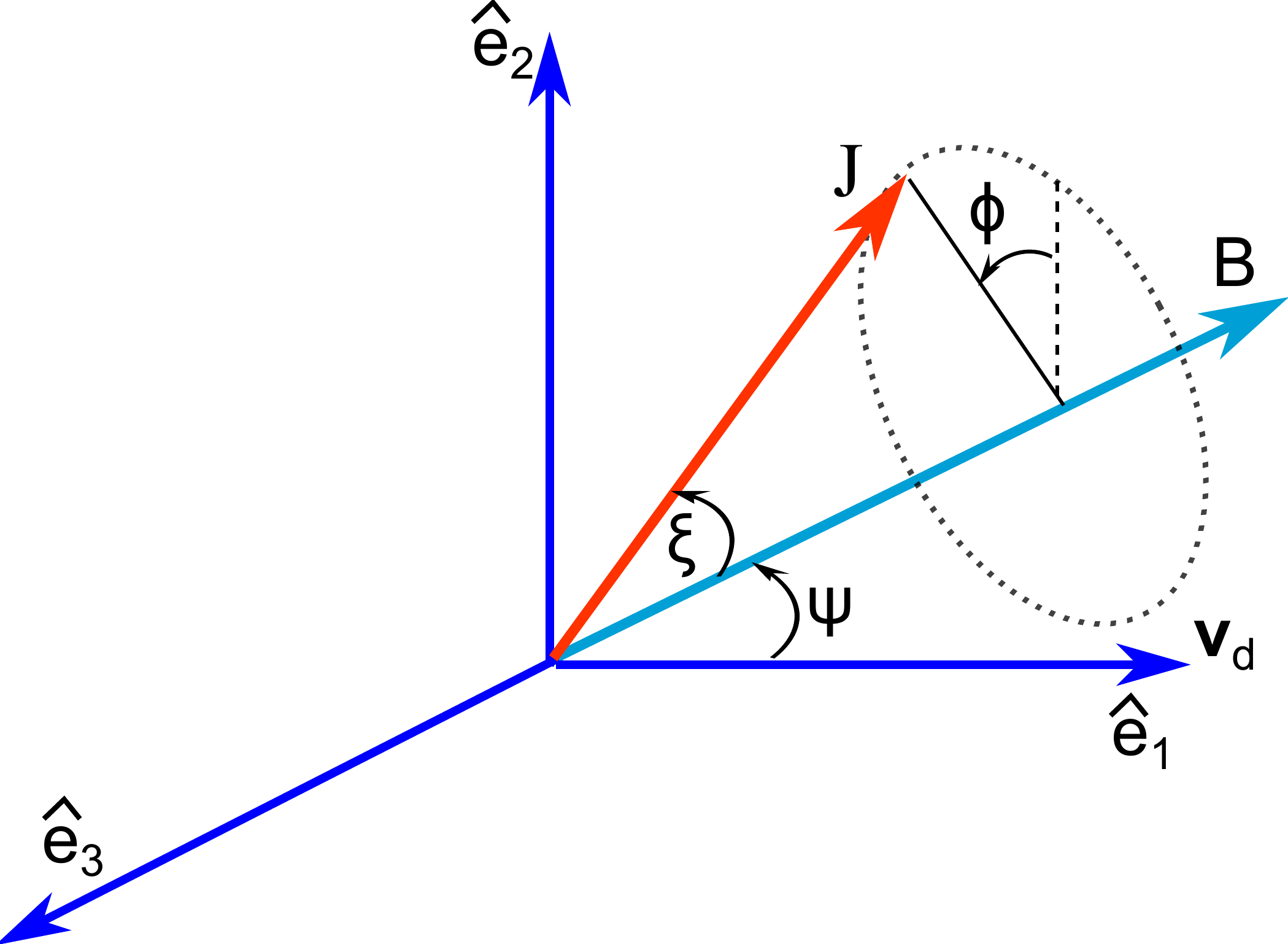}
\caption{Alignment coordinate system $\ehat_{1}\ehat_{2}\ehat_{3}$. The axis $\ehat_{1}$ is chosen to be parallel to $\bv_{d}$, the magnetic field lies in $\ehat_{1}\ehat_{2}$ and makes an angle $\psi$ with respect to $\bv_{d}$. The orientation of $\bJ$ is described by $\xi$ and $\phi$.}
\label{fig:RF_align}
\end{figure}

To study the alignment of the grain angular momentum with respect to the magnetic field, we use the spherical coordinate system ($J,\xi,\phi$) (see Fig. \ref{fig:RF_align}). In this coordinate system, $\bGamma_{m}=-(\hat{\xi}\sin\xi\cos\xi+\ahat_{1}\sin\xi^{2})I_{1}\omega/\tau_{m}$. Thus, the equations of motion for these variables become
\bea
\frac{dJ}{dt}&=&\pi a_{\rm eff}^{3} n_{\H}m_{\H}v_{\rm th}^{2}H-\frac{J}{\tau_{\gas}} - \frac{\sin^{2}\xi}{\tau_{m}}\label{eq:dJdt}\\
\frac{d\xi}{dt}&=&\frac{\pi a_{\rm eff}^{3} n_{\H}m_{\H}v_{\rm th}^{2}}{J}F - \frac{\sin\xi\cos\xi}{\tau_{m}} ,\label{eq:dxidt}\\ 
\frac{\sin\xi d\phi}{dt} &= &\frac{\pi a_{\rm eff}^{3}n_{\H}m_{\H}v_{\rm th}^{2}}{J} G - \frac{2\pi}{\tau_{\rm Lar}},
\ena 
where $F, H$ and $G$ are the aligning, spin-up and precessing torque components of MATs which are the functions of the angles $\xi,\psi,\phi$ (see Appendix \ref{apd:FGH}), and $\tau_{\rm Lar}$ is the Larmor precession timescale around the magnetic field. 

The Larmor precession is usually fast compared to the main dynamical timescales, except in the very dense conditions such as protoplanetary disks (\citealt{2016ApJ...831..159H}). Thus, the averaging over $\phi$ angle is carried out to simplify the equations of motion.  As a result, the equations of motion for $J'=J/I_{1}\omega_{T}$ and $t'=t/\tau_{\gas}$ become:
\bea
\frac{d\xi}{dt'}&=&\frac{M\langle F(\xi, \psi)\rangle}{J'} - \delta_{m}\sin\xi\cos\xi ,\label{eq:dxidt}\\ 
\frac{dJ'}{dt'}&=&M \langle H(\xi,\psi)\rangle-J'\left(1 + \delta_{m}\sin^{2}\xi\right),\label{eq:dJdt}
\ena 
where $\omega_{T}=(2kT_{\gas}/I_{1})^{1/2}$, $\langle F(\xi,\psi)\rangle$ and $\langle H(\xi,\psi)\rangle$ are aligning and spin-up torque components averaged over the Larmor precession (see Appendix \ref{apd:FGH}), and
\bea
M=\frac{n_{\H}m_{\H}v_{\rm th}^{2}\pi a_{\eff}^{3}\tau_{\gas}}{I_{1}\omega_{T}}.
\ena

For the following calculations, we consider MATs from the multiple scattering regime, and the physical parameters for the cold neutral medium (CNM) ($n_{\H}=30\cm^{-3}, T_{\gas}=100\K$) are adopted.

\begin{figure*}
\includegraphics[width=0.33\textwidth]{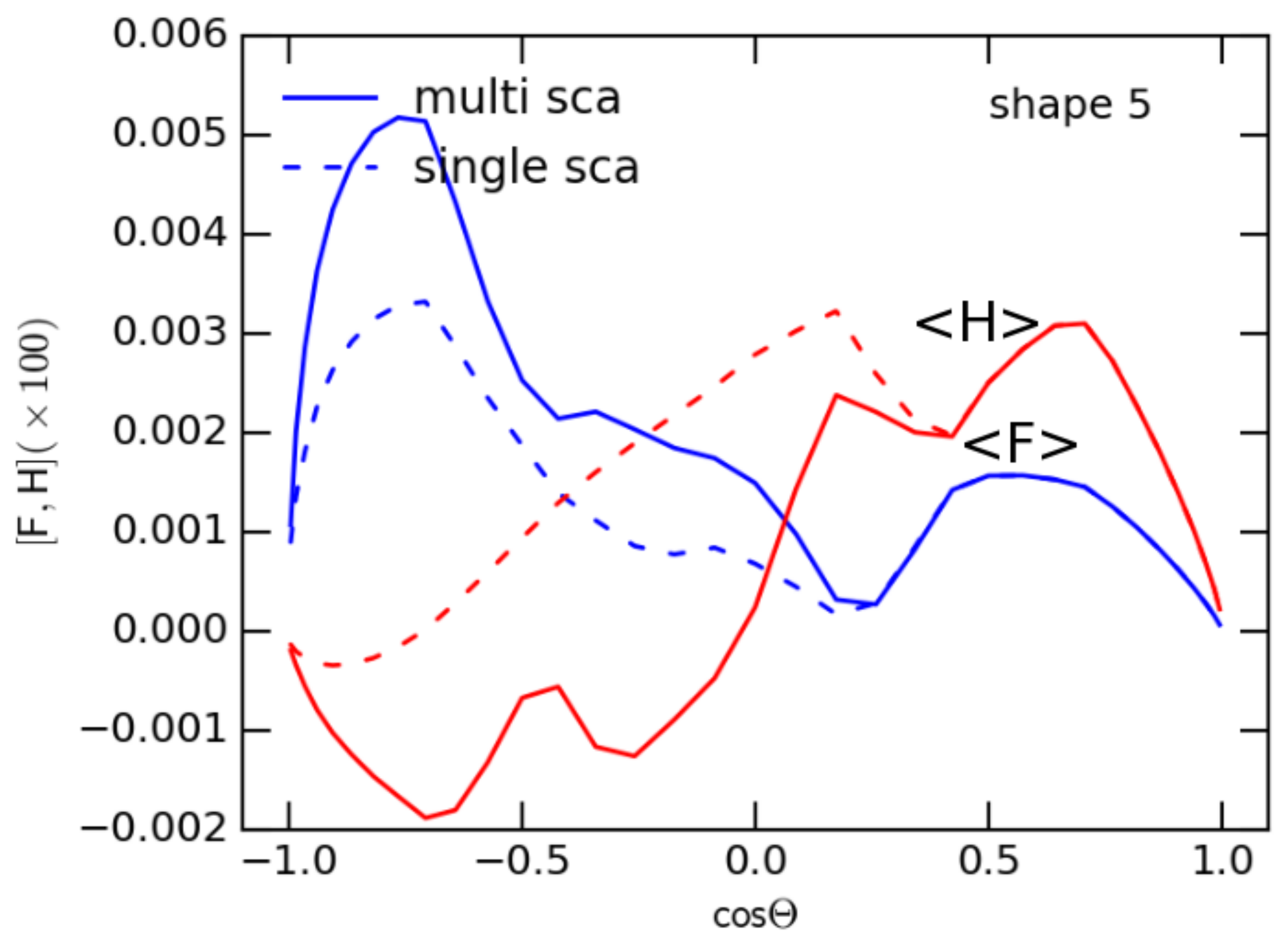}
\includegraphics[width=0.33\textwidth]{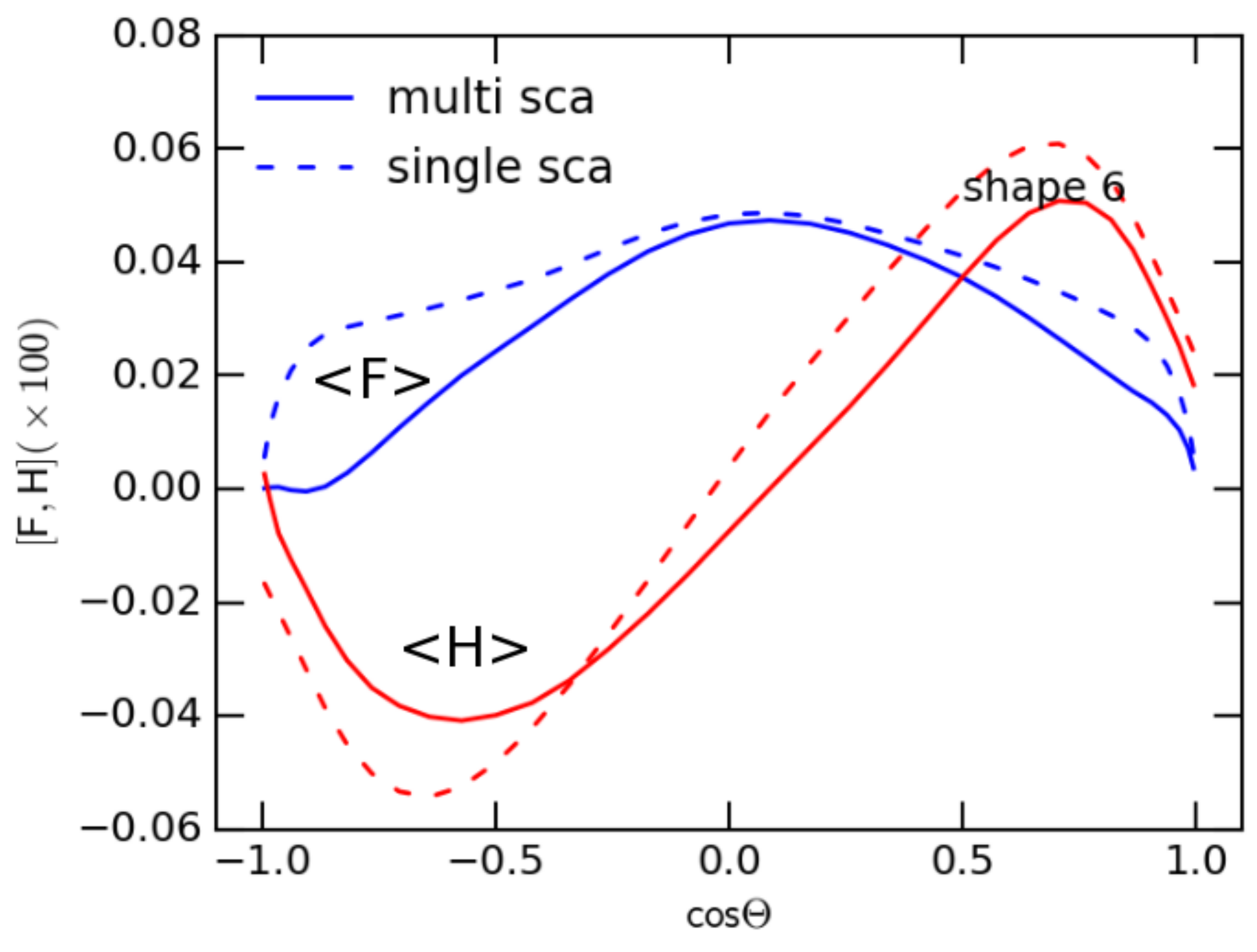}
\includegraphics[width=0.33\textwidth]{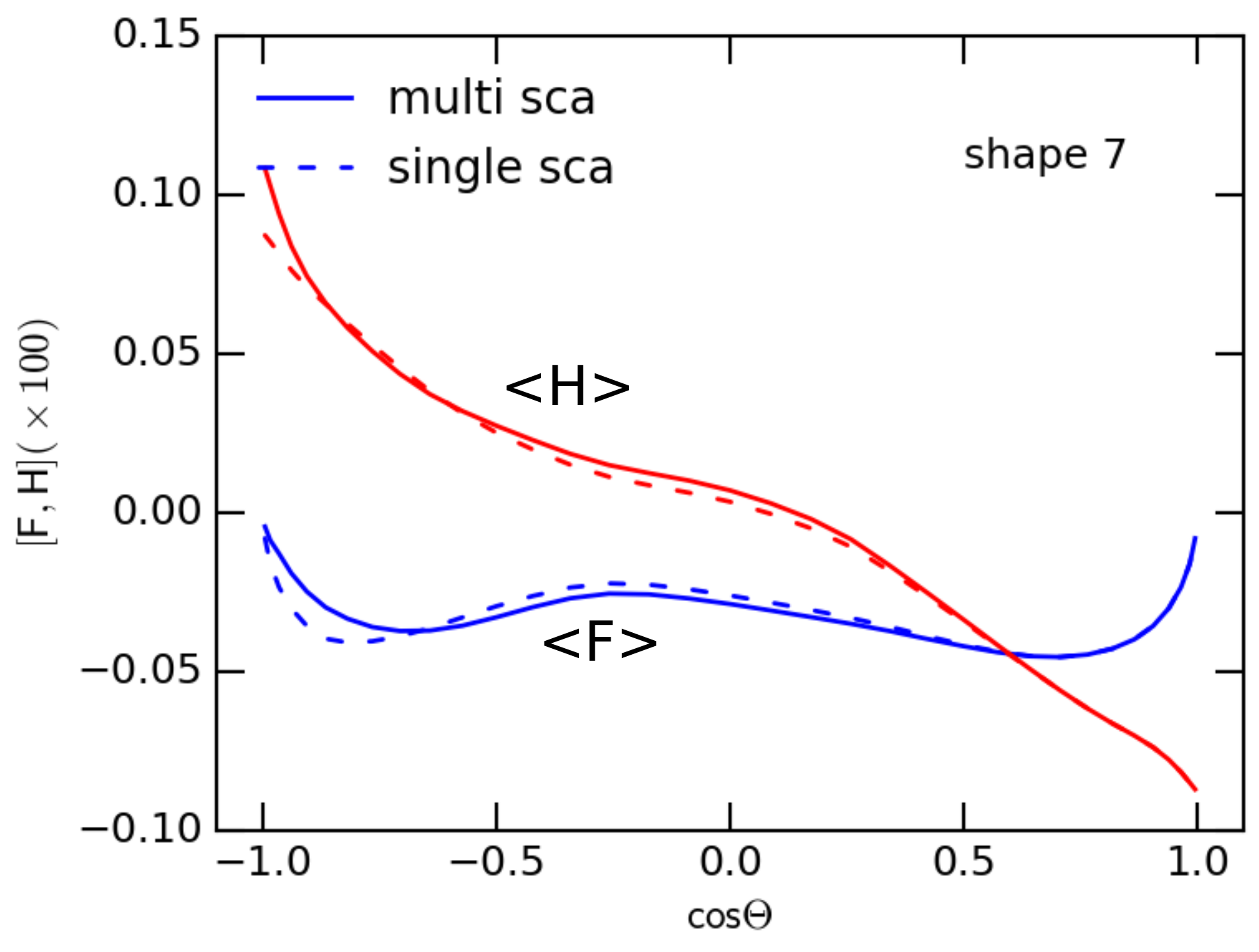}
\includegraphics[width=0.33\textwidth]{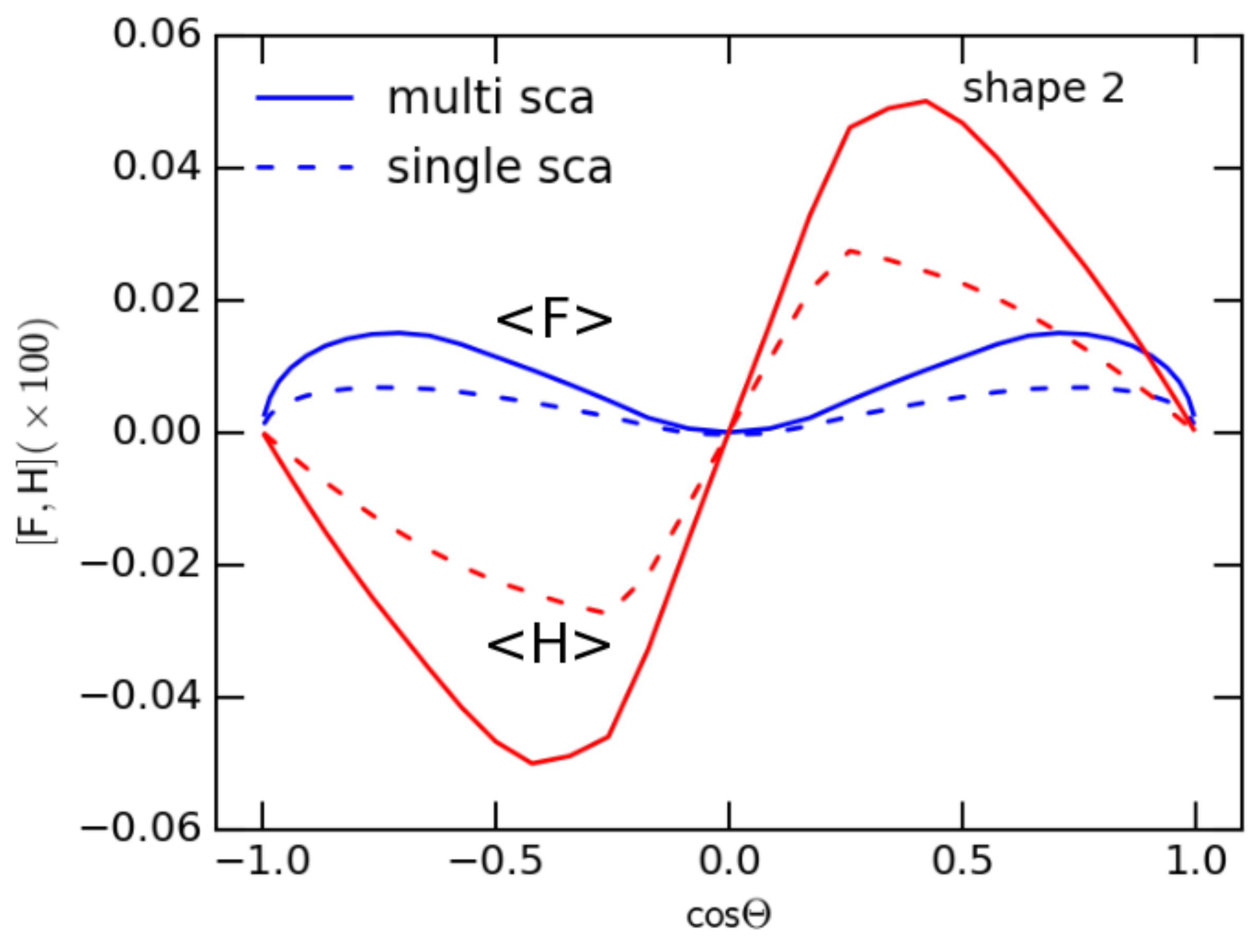}
\includegraphics[width=0.33\textwidth]{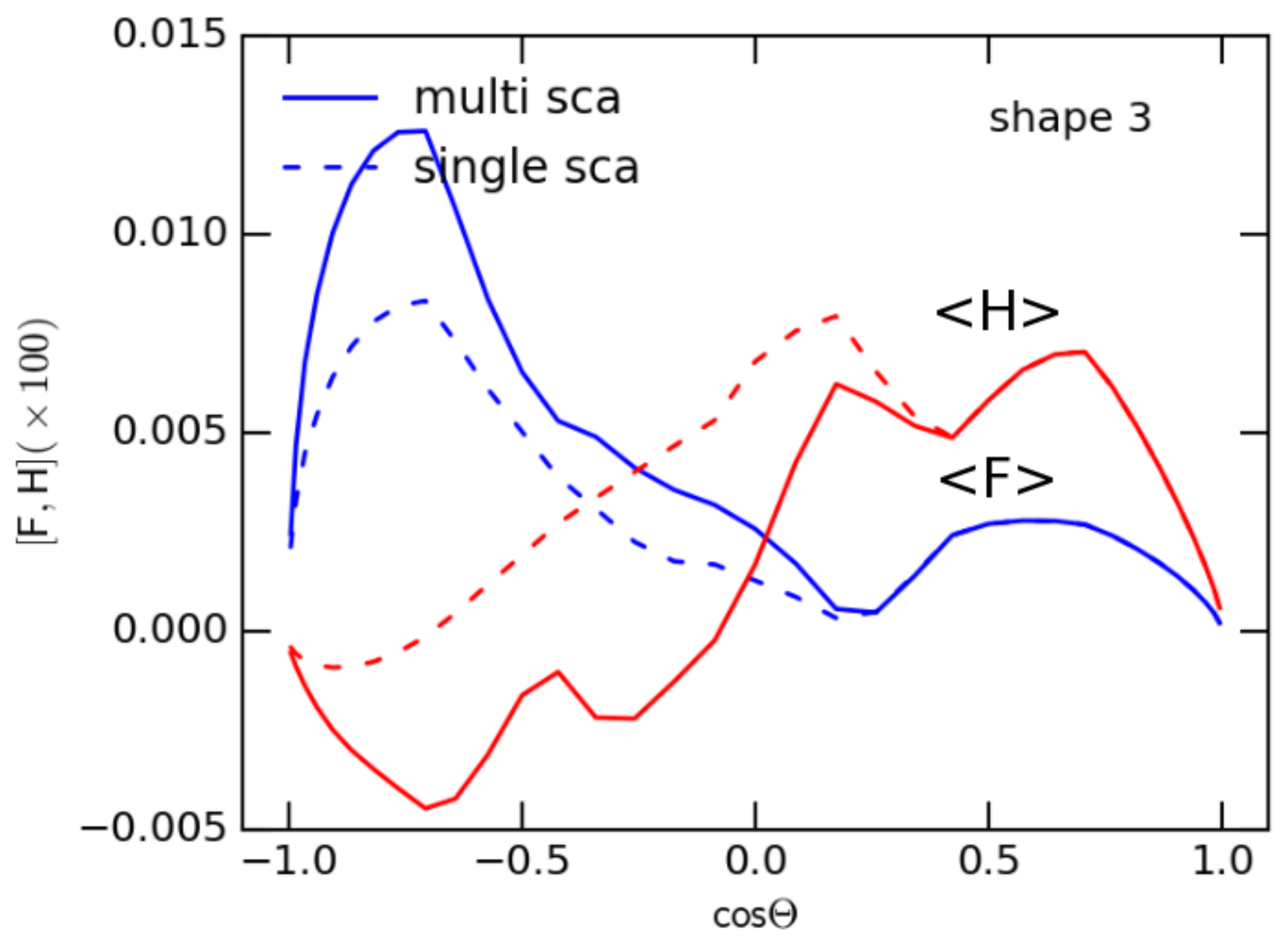}
\includegraphics[width=0.33\textwidth]{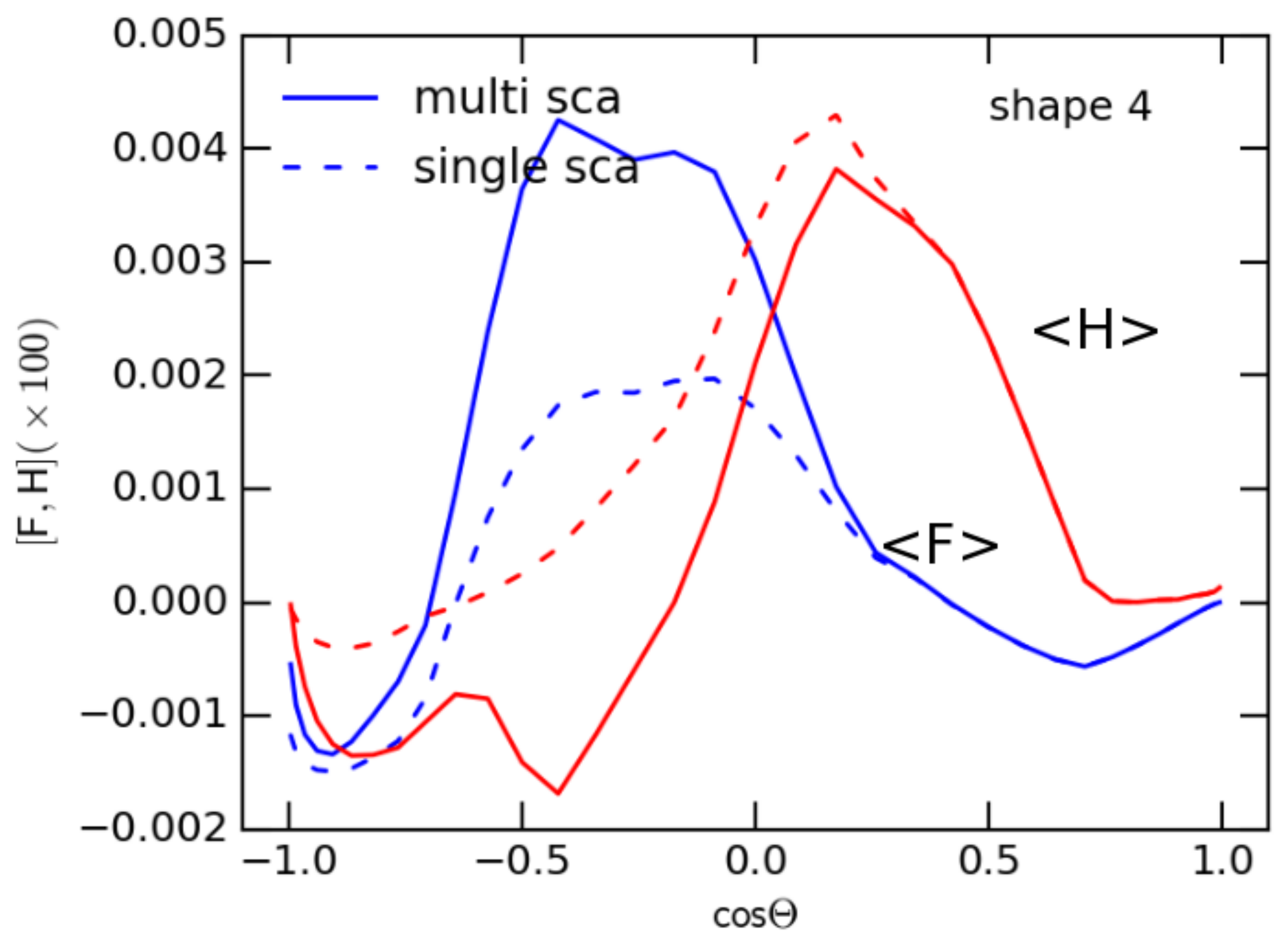}
\caption{The spin-up and aligning torque components for $\psi=0^{\circ}$. Both single scattering (solid lines) and multiple scattering (dashed lines) are considered. The different shapes are shown.}
\label{fig:FH}
\end{figure*}

Figure \ref{fig:FH} shows the spin-up $\langle H\rangle $ and aligning $\langle F\rangle$ torque components for the different shapes when the drift velocity is parallel to the ambient magnetic field. Shapes 3 and 5 have similar forms of $\langle F\rangle$ and $\langle H\rangle$ where zeros of $\langle F\rangle $ (i.e., stationary points) occur at $\sin\Theta=0$.

\subsection{Suprathermal rotation by MATs}
To see whether MATs can spin up grains to suprathermal rotation, we calculate the angular momentum at the stationary point $\sin\xi=0$, as follows (see Equation \ref{eq:dJdt}):
\bea
\frac{J_{\max}^{\rm MAT}(\psi)}{I_{1}\omega_{T}}=M\langle H(\xi=0,\psi)\rangle,\label{eq:Jmax}
\ena
which corresponds to the maximum angular momentum spun-up by MATs when $\bJ$ is parallel to the magnetic field. We compute $J_{\max}^{\rm MAT}$ for the different grain sizes and the drift direction $\psi$. 

\begin{figure*}
\centering
\includegraphics[width=0.45\textwidth]{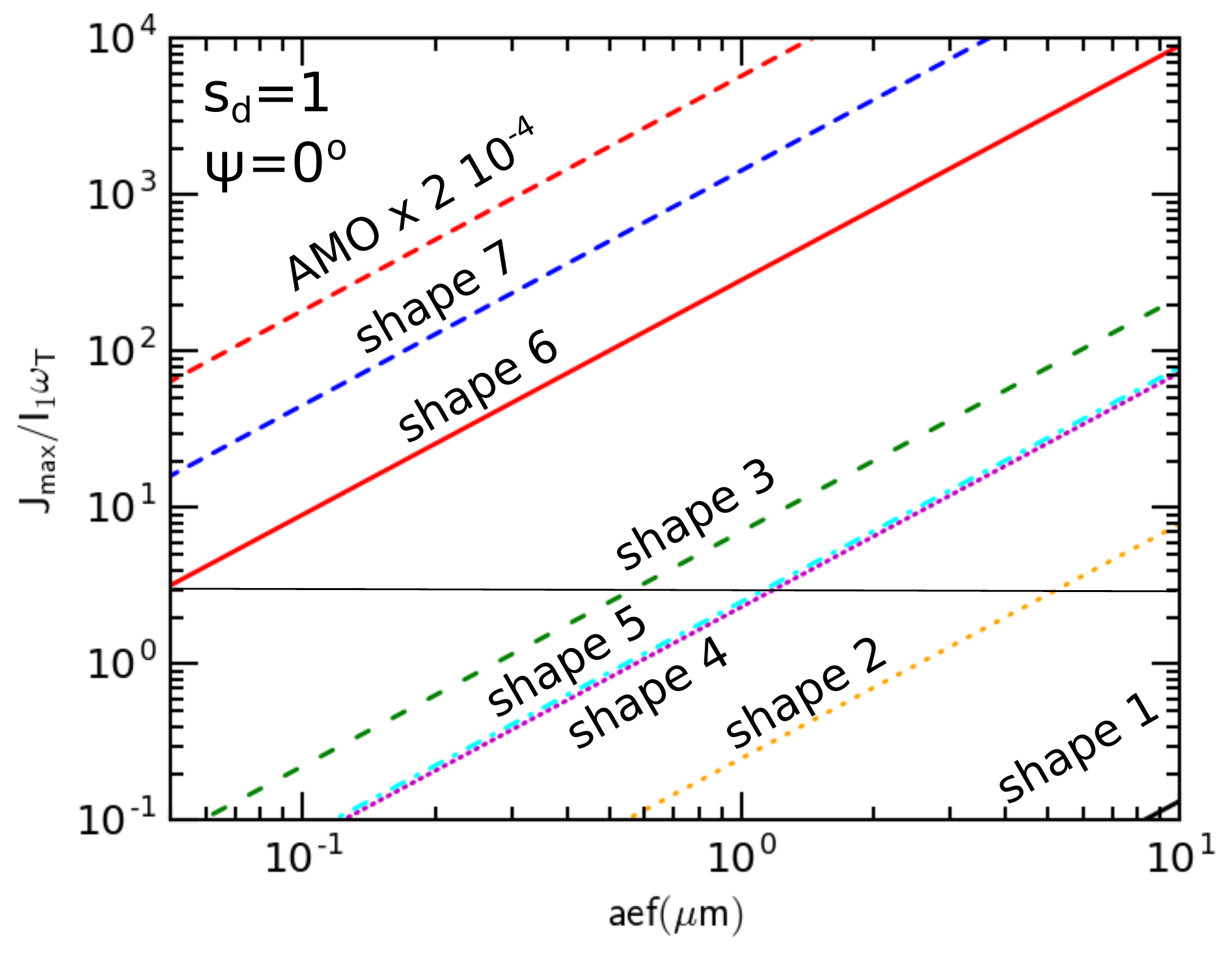}
\includegraphics[width=0.45\textwidth]{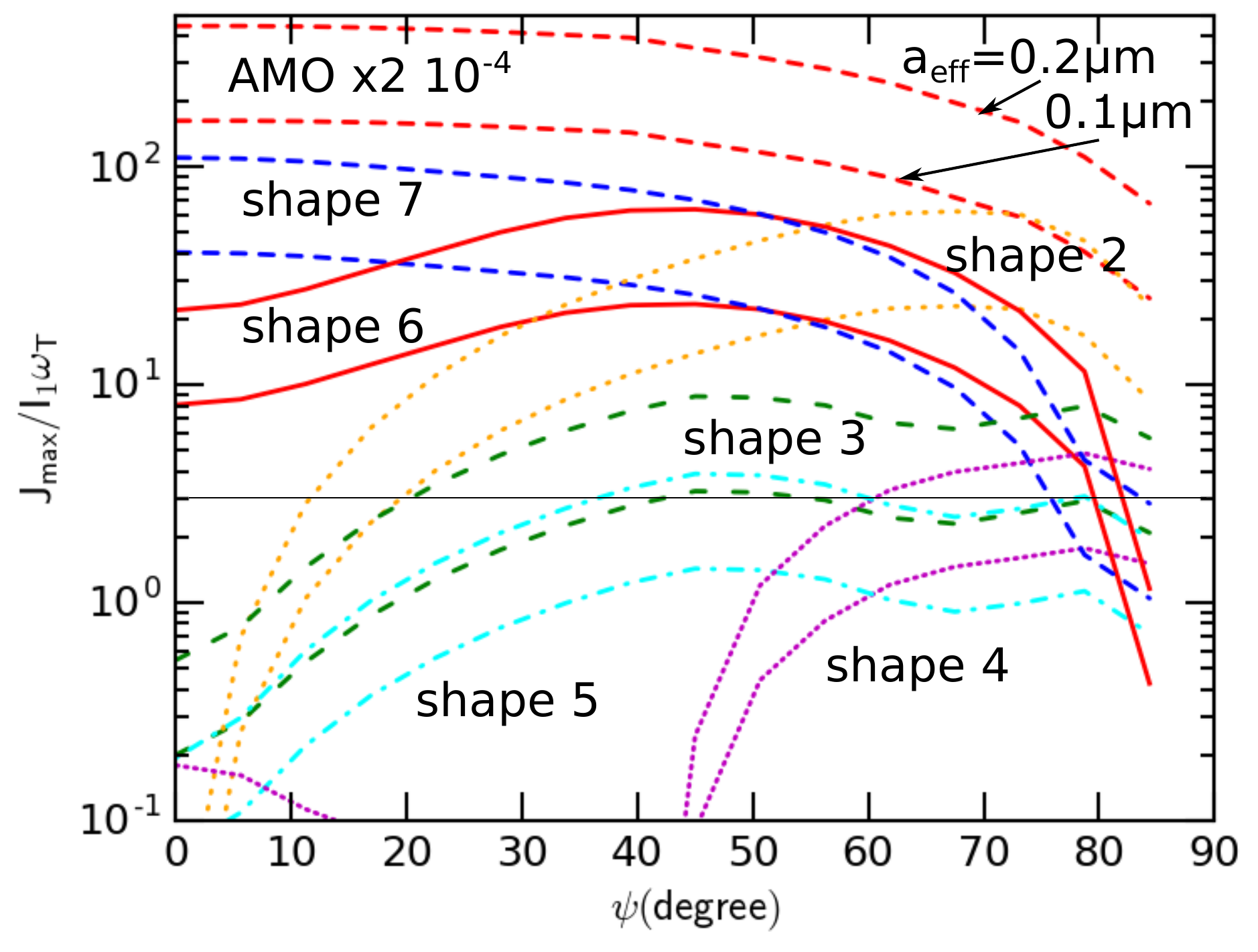}
\caption{Left panel: Maximum angular momentum $J_{\max}$ spun up by MATs from AMO and irregular grains vs. grain size for the drift angle $\psi=0^{\circ}$. Right panel: $J_{\max}$ vs. $\psi$ computed for $a_{\eff}=0.1\mum$ and $0.2\mum$. The horizontal line marks the suprathermal rotation limit of $3I_{1}\omega_{T}$. The results for $s_{d}=1$ are shown.}
\label{fig:Jmax}
\end{figure*}

Figure \ref{fig:Jmax} (left panel) shows $J_{\max}^{\rm MAT}$ as a function of the grain size for the different grain shapes and $s_{d}=1$. {Results predicted by AMO where its magnitude is multiplied by a factor of $2\times 10^{-4}$ is also shown in red solid line for comparison.} 
Since the MAT efficiency is linearly proportional to $s_{d}$, the value of $J_{\max}^{\rm MAT}$ for an arbitrary $s_{d}$ is easily evaluated. The value of $J_{\max}^{\rm MAT}$ varies significantly with the grain shape, with largest torques for shape 7 and smallest torques for shape 1 of mirror symmetry as expected. The value of $J_{\max}^{\rm MAT}$ scales as $a_{\eff}$.

Figure \ref{fig:Jmax} (right panel) shows the value of $J_{\max}^{\rm MAT}$ as a function of $\psi$ for the different shapes and two typical grain sizes. For shapes 6 and 7, $J_{\max}^{\rm MAT}$ tends to decrease with increasing $\psi$ and reach minimum at $\psi=90^{\circ}$. In contrast, for shapes 2-5, $J_{\max}^{\rm MAT}$ is minimum at $\psi=0^{\circ}$ and tends to increase with increasing $\psi$ up to $\psi\sim 50^{\circ}$. Thus, shapes 2-5 can still be driven to suprathermal rotation for large drift angles (see Figure \ref{fig:Jmax}). 

\begin{figure*}
\centering
\includegraphics[width=0.45\textwidth]{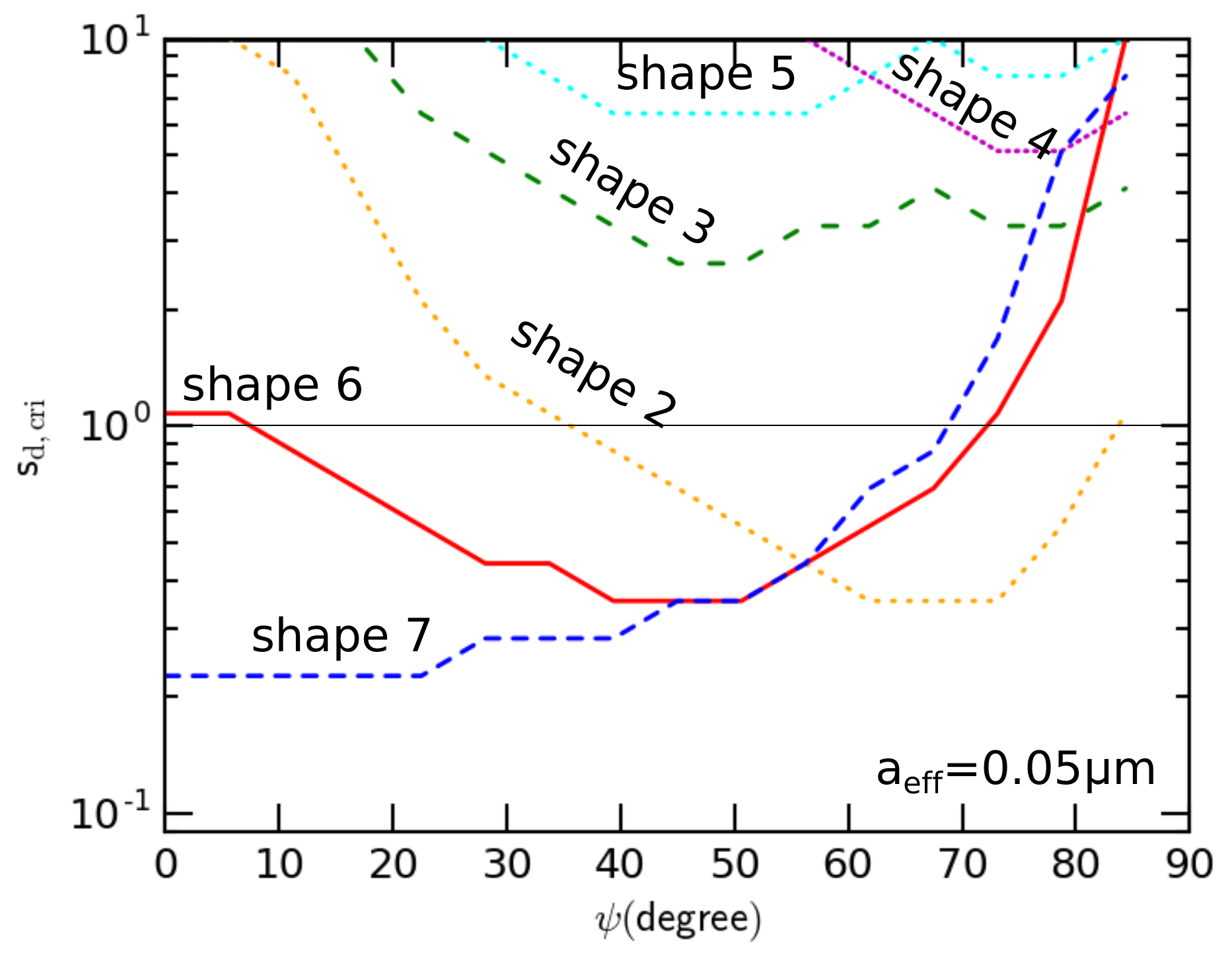}
\includegraphics[width=0.45\textwidth]{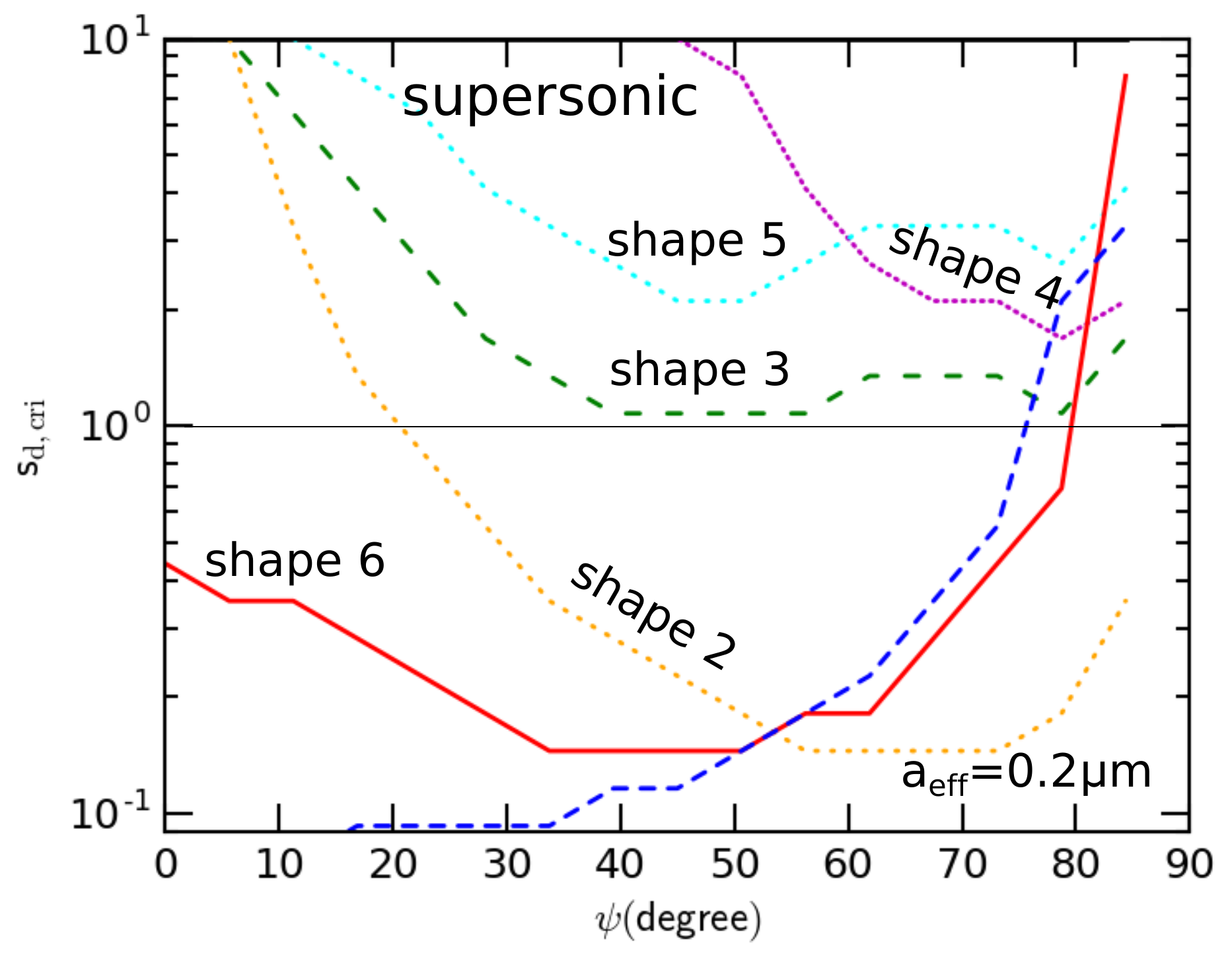}
\caption{Critical drift velocity for the suprathermal rotation vs. the drift angle for $a_{\eff}=0.05\mum$ (left panel) and $a_{\eff}=0.1\mum$ (right panel). Grains of shapes 2, 6 and 7 can be suprathermal rotation at subsonic drift, i.e., $s_{d}< 1$, while shapes 3-5 require supersonic drift to reach suprathermal rotation.} 
\label{fig:sdsup_psi}
\end{figure*}

To derive the critical value of the drift velocity, $s_{d,\rm cri}$, that a given grain can be driven to suprathermal rotation by MATs, we compute $J_{\max}^{\rm MAT}$ for a wide range of $s_{d}$ and grain sizes. Figure \ref{fig:sdsup_psi} shows the obtained value $s_{d,\rm cri}$ as a function of the drift angle for the two values of $a_{\eff}$. Grains of HIS (shapes 2, 6, and 7) can be driven to suprathermal rotation even with subsonic velocity of $s_{d}\sim 0.1$. In contrast, grains of WIS (shapes 3-5) only achieve suprathermal rotation when grains are moving at supersonic speeds ($s_{d}\ge 1$).

\subsection{Phase trajectory map of MAT alignment} 
In the following, we will study the trajectory phase map of these shapes. The instantaneous orientation of the grain in the lab frame $\ehat_{j}$ for $j=1,2,3$ can be characterized by $J$ and $\xi$. To study MAT alignment, we first solve Equations (\ref{eq:dxidt}) and (\ref{eq:dJdt}) numerically with timestep $dt'={\rm min}[10^{-3}, 0.1/\delta_{m}]$ where $t'=t/\tau_{\gas}$ and $\delta_{m} = \tau_{\gas}/\tau_{m}$. There, we visualize MAT alignment in terms of phase trajectory map $J(t),\xi(t)$. We consider an ensemble of grains with the different initial orientations $\xi$ drawn from a uniform distribution and $J=J_{0}$. We assume that the axis $\ahat_{1}$ is parallel to $\bJ$ and shows the results for this positive flipping state. Grains may be in the negative flipping state of $\ahat_{1}$ anti-parallel to $\bJ$. However, when the thermal flipping is taken into account, the trajectory maps of the two flipping states are identical (\citealt{Hoang:2008gb}; \citealt{2009ApJ...695.1457H}), thus we show the maps for the positive flipping case only. Shape 1 has very small MATs, thus, it is not of interest to study the trajectory map for this shape.

\subsubsection{Ordinary paramagnetic grains}
\begin{figure*}
\includegraphics[width=0.33\textwidth]{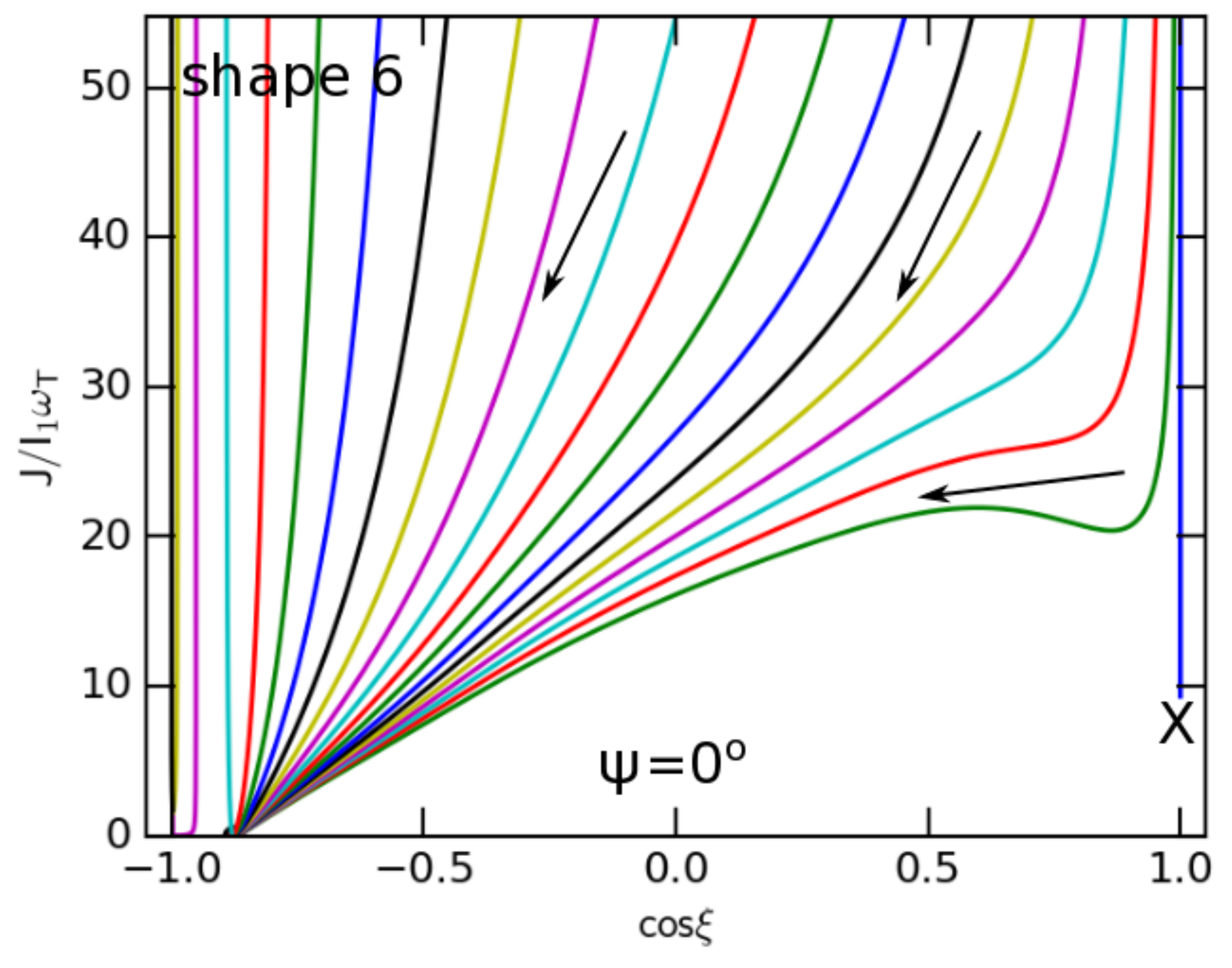}
\includegraphics[width=0.33\textwidth]{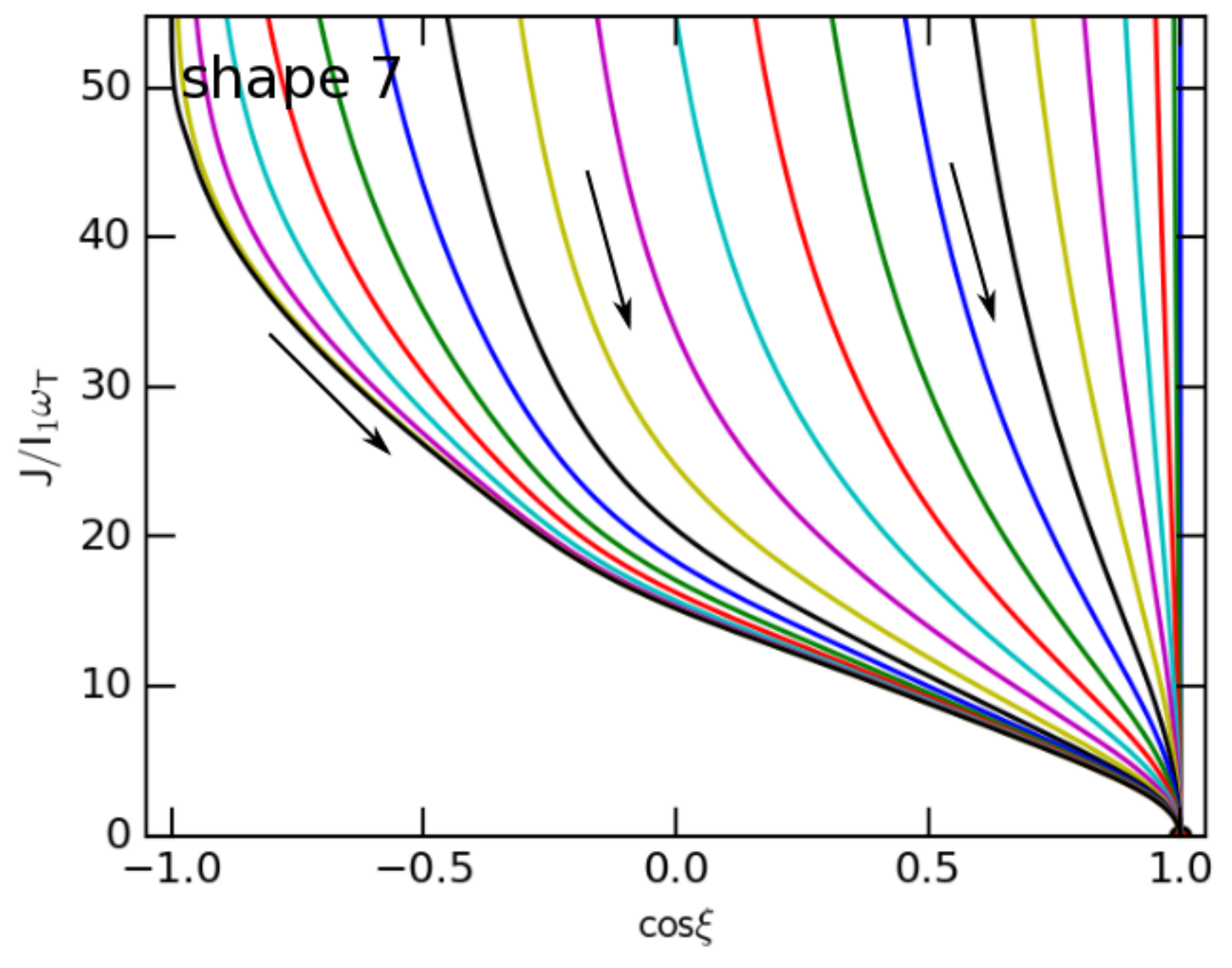}
\includegraphics[width=0.33\textwidth]{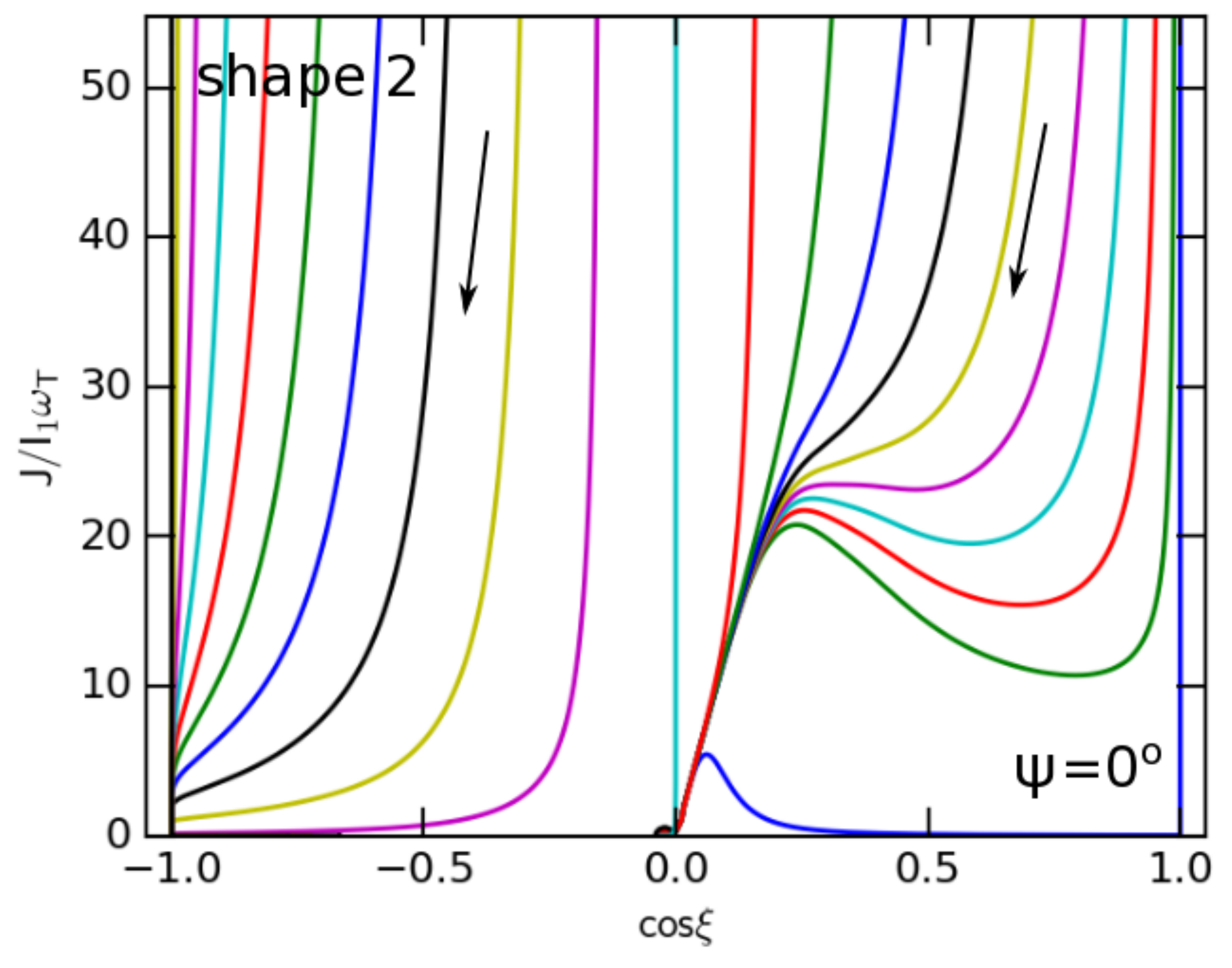}
\includegraphics[width=0.33\textwidth]{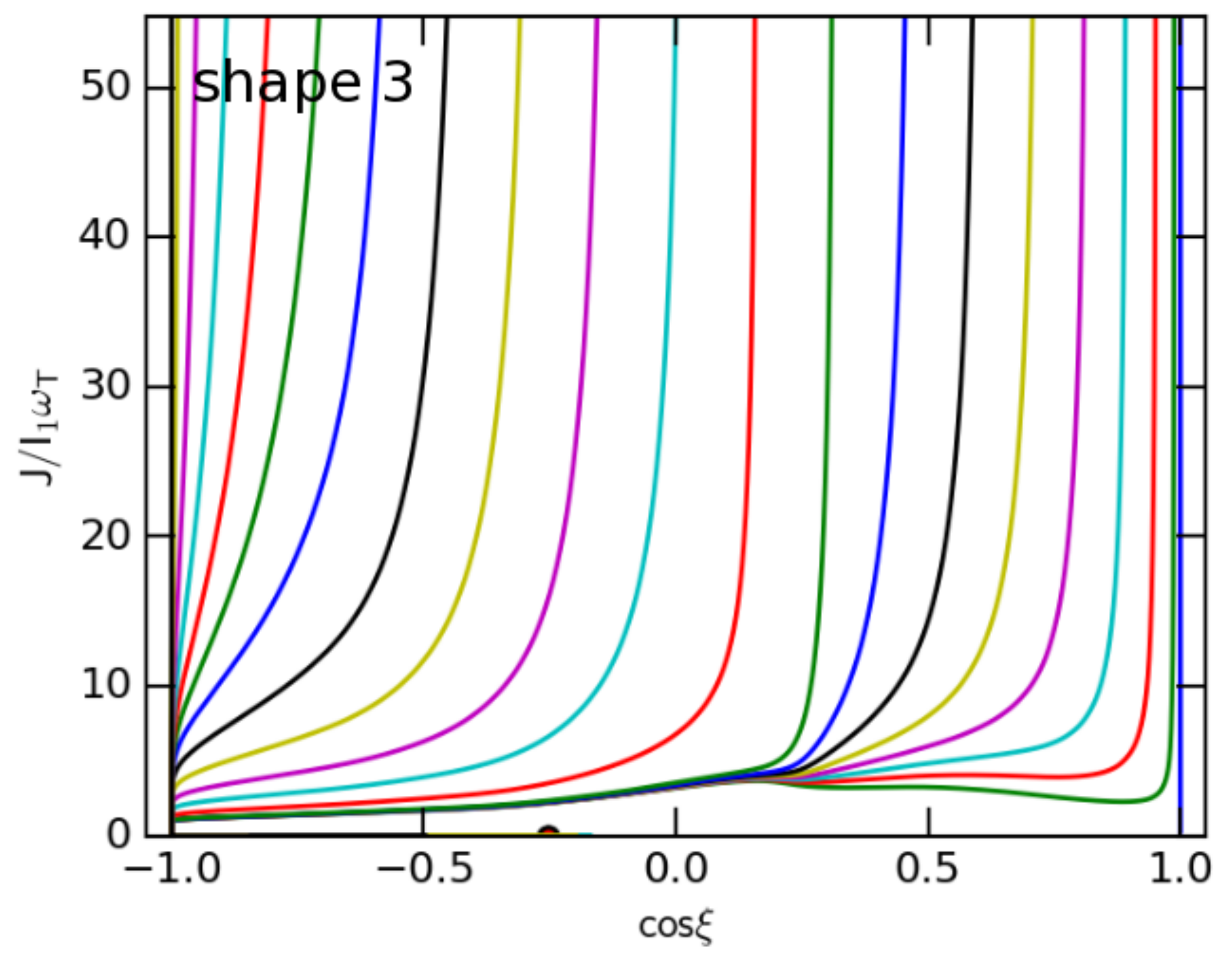}
\includegraphics[width=0.33\textwidth]{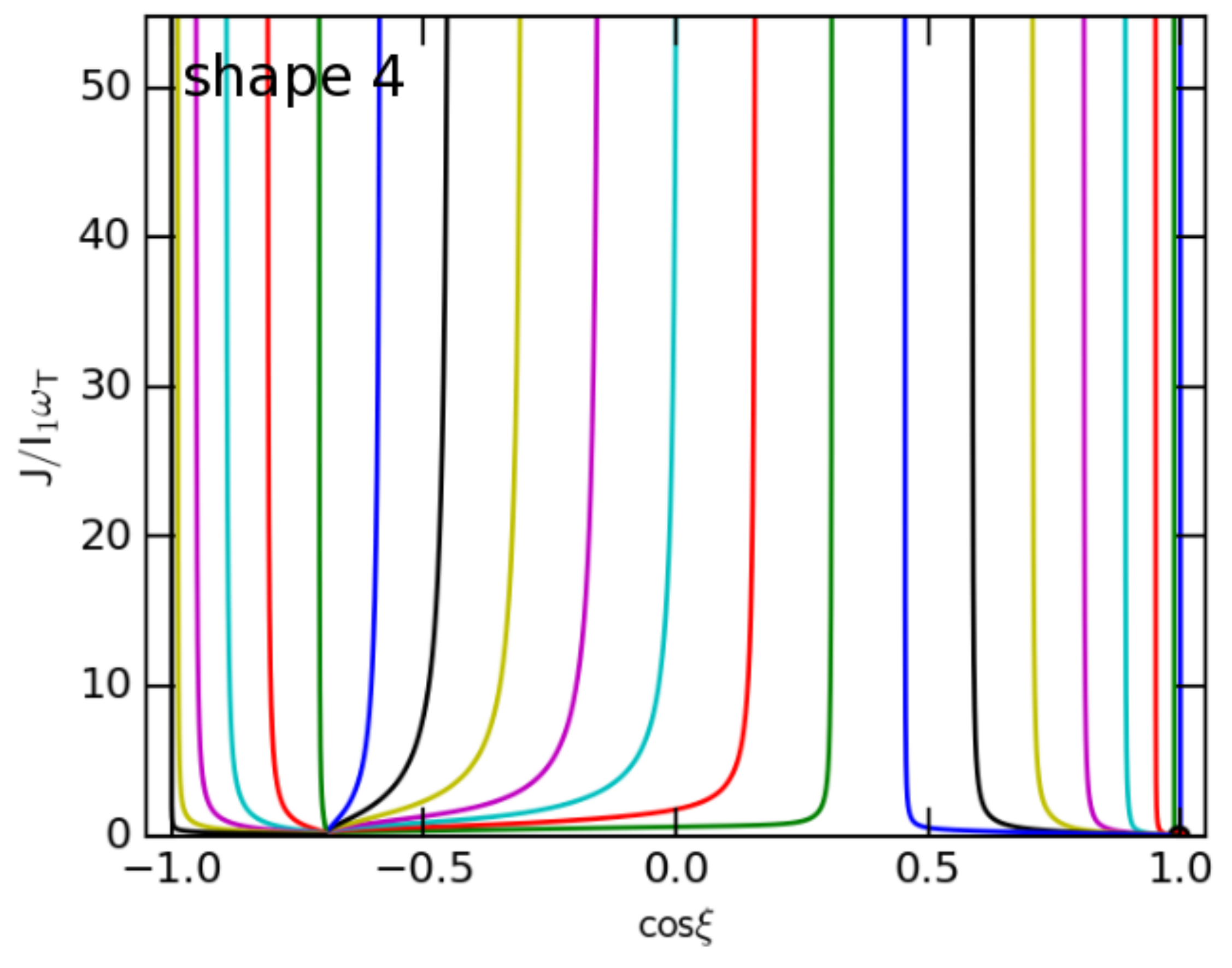}
\includegraphics[width=0.33\textwidth]{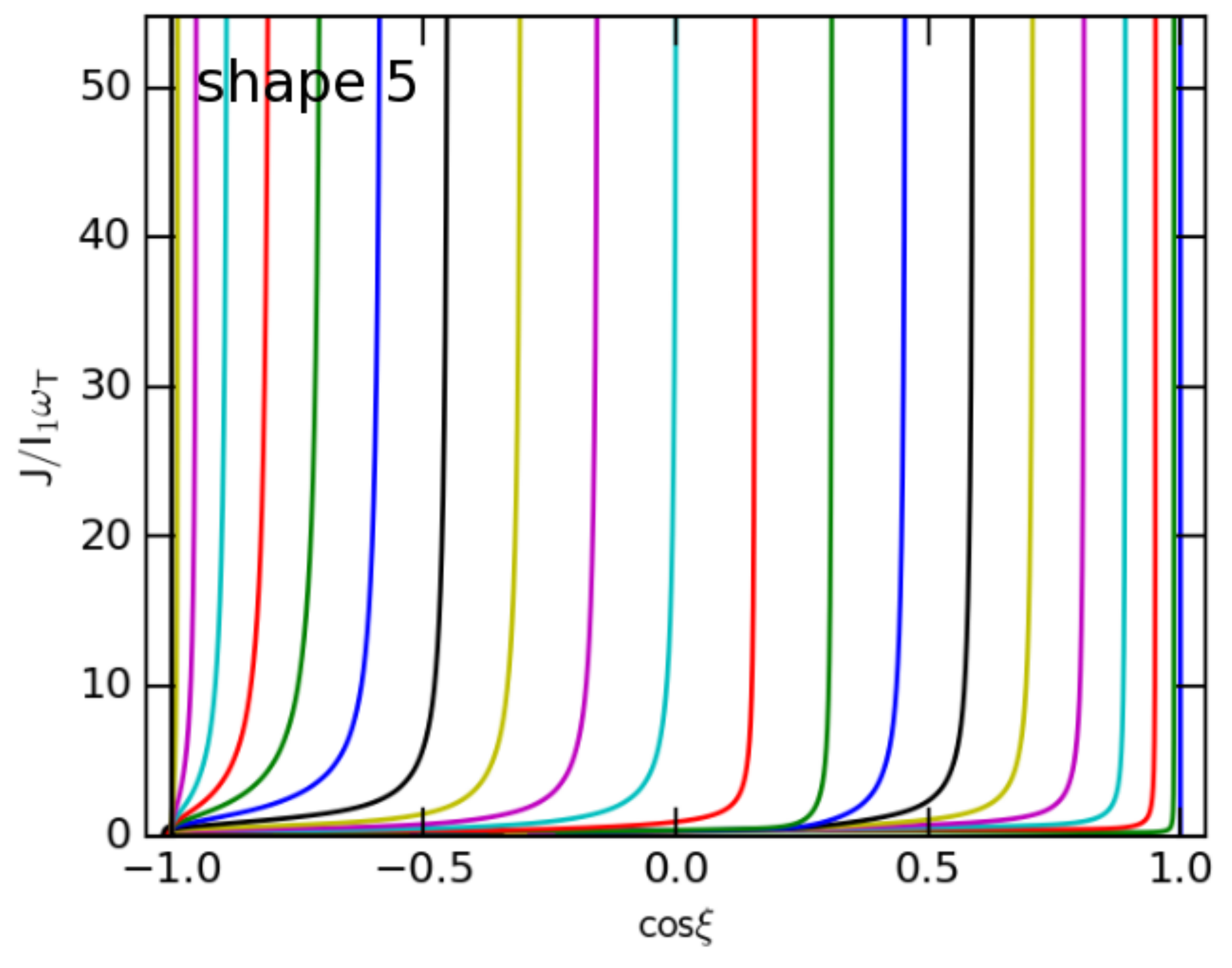}
\caption{Trajectory maps of MAT alignment for ordinary paramagnetic grains of the various shapes drifting parallel to the magnetic field ($\psi=0^{\circ}$). Arrows indicate the change of the grain orientation over time. Two repellor points (marked by X) are seen for shapes 6 and 7, and grains are driven to low-J attractor points.}
\label{fig:MATmap_ODP}
\end{figure*}

We first study alignment for ordinary paramagnetic grains. Figure \ref{fig:MATmap_ODP} shows trajectory maps for the different shapes and $\psi=0^{\circ}$. Grains are driven to low-J attractors, and only shapes 6 and 7 have the high-J repellors (denoted by the cross). For most of the shapes, the low-J attractors occur at $\cos\xi\approx \pm 1$. Due to the low MATs, the orientation of grains of shapes 3-5 are hardly changed for $J> I_{1}\omega_{T}$.

Grain alignment at low-J attractors is unstable because of the low angular momentum \citep{Hoang:2008gb}. When stochastic excitations by gas collisions are taken into account, grains aligned at the low-J attractors will be disturbed substantially. Thus, the degree of MAT alignment is low in the absence of high-J attractors \citep{2016ApJ...831..159H}.

\subsubsection{Superparamagnetic grains with iron inclusions}
Next, we investigate the MAT alignment for superparamagnetic grains containing iron inclusions. Although a large value of $\delta_{m}$ up to $100$ can be achieved when grains contain big iron clusters, the essential effect is not much different due to the saturation of grain alignment \citep{2016ApJ...831..159H}. Thus, we show the results for a typical value of $\delta_{m}=2$. 

\begin{figure*}
\includegraphics[width=0.33\textwidth]{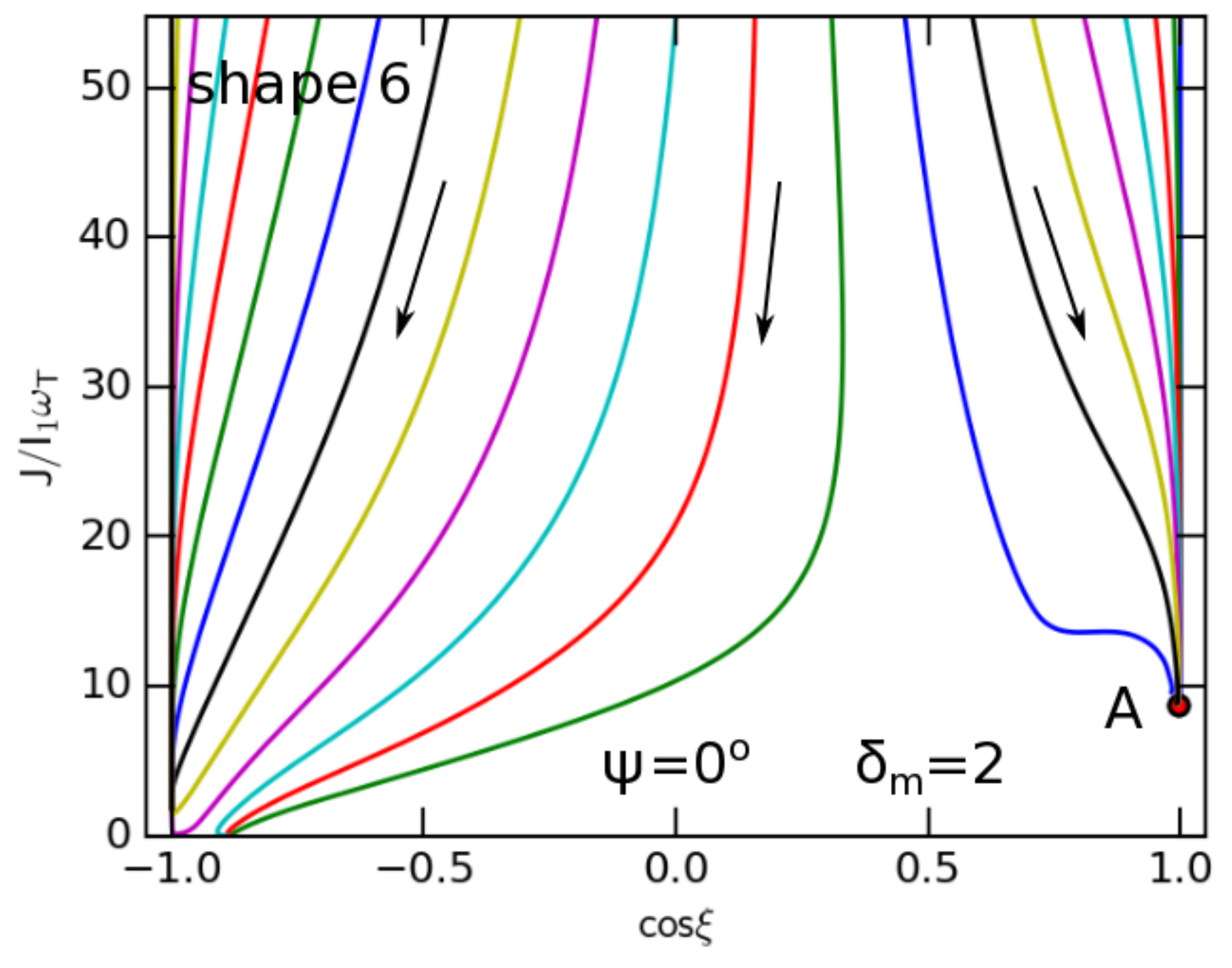}
\includegraphics[width=0.33\textwidth]{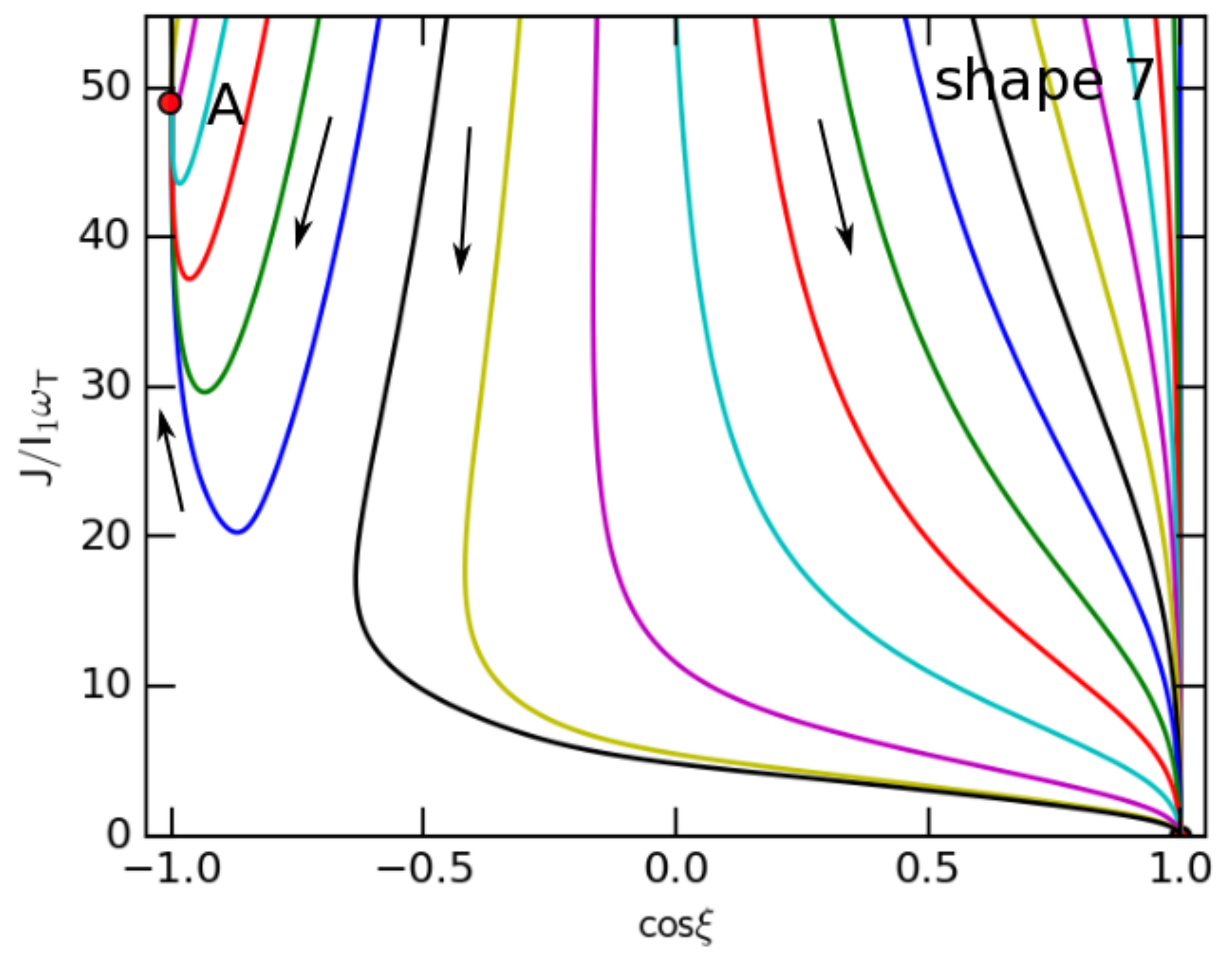}
\includegraphics[width=0.33\textwidth]{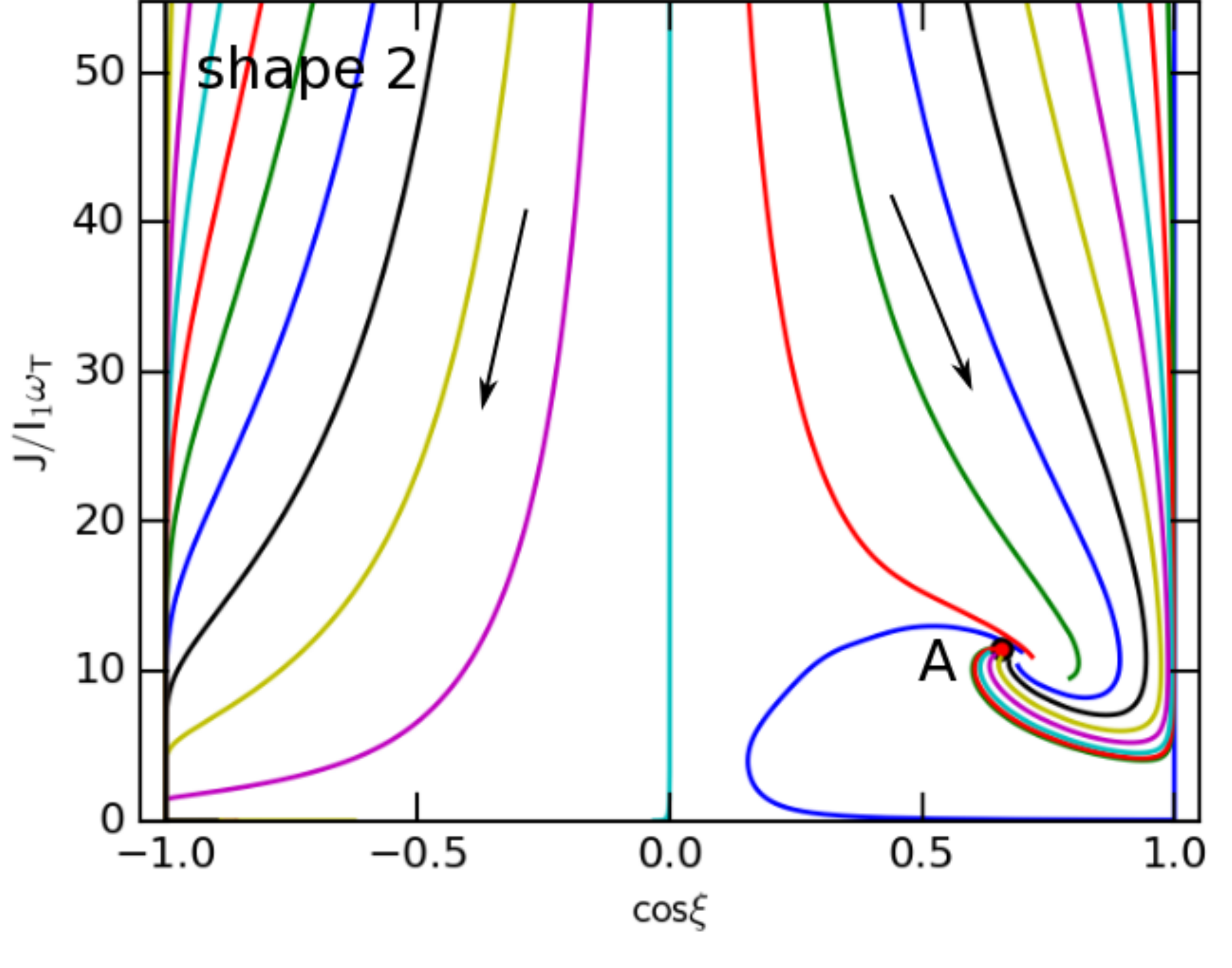}
\includegraphics[width=0.33\textwidth]{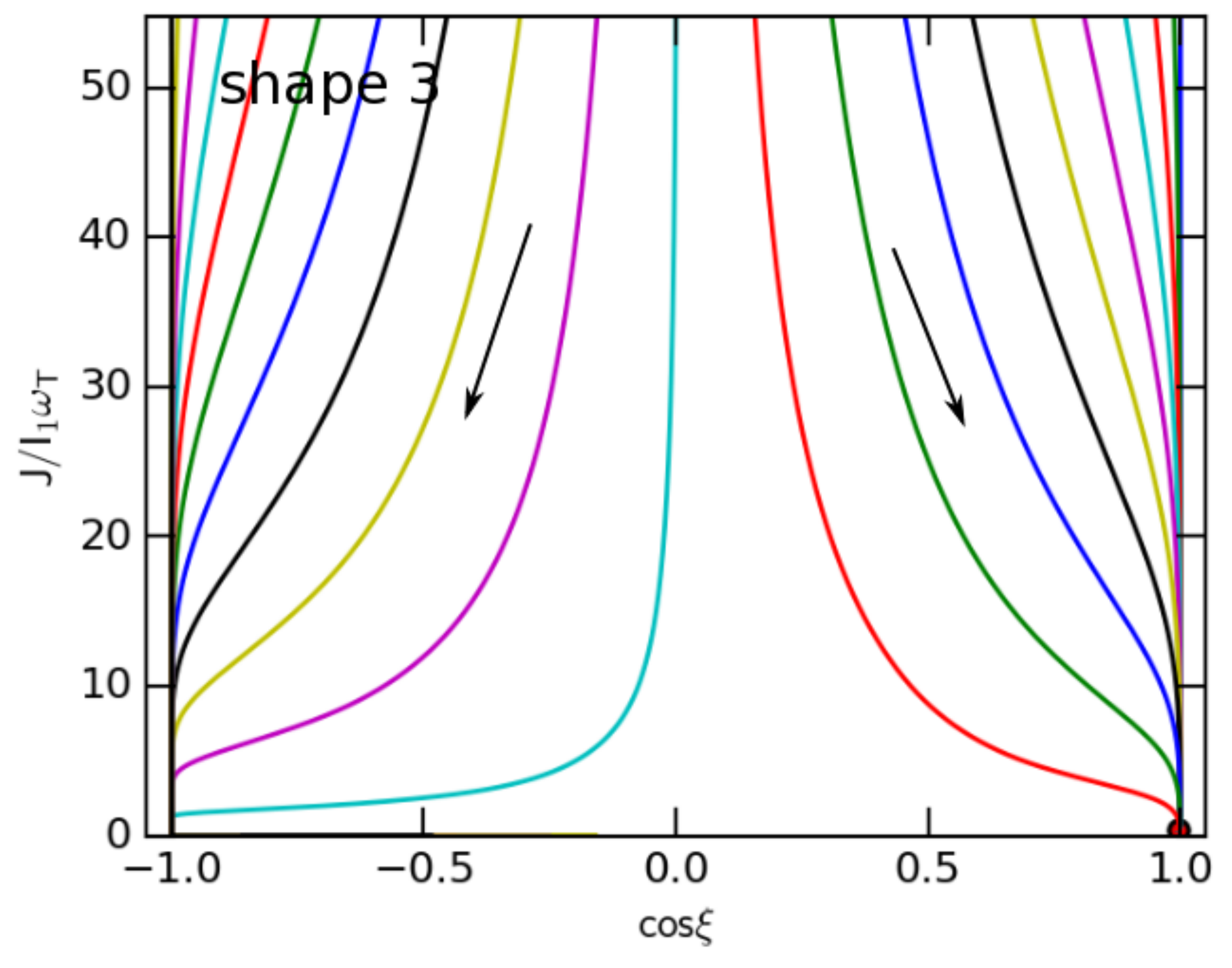}
\includegraphics[width=0.33\textwidth]{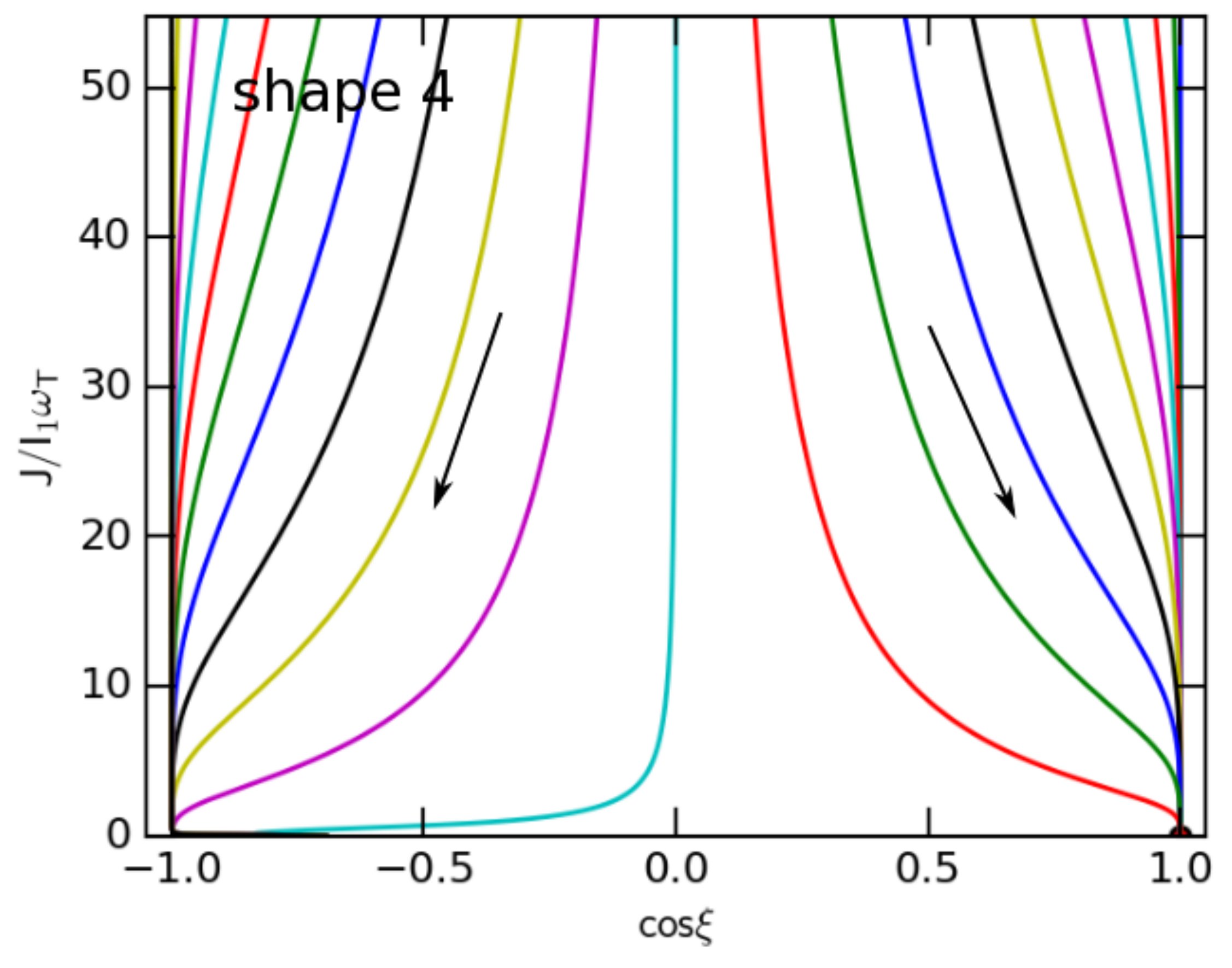}
\includegraphics[width=0.33\textwidth]{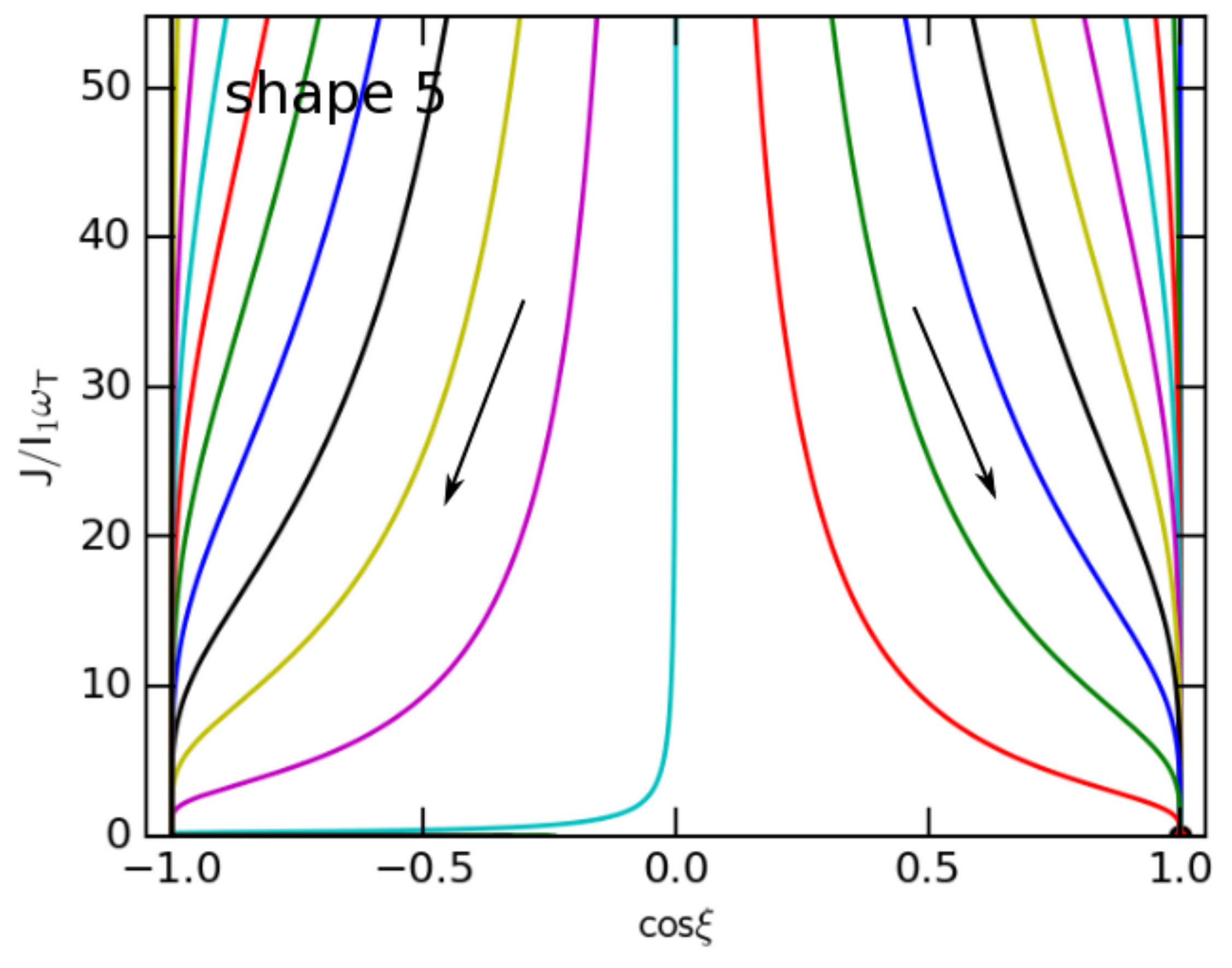}
\caption{Same as Figure \ref{fig:MATmap_ODP}, but for superparamagnetic grains with $\delta_{m}=2$. Shapes 2, 6, and 7 have high-J attractor points (denoted by a filled circle, point A) due to enhanced magnetic relaxation and MATs, while other shapes exhibit only low-J attractor points.}
\label{fig:MATmap_SUP}
\end{figure*}

Figure \ref{fig:MATmap_SUP} shows the phase trajectory maps for superparamagnetic grains at $\psi=0^{\circ}$. The high-J repellor point in shapes 2, 6, and 7 is converted to an attractor point due to superparamagnetic effect. Due to opposite helicity, the attractor point in shape 7 occurs at $\cos\xi=-1$, corresponding the the angular momentum anti-parallel to $\Bv$. Other shapes have only low-J attractors. 
We also run simulations for $\delta_{m}=10$ and find that the trajectory maps are similar to those with $\delta_{m}=2$, except the fraction of grains driven to high-J attractors is increased. 

While accounting for stochastic excitation by gas collisions, grains are eventually driven to the high-J attractors (\citealt{2016ApJ...831..159H}), leading to perfect alignment. Therefore, the degree of MAT alignment of shapes 2, 6 and 7 is expected to be perfect thanks to the presence of high-J attractors, while shapes 1, 3-5 are weakly aligned due to the absence of high-J attractors.

\subsection{Dependence of MAT alignment on the drift direction}
To see the dependence of MAT alignment on the drift direction with respect to the ambient magnetic field, we solve equations of motion for the trajectory map for several angles $\psi=30, 60, 80^{\circ}$.

\begin{figure*}
\includegraphics[width=0.3\textwidth]{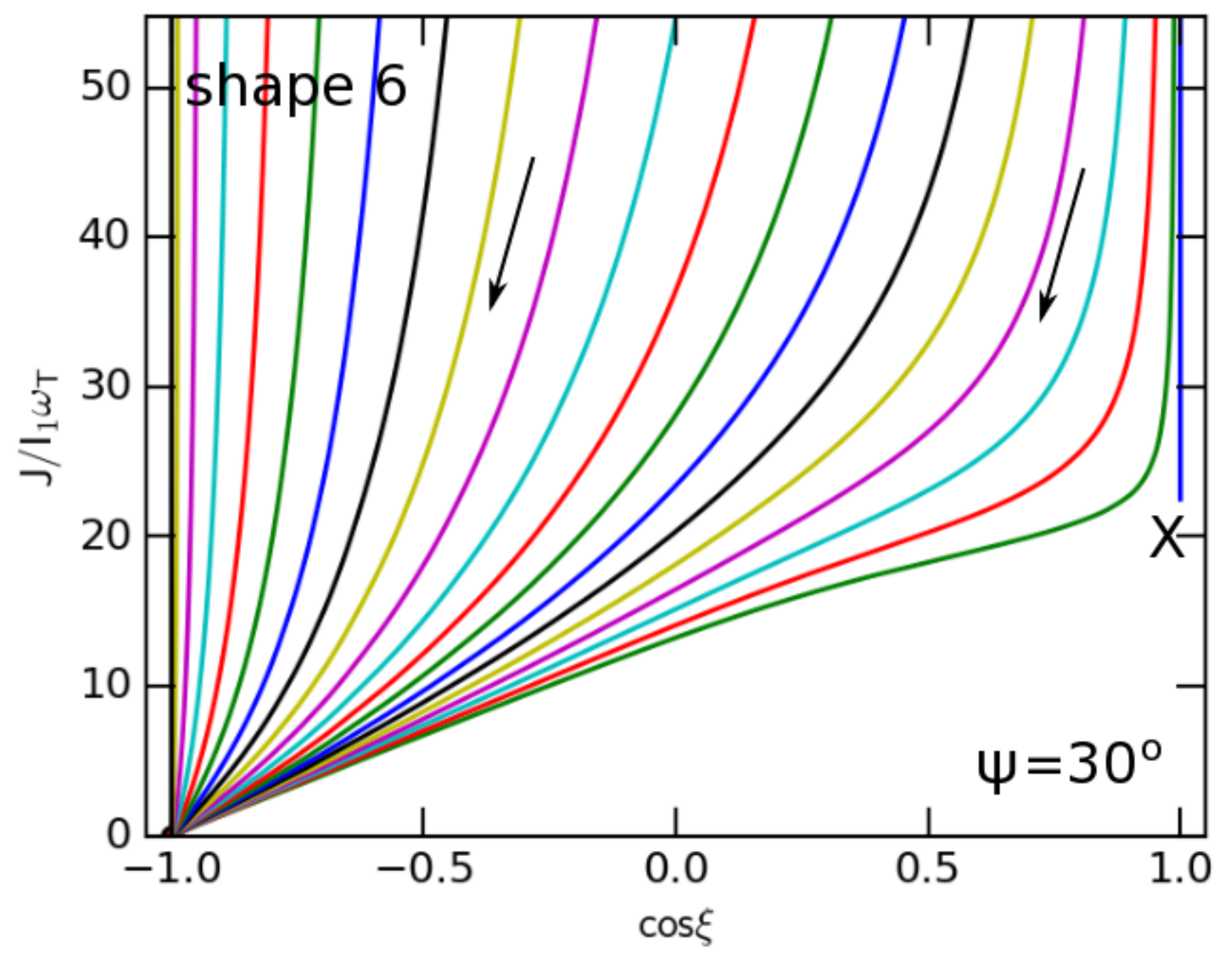}
\includegraphics[width=0.3\textwidth]{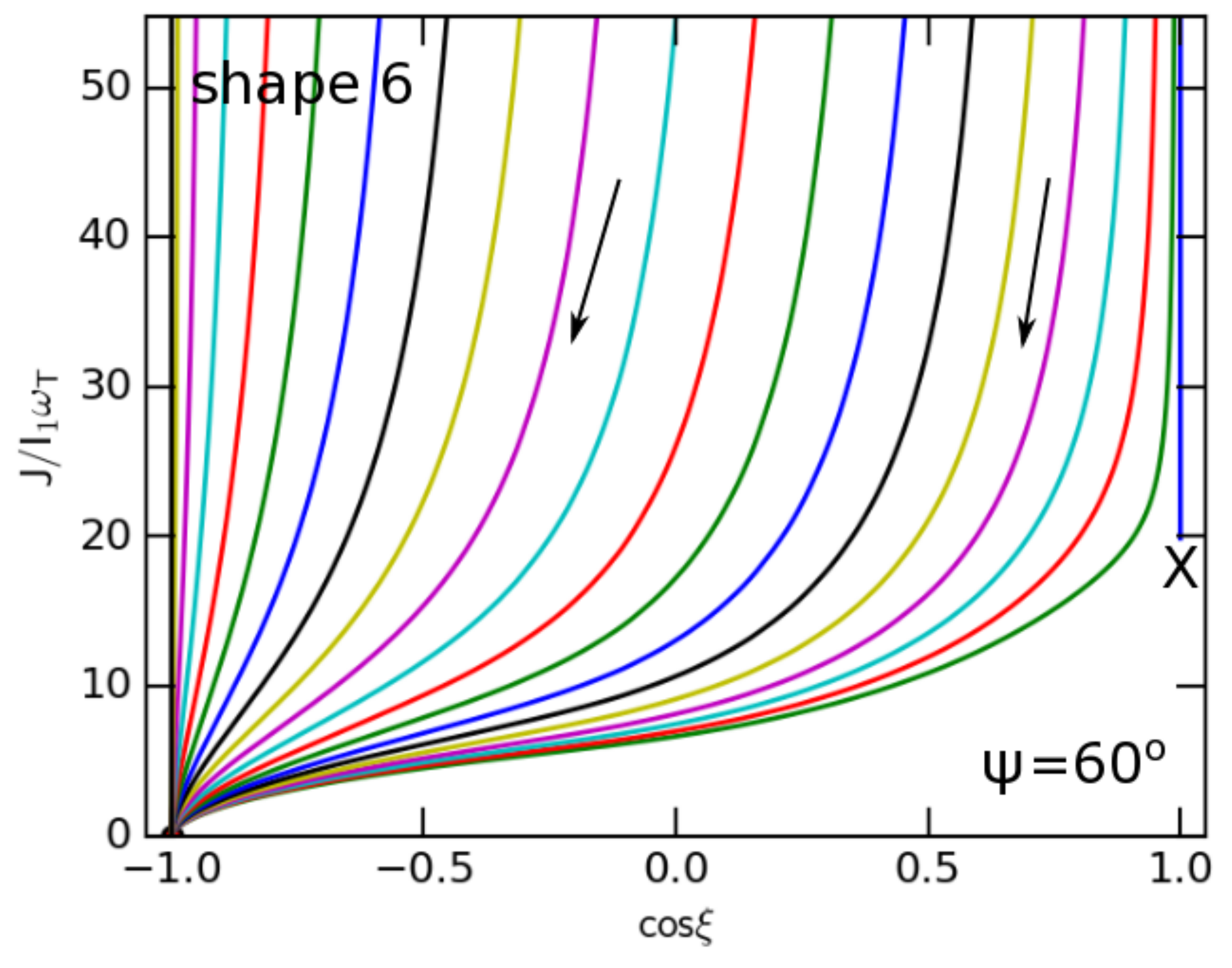}
\includegraphics[width=0.3\textwidth]{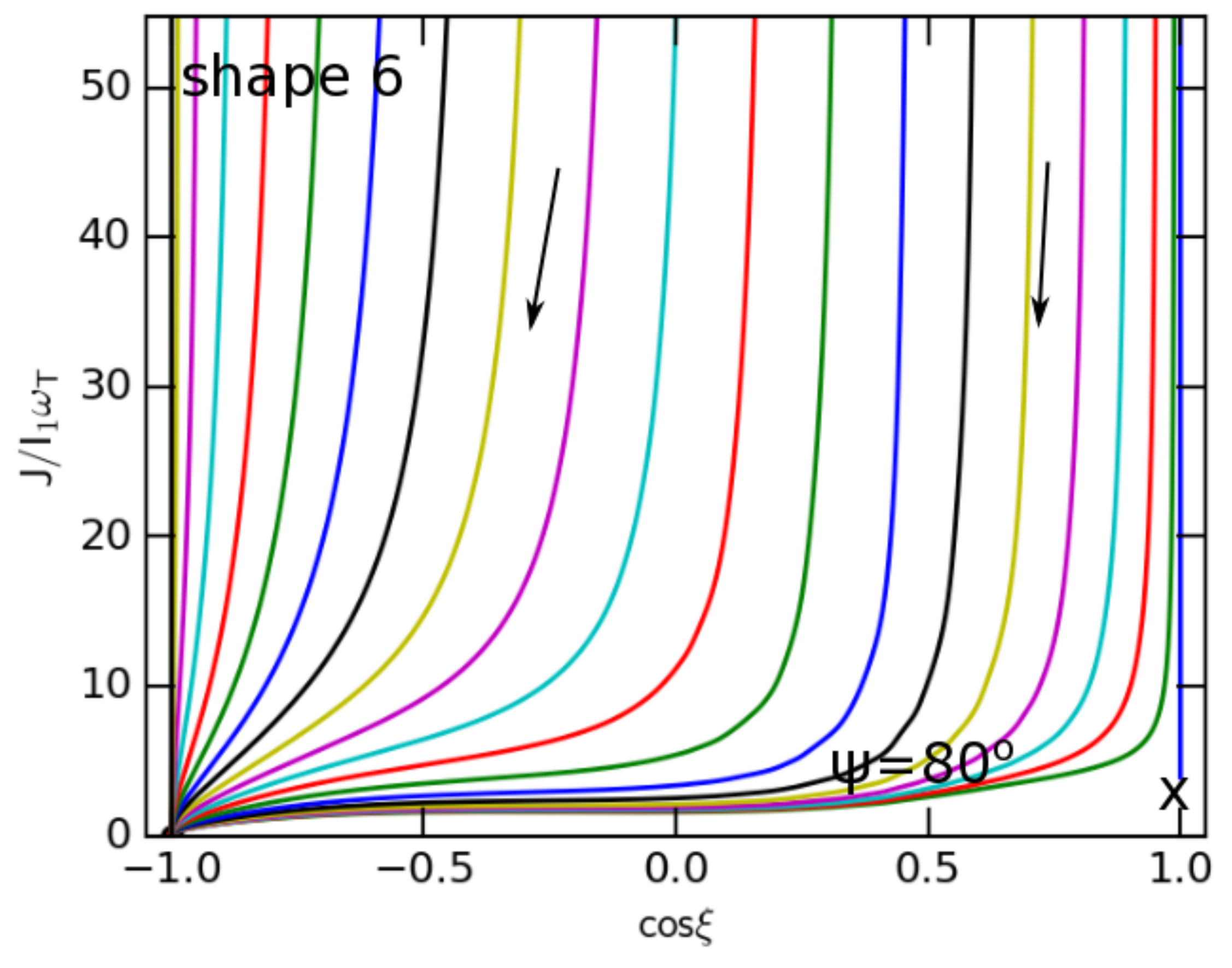}

\includegraphics[width=0.3\textwidth]{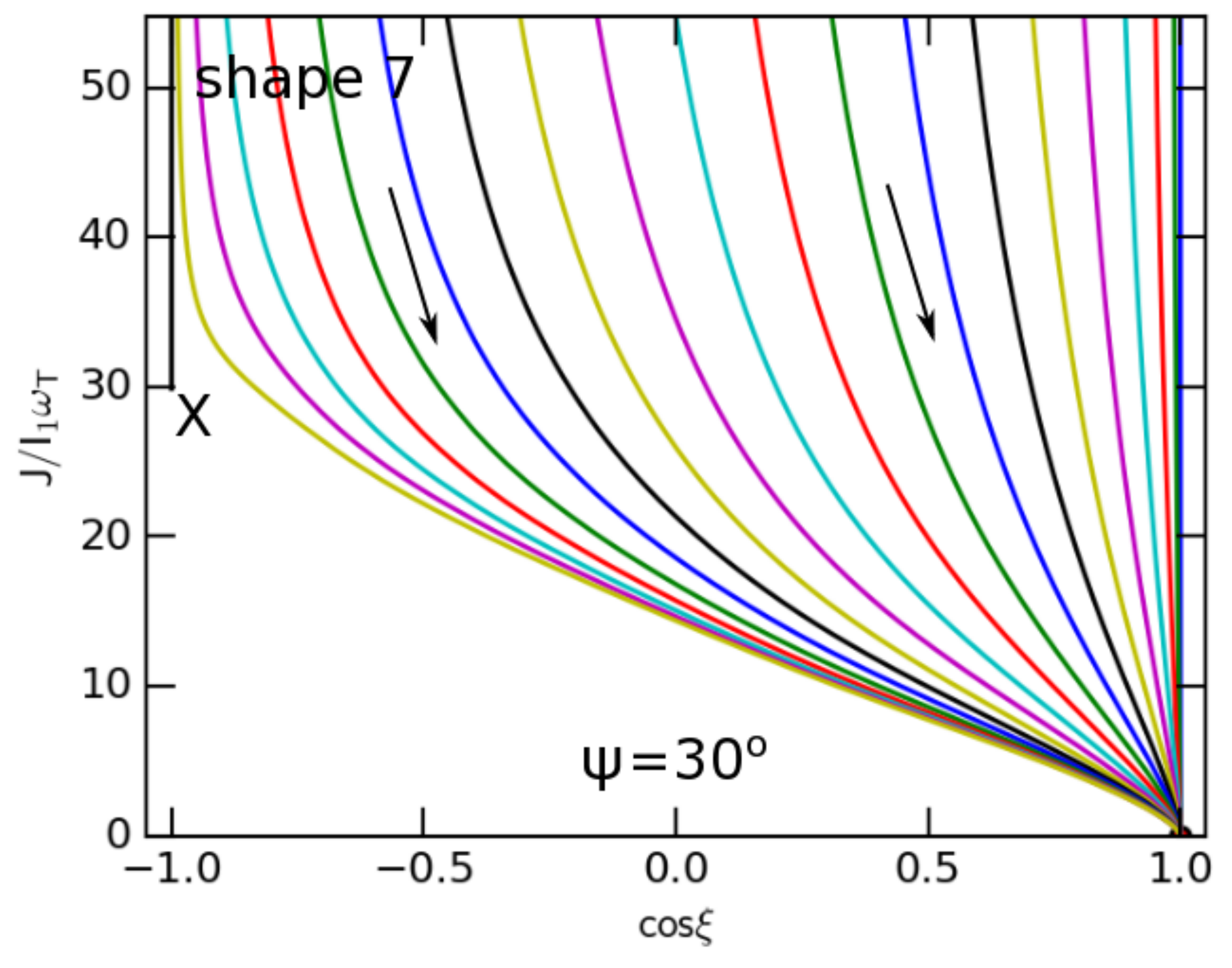}
\includegraphics[width=0.3\textwidth]{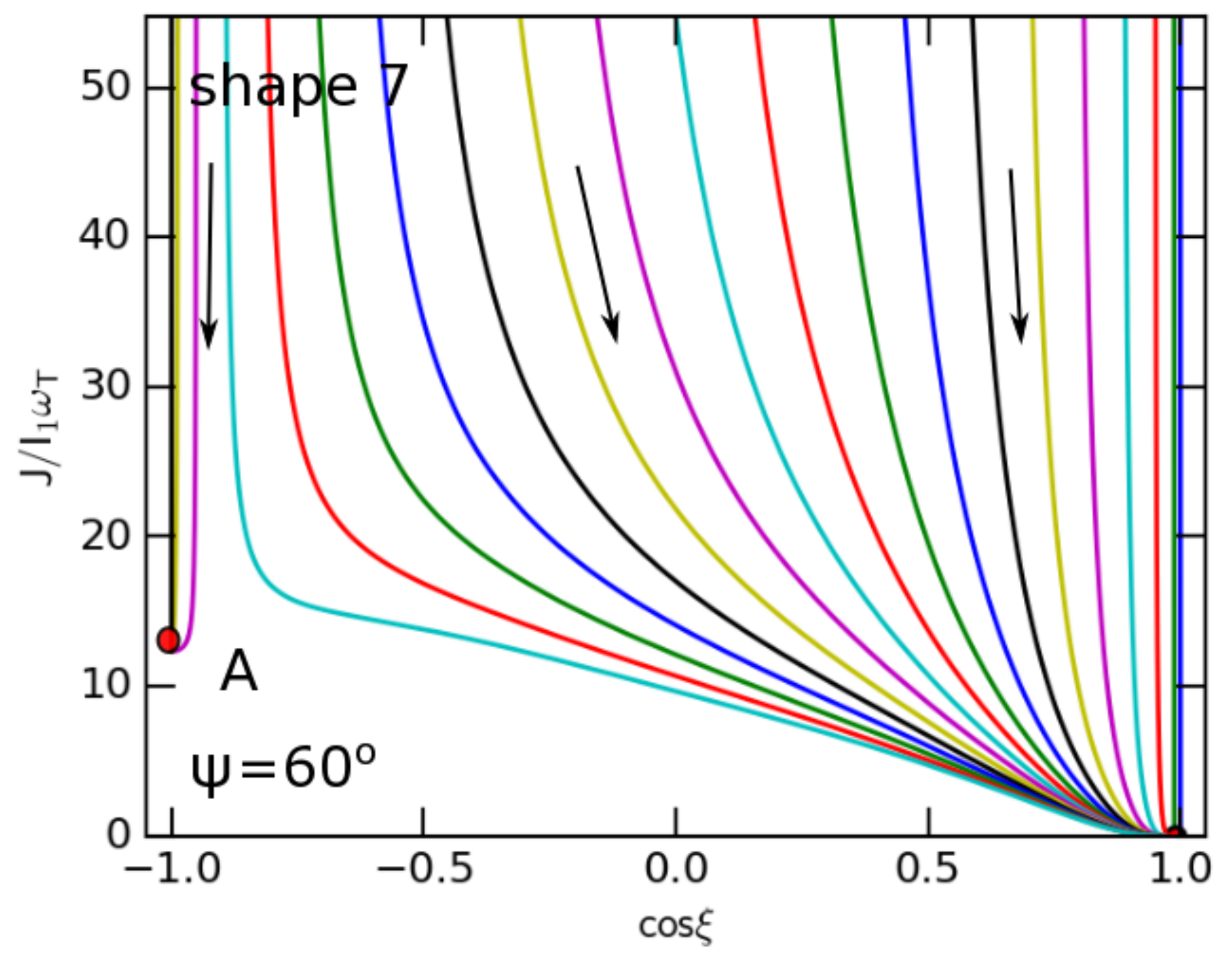}
\includegraphics[width=0.3\textwidth]{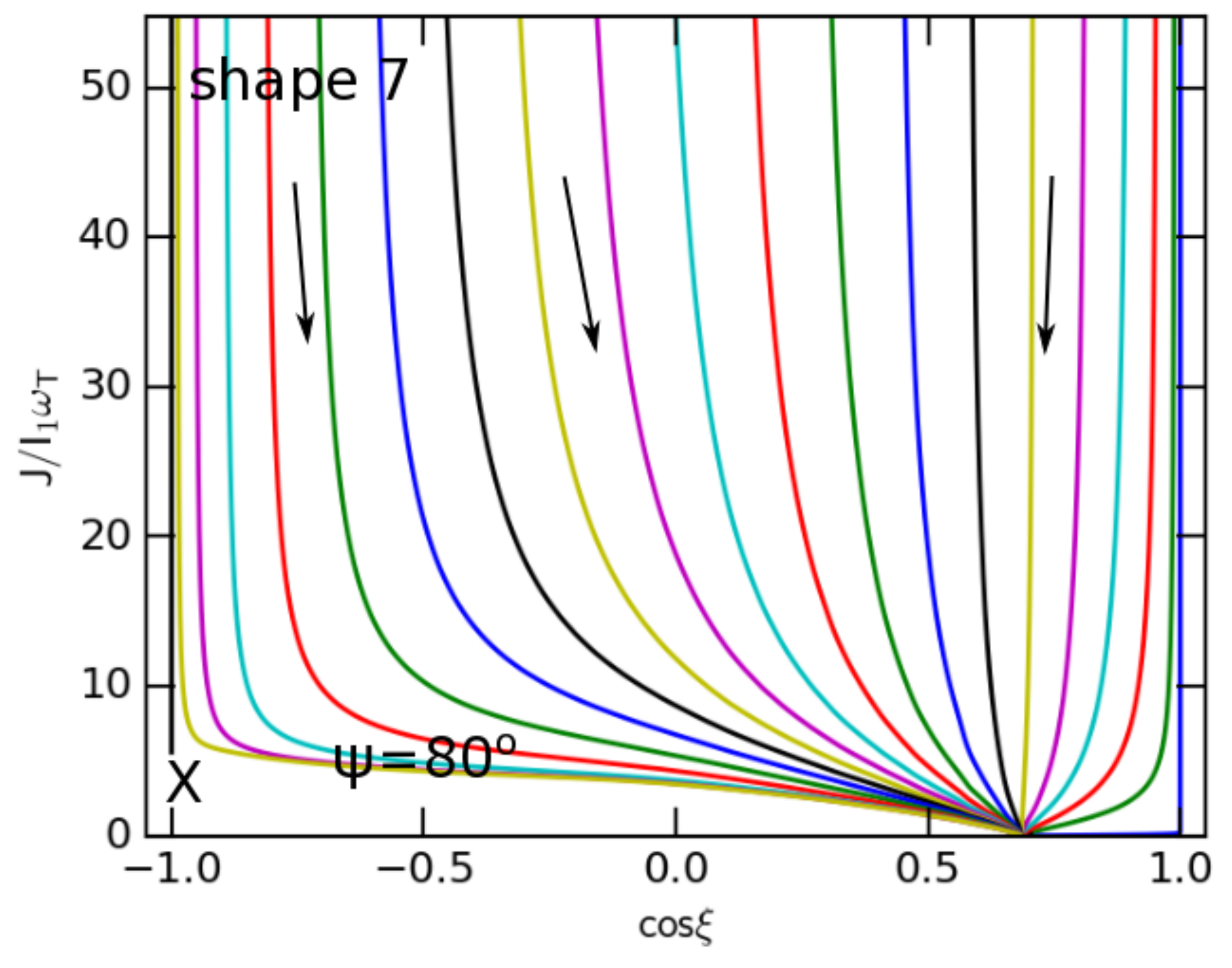}

\includegraphics[width=0.3\textwidth]{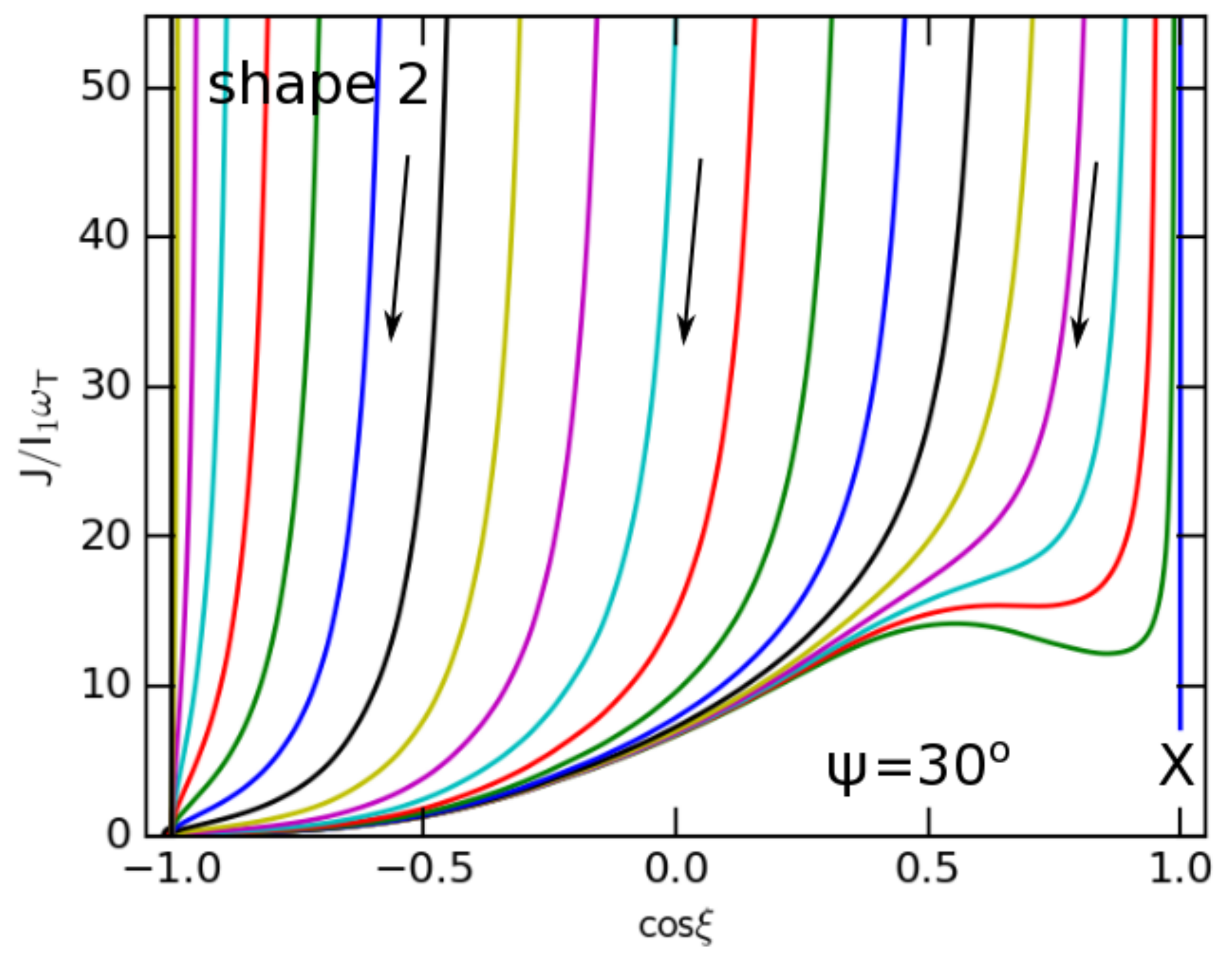}
\includegraphics[width=0.3\textwidth]{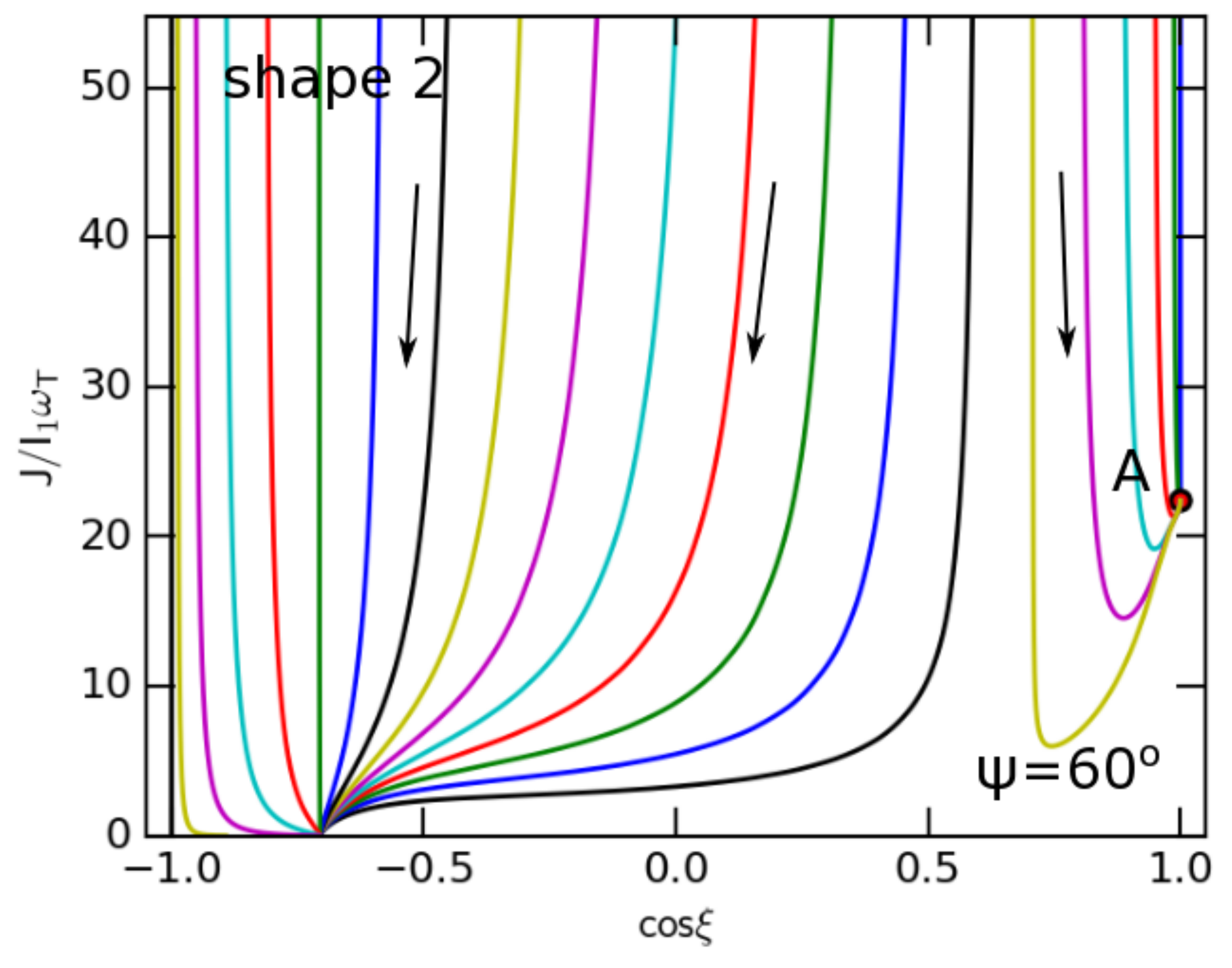}
\includegraphics[width=0.3\textwidth]{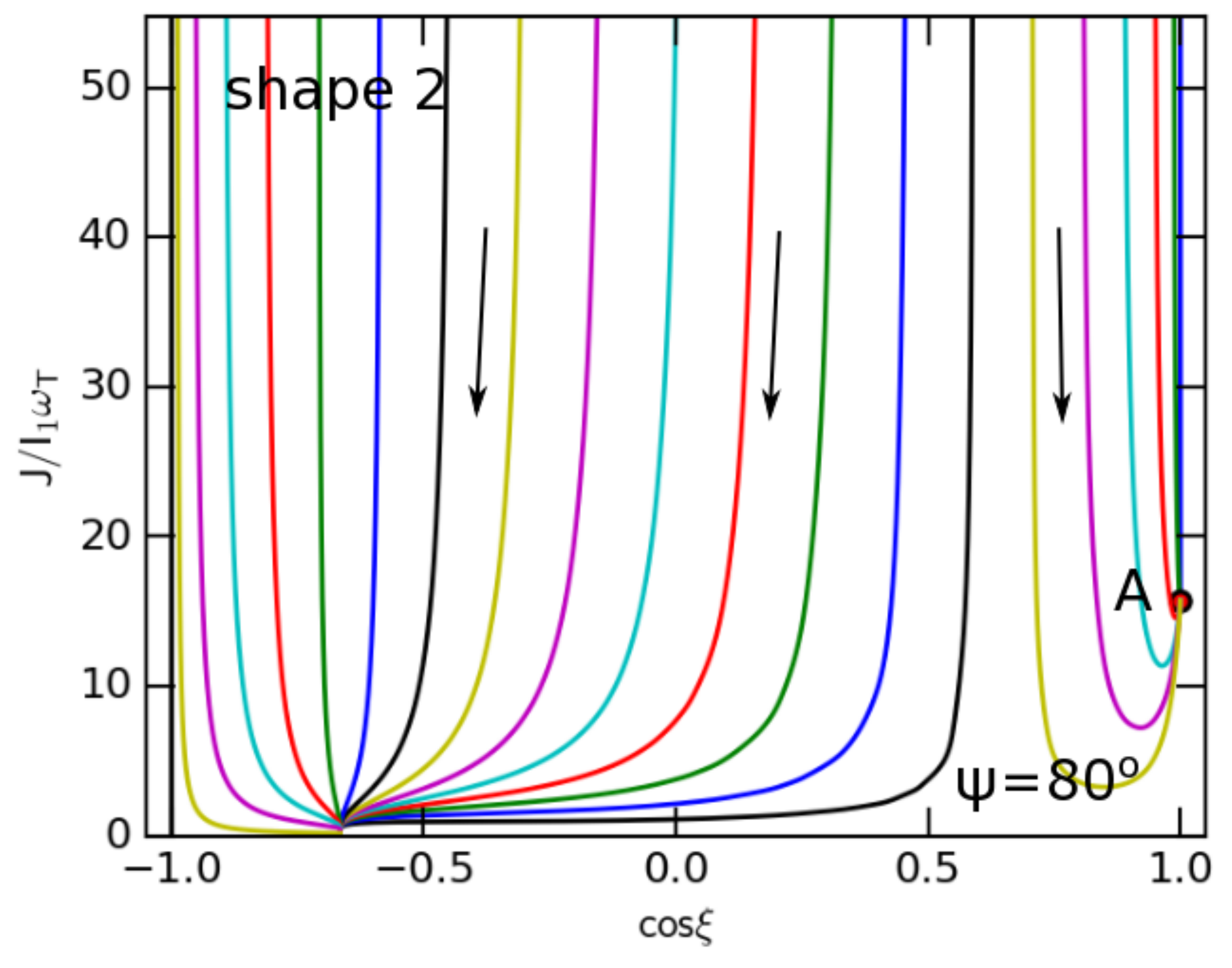}

\caption{Phase maps for the different angles $\psi=30, 60, 80^{\circ}$ for shape 6 (top panels), shape 7 (middle panels) and shape 2 (bottom panels). Shape 2 has high-J attractor points, while shape 6 has only low-J attractor points. The drift speed of $s_{d}=1$ is considered.}
\label{fig:MATmap_ODP_psi}
\end{figure*}

Figure \ref{fig:MATmap_ODP_psi} shows the trajectory maps for the case of ordinary paramagnetic grains of shapes 2, 6 and 7. For shape 6, MAT alignment has only low-J attractors (with high-J repellor points). The high-J attractor is present for $\psi=60^{\circ}$ in the trajectory map of shape 7. For shape 2, MAT alignment can occur with high-J attractors (filled circles) for the $\psi=60^{\circ}$ and $80^{\circ}$.

\begin{figure*}
\centering
\includegraphics[width=0.3\textwidth]{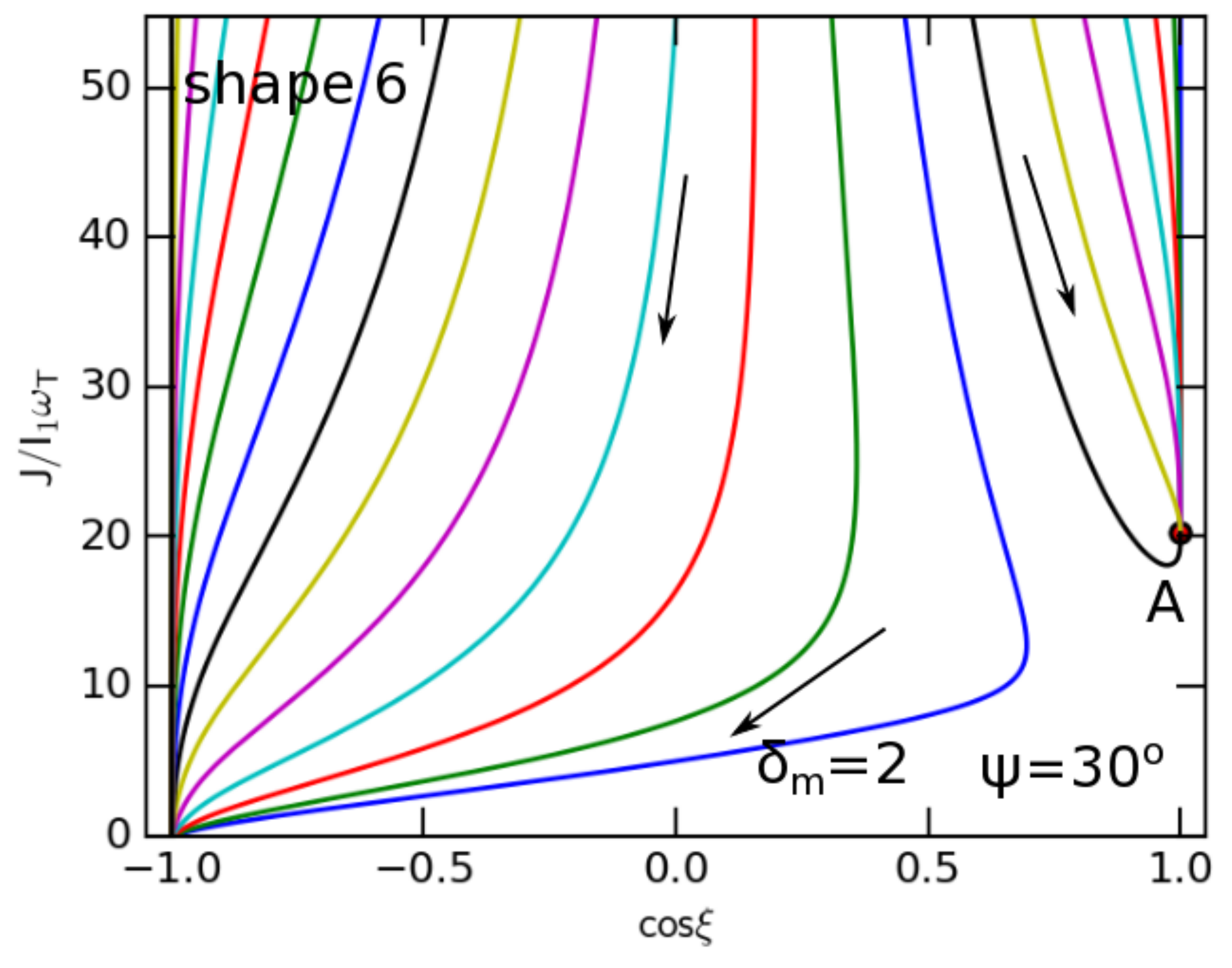}
\includegraphics[width=0.3\textwidth]{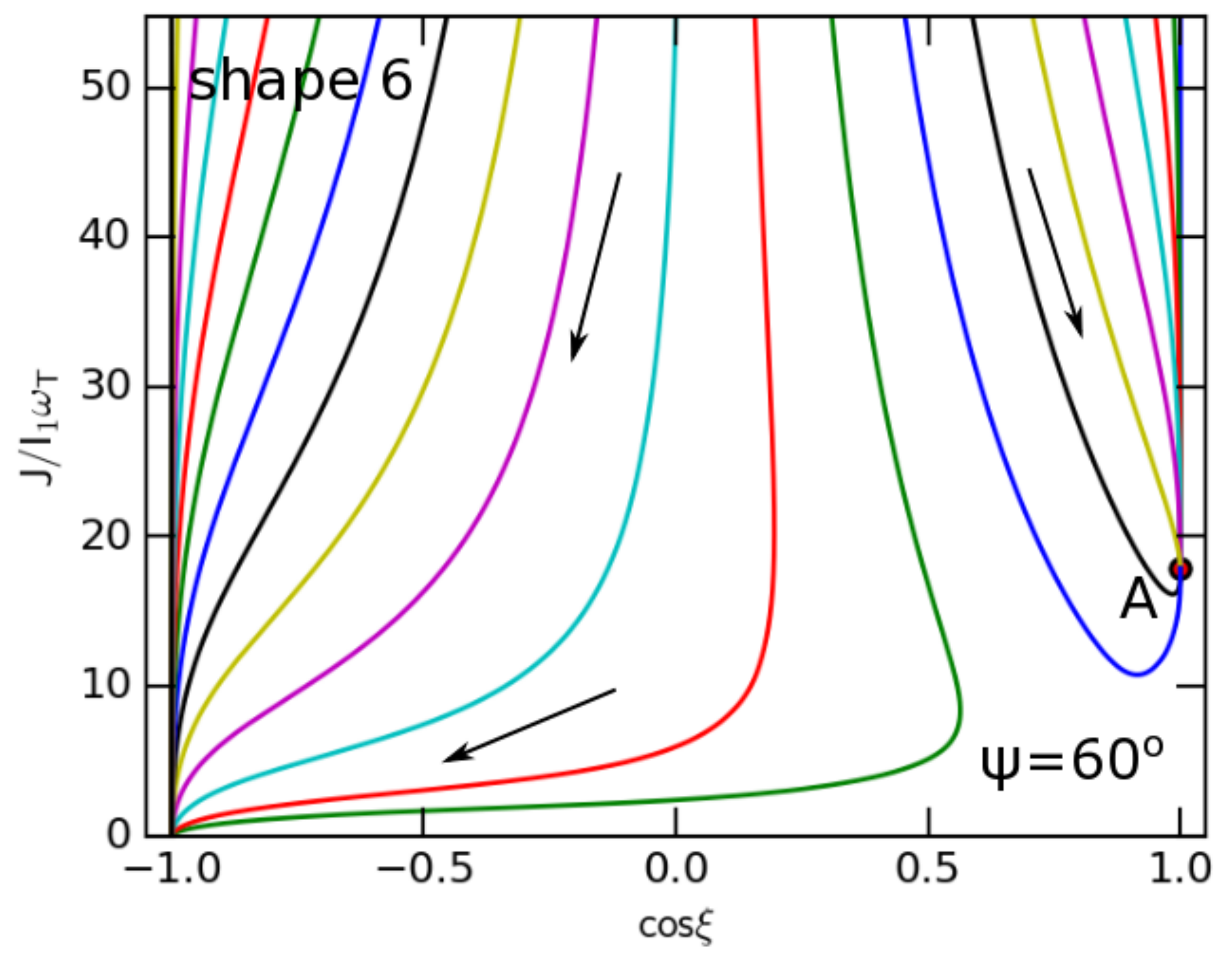}
\includegraphics[width=0.3\textwidth]{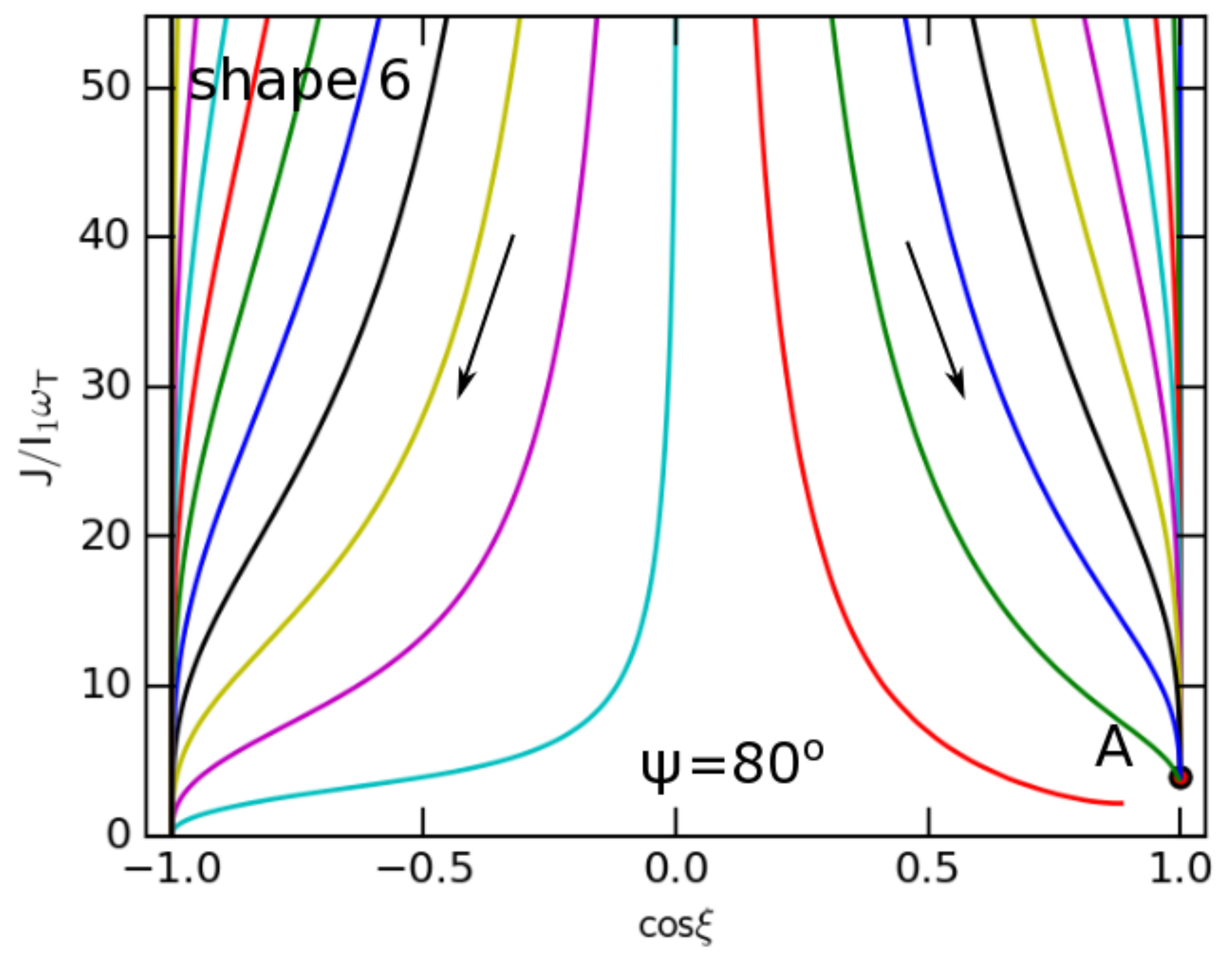}

\includegraphics[width=0.3\textwidth]{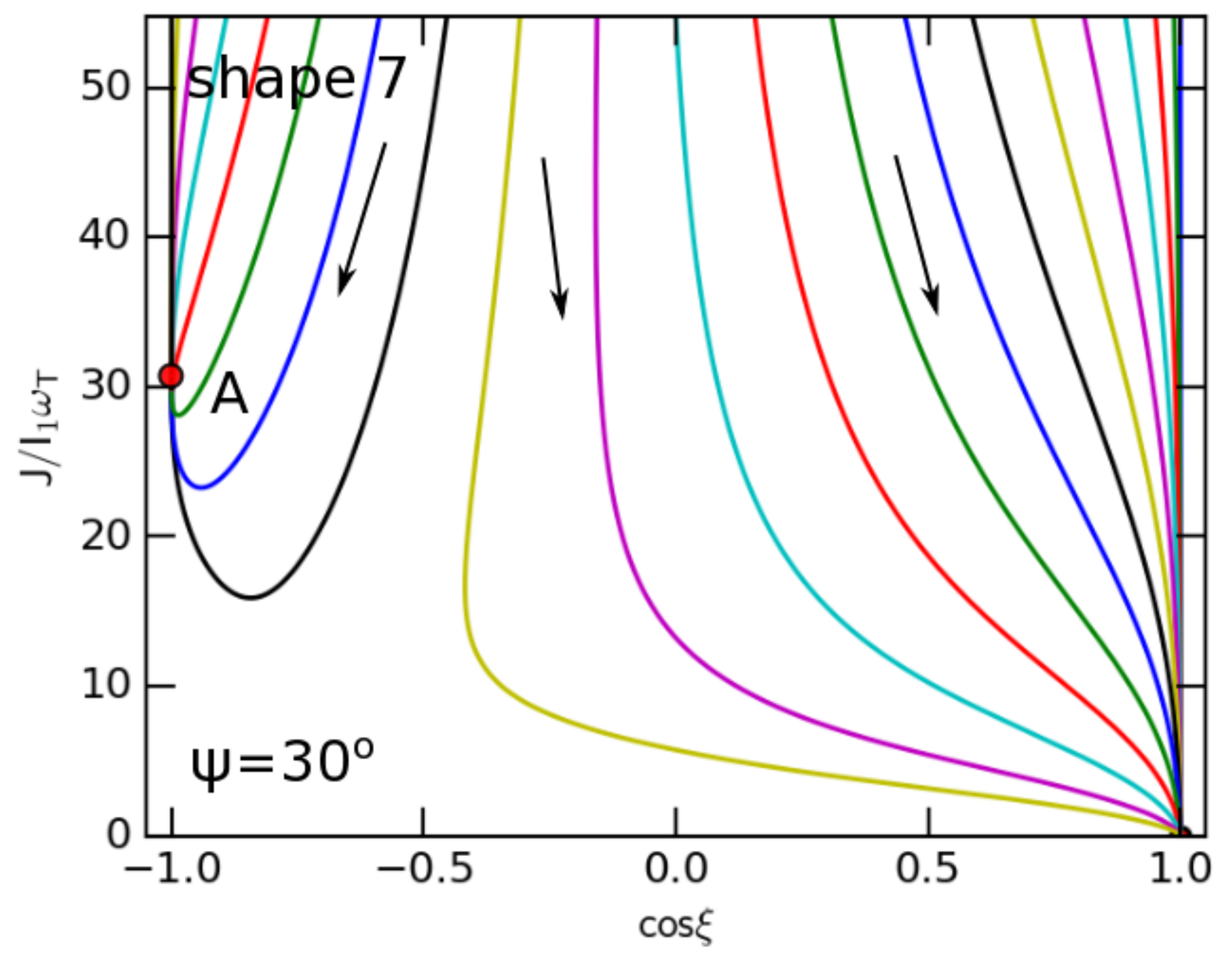}
\includegraphics[width=0.3\textwidth]{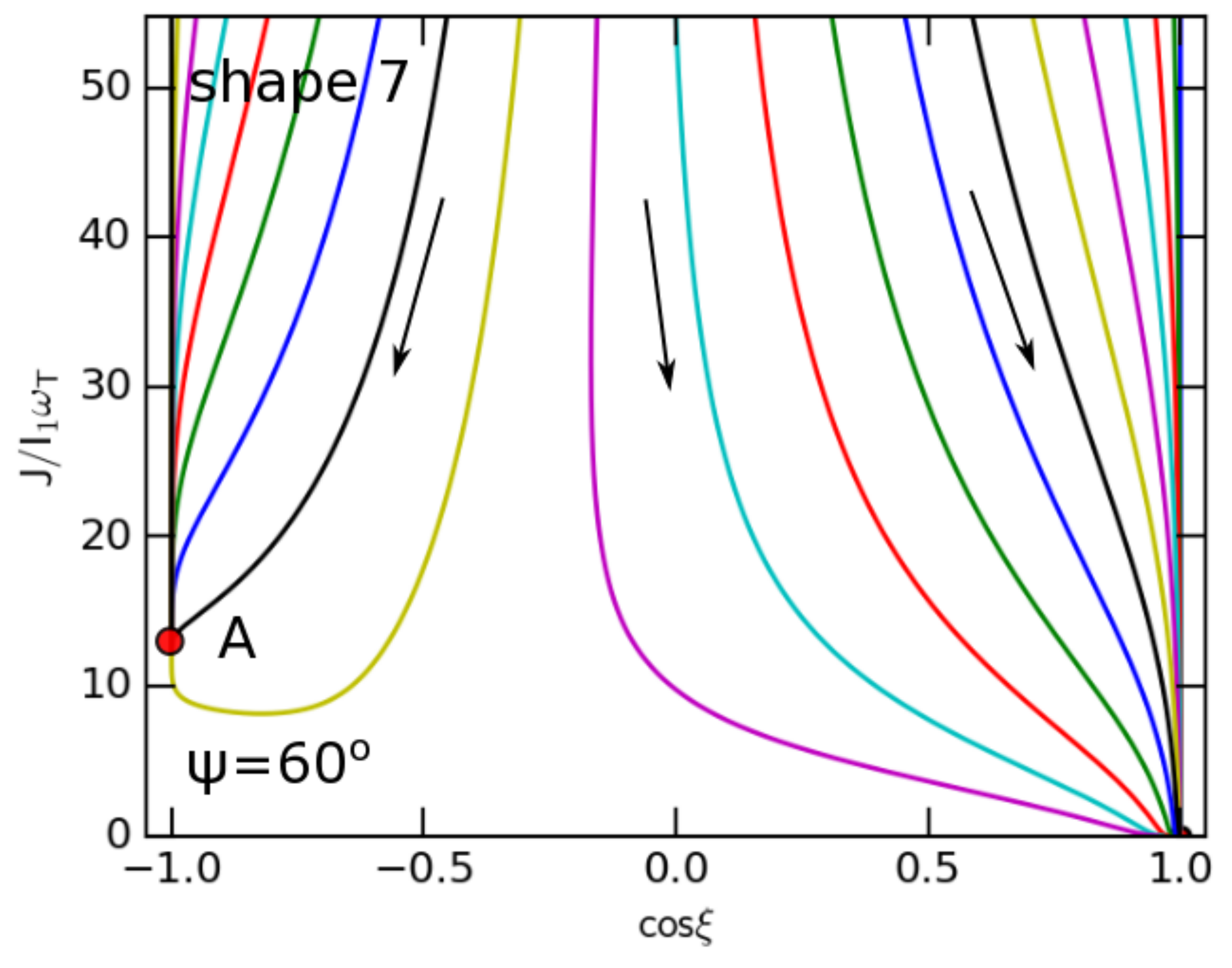}
\includegraphics[width=0.3\textwidth]{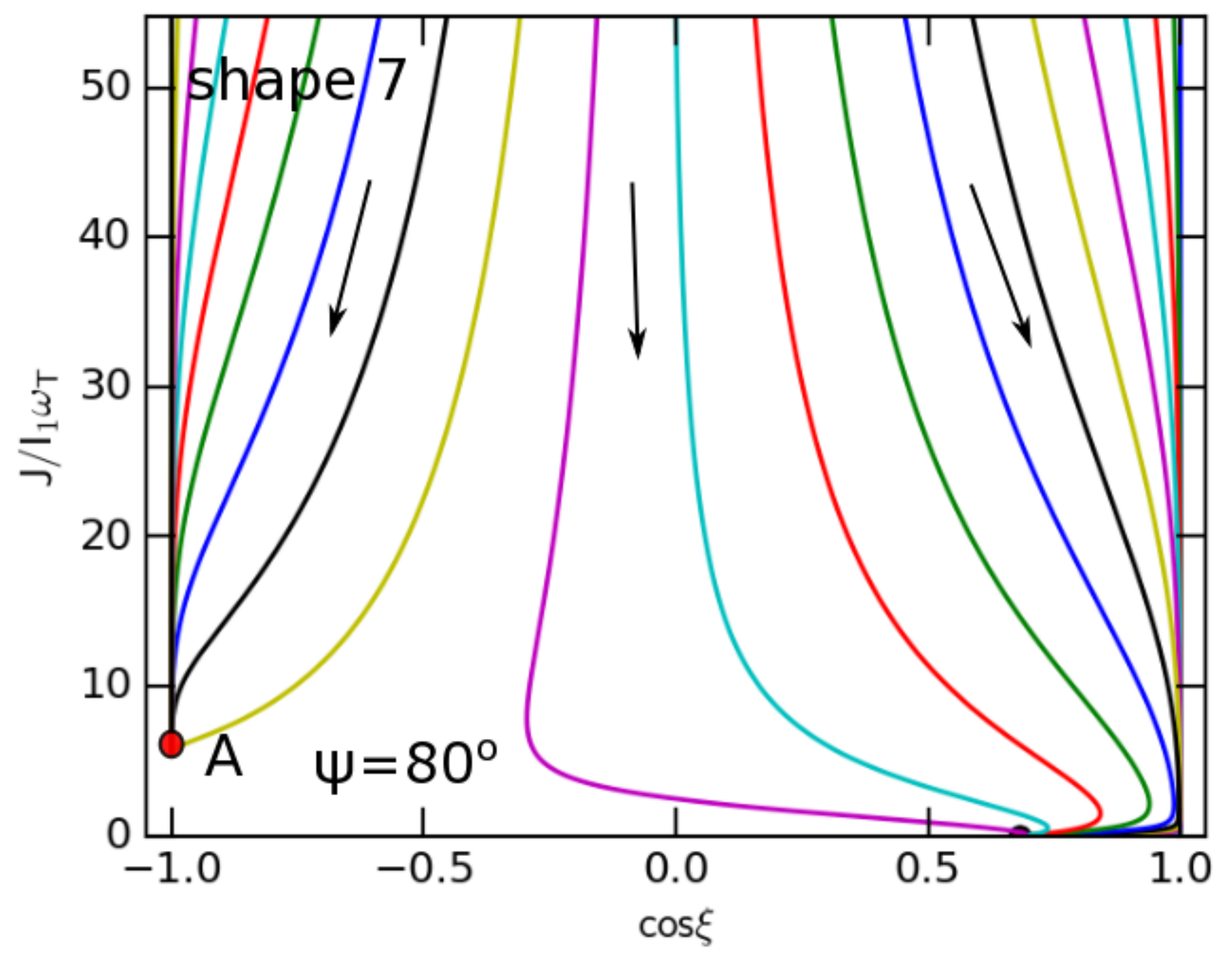}

\includegraphics[width=0.3\textwidth]{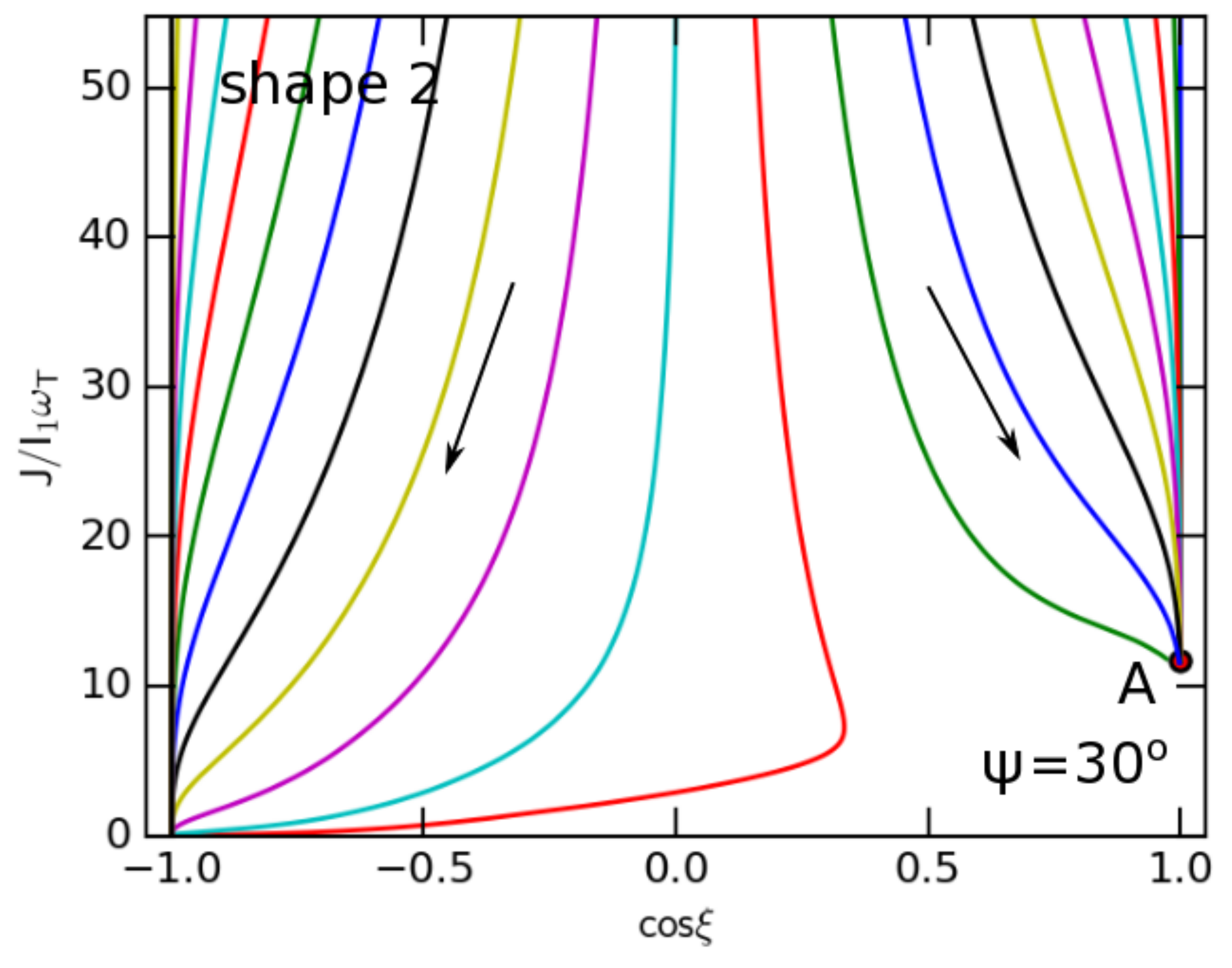}
\includegraphics[width=0.3\textwidth]{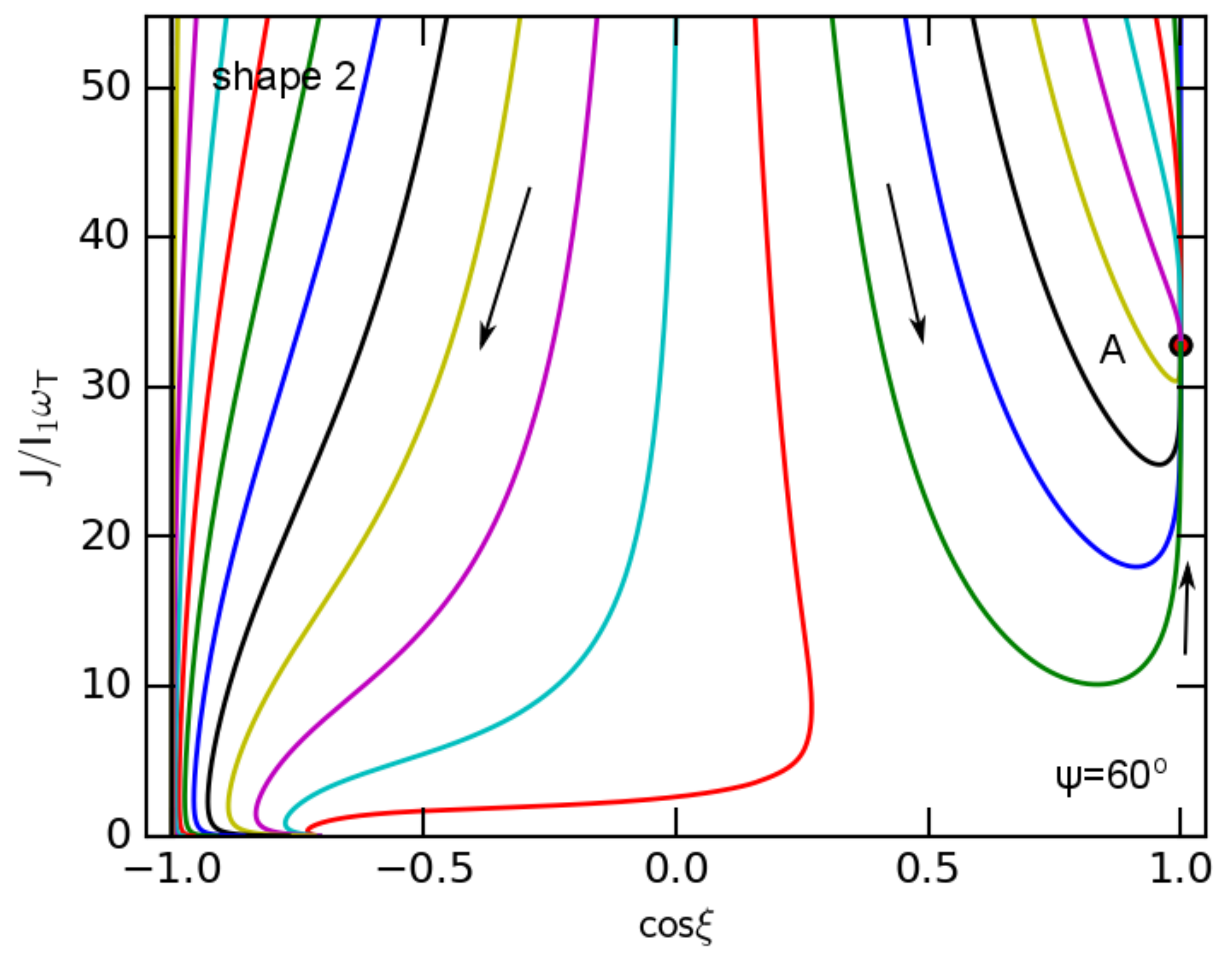}
\includegraphics[width=0.3\textwidth]{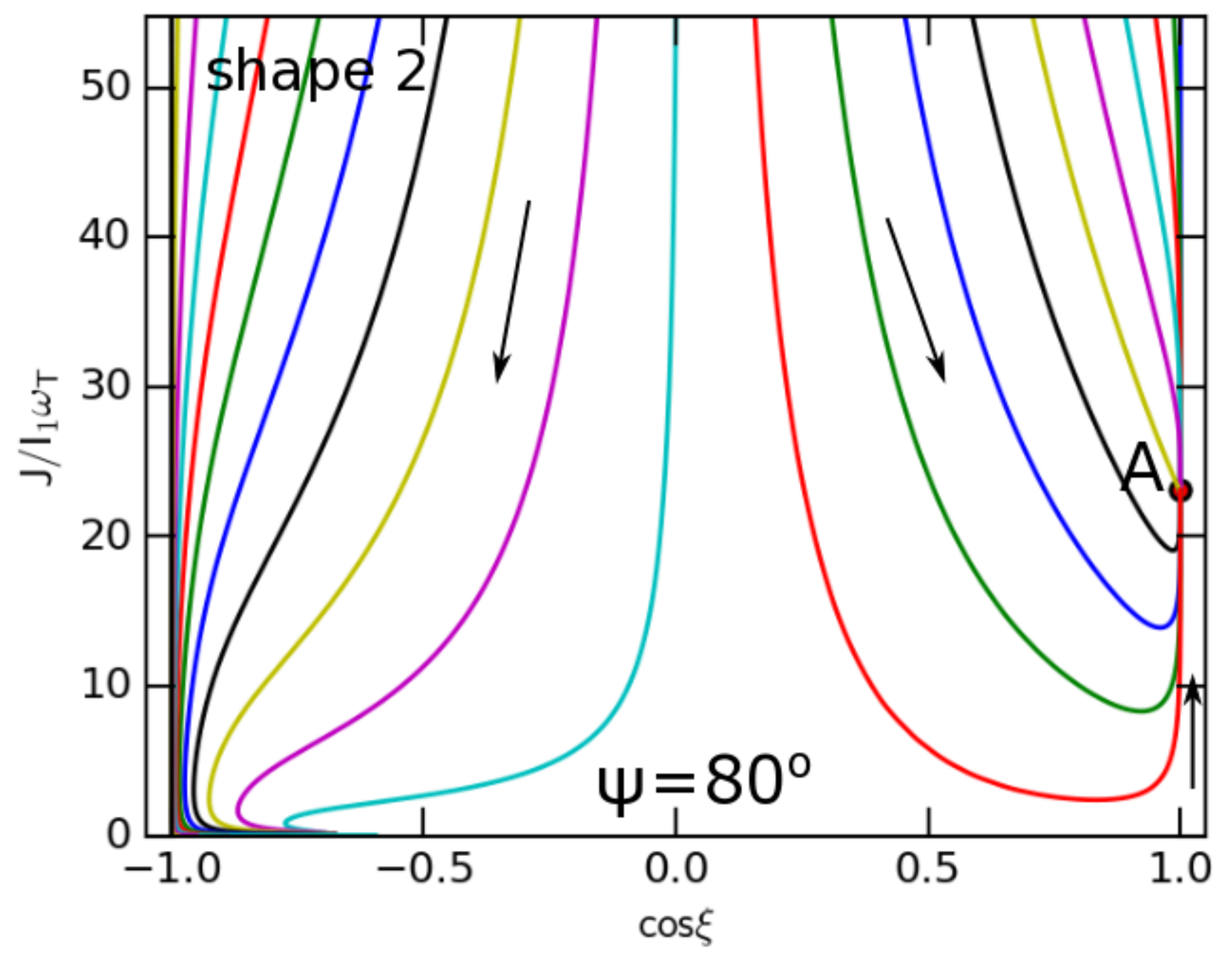}
\caption{Same as Figure \ref{fig:MATmap_ODP_psi}, but for superparamagnetic grains with $\delta_{m}=2$. Grains are perfectly aligned at high-J attractor points and low-J attractor points.}
\label{fig:MATmap_SUP_psi}
\end{figure*}

Similarly, Figure \ref{fig:MATmap_SUP_psi} shows the results for superparamagnetic grains. For three HIS (shapes 2, 6, and 7), the high-J attractors appear for all angles $\psi$. Therefore, MAT alignment for superparamagnetic grains can achieve perfect alignment when random collisional excitations are accounted for. For other WIS (shapes 3-5), we expect the effect of iron inclusions is negligible for small drift angles (i.e., $\psi<30^{\circ}$) because MATs are insufficient to produce suprathermal rotation, such that the superparamagnetic torque can stabilize its alignment (see Figure \ref{fig:Jmax}). For large drift angles, iron inclusions can induce perfect alignment as shapes 2, 6, and 7.

\section{Discussion}\label{sec:dis}

\subsection{MAT efficiency of irregular grains and comparison with Lazarian \& Hoang (2007b) analytical model}

First, MATs of irregular grains introduced in LH07b are very different from the torques that act in the original mechanism \cite{1952MNRAS.112..215G} as well as in alternative mechanical alignment mechanisms proposed in the last century, e.g. from the cross-sectional and crossover mechanical alignment suggested in \cite{Lazarian:1995p3034}. The MATs from LH07b act analogously to radiative torques by increasing the angular momentum of a grain in proportion of time. This makes them radically different from the Gold stochastic torques. The latter act on regular oblate or prolate grains, but grain irregularities render grains with helicity in terms of their interaction with the gaseous flow. This helicity induces the new type of torque and alignment. 

We find that the grain shape that exhibits mirror symmetry produces negligible MATs. Interestingly, for highly irregular shapes (HIS), including shapes 2, 6, and 7, we find that MATs are strong and able to drive grains to suprathermal rotation with subsonic drift. Moreover, MATs of these shapes also exhibit generic properties as predicted by the AMO, including symmetric $Q_{e1}$ and zeros of $Q_{e2}$ when the drift direction is parallel to the axis of maximum moment of inertia $\ahat_{1}$. For weakly irregular shapes (WIS, shapes 3-5), the magnitude of MATs is found to be lower. Our numerical results demonstrate the role of the grain surface properties on MATs.

{In LH07b, we discussed two possibilities that can reduce MATs from our AMO, including the refection coefficient $E$ and degree of helicity $D$. In that sense, the AMO can be considered to have perfect helicity, i.e., $D=1$. Naturally, most of the grains do not have clear-cut facets as the grain model of LH07b, but have numerous irregularities which can be considered of numerous facets of different orientations, leading to the reduction of helicity $D$. 

For a simple shape considered in Section \ref{sec:AMO}, we have shown that the MAT efficiency decreases with increasing the number of facets $N_{\rm facet}$. We note that the random walk formula adopted for our analytical estimates is applicable only for $N_{\rm facet}\gg 1$, i.e., not applicable for shapes 3-5 with a few facets.}

{We then can quantify the degree of helicity of an arbitrary irregular shape as follows:
\bea
D \equiv \frac{\rm max(J_{\rm max}^{\rm irr}(\psi))}{\rm max (J_{\rm max}^{\rm AMO}(\psi))},
\ena
where $J_{\rm max}^{\rm irr}(\psi)$ and $J_{\rm max}^{\rm AMO}(\psi)$ are the maximum spin-up momentum evaluated with MATs from irregular shapes and AMO, respectively. The obtained value $D$ for the selected shapes is shown in Table \ref{tab:MAT_RAT}. HIS have larger values $D$ than WIS, and the symmetry shape 1 has a negligible degree of helicity as expected.}

\subsection{MAT alignment and comparison to RAT alignment}
We note that shapes 6 and 7 were created to calculate RATs, and they found strong RATs for these shapes. Interestingly, we found that MATs for these shapes are also strong and exhibit the same helicity (right helicity for shape 6 and left helicity for shape 7) and generic properties as RATs. We also find that our new shape 2 can produce strong MATs, comparable to those of shapes 6 and 7.

For HIS grains, we find that subsonic drift can spin-up the typical $0.1\mu m$ grains to suprathermal rotation. For such grain shapes, MAT alignment exhibits low-J and high-J repellor/attractor points in the phase trajectory map and has long axis perpendicular to the magnetic field, which is analogous to RAT alignment. In the presence of iron inclusions, repellor points are converted to high-J attractors due to enhanced magnetic relaxation. For WIS grains (shapes 3-5), supersonic drift is required to achieve suprathermal rotation. Thus, in the typical ISM conditions, all grains are driven to thermal rotation by gas damping. 

MAT alignment also depends on the drift direction. For shapes 6 and 7, we find that the grain maximum angular momentum spun-up by MATs, $J_{\rm max}^{\rm MAT}$, tends to decrease with the angle $\psi$. However, MAT alignment of shapes 2-5 tends to increase with $\psi$ and become flat (see Figure \ref{fig:Jmax}). 

Table \ref{tab:MAT_RAT} compares the MAT and RAT alignment for seven irregular shapes where RH and LH denote right helicity and left helicity. Both MAT alignment and RAT alignment with low-J/high-J attractors and repellors are indicated. The presence of high-J attractors for shape 3 is only achieved at some large drift angles (see Figure \ref{fig:Jmax}).
 
\begin{table*}
\centering
\caption{Characteristics of MAT alignment vs. RAT alignment for the $a_{\eff}=0.1\mum$ grains in the typical ISM conditions and $s_{d}\sim 1$}\label{tab:MAT_RAT}
\begin{tabular}{l l l l l l l l l l } \hline\hline\\
Shape & Helicity & MATs & $D$  & MAT Align & SMAT Align & Helicity & RATs & RAT Align & SRAT Align \cr
\hline\cr
1 &  Sym &negligible & $1.3\times 10^{-10}$ & NA &NA & NA\cr
2 &  RH & strong &$2.5\times 10^{-5}$& high-J rep./attr. & high-J attr. & NA \cr
3 &  RH & moderate &$3.6\times 10^{-6}$ & high-J rep./attr. & high-J attr. & NA \cr
4 &  RH & moderate &$1.9\times 10^{-6}$ & low-J rep./attr.  & low-J attr. & NA \cr
5 &  RH & moderate &$1.6\times 10^{-6}$ & low-J rep./attr.  & low-J attr. & NA \cr
6 & RH & strong & $2.6\times 10^{-5}$ & high-J rep./attr.  & high-J attr. & RH & strong & high-J rep./attr. & high-J attr.\cr
7 & LH & strong &$4.5\times 10^{-5}$ & high-J rep./attr. & high-J attr. & LH & strong & high-J rep./attr. & high-J attr.\cr

\hline\hline\\
\end{tabular}
\end{table*}

{Numerous observations of dust polarization show the correspondence to the predictions by RAT alignment for molecular clouds (\citealt{2008ApJ...674..304W}; \citealt{2011A&A...534A..19A}), starless cores (\citealt{2014A&A...569L...1A}; \citealt{2015AJ....149...31J}). This reveals that MAT alignment is not a dominant mechanism in these environments. Therefore, for the CMB polarization studies which are involved in the diffuse ISM, the modeling of foreground polarization is not going to be further complicated by the presence of MAT alignment due to the dominance of RAT alignment.}

\subsection{Astrophysical environments with potential MAT alignment}
We now discuss the potential environment conditions where MAT alignment is important. In general, dust-gas drift can be triggered by cloud-cloud collisions \citep{1952MNRAS.112..215G}, radiation pressure,  ambipolar diffusion, and gravitational sedimentation (see \citealt{2007JQSRT.106..225L}). 

First, interstellar cloud-cloud collisions are usually triggered by strong radiation pressure from supernova explosions, producing supersonically drifting grains. Thus, MAT alignment is expected to be efficient in such clouds. The line of sight to SN 2014J is thought to encounter numerous individual clouds \citep{Patat:2015bb}. Therefore, observations of polarization towards SNe Ia can provide useful tests for MAT alignment \citep{2017ApJ...836...13H}.

{Second, in outflows around the late-type stars (e.g., Asymptotic Giant Branch (AGB), post-AGBs, and planetary nebulae (PNe)), grains are found to be supersonically drifting through the gas due to strong radiation pressure (see \citealt{1993ApJ...410..701N}). Thus, we expect MAT alignment to be important in these conditions. Grains are also expected to drift in the innermost outflow regions around high-mass young stellar objects where the outflow is produced by radiation pressure instead of hydromagnetic effects.}

Third, \cite{1995ApJ...453..238R} found that for supersonically drifting grains in a weakly magnetized molecular cloud by ambipolar diffusion, Gold alignment can produce the alignment efficiency of $R\sim 0.3$ for $s_{d}\gg 1$ and $\delta_{m}=10$ (i.e., superparamagnetic grains). Our results show that for grains of HIS, MAT alignment can be efficient, and superparamagnetic grains can be perfectly aligned. This unique feature allows us to differentiate classical Gold alignment from MAT alignment.

Recently, fast modes of MHD turbulence are found to accelerate charged dust grains to supersonic speeds (\citealt{2003ApJ...592L..33Y}; \citealt{Yan:2004ko}; \citealt{Hoang:2012cx}). Moreover, transit time damping is found to be an efficient acceleration mechanism \citep{Hoang:2012cx}. For the CNM, the MHD turbulence can produce the drift speed of $v_{d}\sim 1.5$km/s, corresponding to $s_{d}\sim 1.2$. In molecular and dark clouds, \cite{Yan:2004ko} obtained $s_{d}\sim 1.1-2.2$ and $\sim 2.7-3.5$ for $a=0.1-1.0\mum$. Compared to the critical drift speed for MAT alignment in Figure \ref{fig:sdsup_psi}, we see that the MAT alignment is important in dark clouds for the drift angles of $\psi>20^{\circ}$ with the magnetic field. Interestingly, it is found that MHD turbulence tend to accelerate grains in the direction that makes a large angle with the mean magnetic field (\citealt{Yan:2004ko}). In protoplanetary disks where MRI activity is active, turbulence is perhaps sufficient to generate drift motion of grains and trigger MAT alignment by superparamagnetic grains.

Other processes, such as photoelectric force due to the ejection of photoelectrons by anisotropic radiation, photodesorption of atoms, can also induce relative motion of grains with the ambient gas. However, these processes are unable to drive grains to supersonic motion (\citealt{Weingartner:2001p3480}).

{Gold mechanical alignment was referred to explain polarization observed in outflows around protostars, i.e., early stages of star formation (\citealt{1998ApJ...502L..75R}; \citealt{2006ApJ...650..246C}; \citealt{2009ApJ...695.1399T}). In light of the new MAT mechanism, we expect the efficient alignment of HIS grains with the magnetic fields by MATs in outflows. 
The MAT alignment mechanism helps to probe magnetic fields in special conditions where radiative alignment fails due to lack of radiation source.}

\subsection{Effect of grain precession around the drift direction vs. Larmor precession}
For grains in the interstellar diffuse medium, the Larmor precession of the grain magnetic moment around the ambient magnetic field is usually fast due to the large magnetic moments rendered by grains as a result of the Barnett effect (\citealt{1976Ap&SS..43..291D}; see also \citealt{LAH15}). Grains are aligned with the magnetic fields. There are special situations when the axis of alignment is not the magnetic field but the direction of the grain drift. Physically, this means that the grain precession induced by the gaseous bombardment is faster than the Larmor precession. This is shown to occur for ordinary grains in the protoplanetary disks with very high density, in which the Larmor precession is slower than the gas damping. In these conditions, if the grain drift is supersonic, the alignment can occur with the drift direction due to the MAT. If this is the case, then it can allow us to trace the direction of outflows using the polarization mapping. 

\subsection{MAT alignment in the presence of pinwheel torques, strong and weak internal relaxation}
Finally, let us discuss MAT alignment of grains in the presence of pinwheel torques and alignment of grains without internal relaxation.

\cite{1979ApJ...231..404P} suggested several processes that can spin-up grains to suprathermal rotation, including formation of hydrogen molecules on the grain surface, the variation of accommodation coefficient and emissivity on the grain surface. Although for small grains the effect of pinwheel torques is suppressed due to rapidly flipping and thermal trapping \citep{1999ApJ...516L..37L}, for for large grains, it is still effective. \cite{2009ApJ...695.1457H} found that RAT alignment is shown to be enhanced by pinwheel torques. The effect of alignment by H$_2$ torques is observed in reflection nebula IC 63 \citep{2013ApJ...775...84A}, and it is successfully modelled by \cite{2015MNRAS.448.1178H}. Since MAT alignment of irregular shapes is essentially similar to RAT alignment, we predict that pinwheel torques also act to enhance the alignment by lifting the low-J attractor points and create new high-J attractor points.

In the presence of strong thermal fluctuations with efficient internal relaxation, we expect the crossover becomes low-J attractor points. Therefore, MAT alignment occurs with low-J and high-J attractors as RAT alignment. For weak internal relaxation present in very large grains, MAT alignment is expected to be more complicated. In addition to low-J and high-J attractors, some low-J attractor with "wrong" alignment may be present \citep{2009ApJ...697.1316H}. However, due to random collisions with the gas, such a wrong low-J alignment will be reduced, leading to a moderate degree of alignment.

\section{Summary}\label{sec:sum}
We have studied the alignment of grains by mechanical torques for seven different irregular shapes. Our results are summarized as follows:
\begin{itemize}

\item[1.] Among seven considered shapes, shape 1 induces negligible mechanical torques due to its mirror symmetry. Highly irregular shapes (HIS, shapes 2, 6 and 7) can produce strong mechanical torques that exhibit some generic properties as seen in RATs. Weakly irregular shapes (WIS, shapes 3-5) produce smaller mechanical torques. Such a dramatic difference in MATs for the considered shapes is expected from the physics of mechanical torques that only depends on the grain surface. 

\item[2.] HIS grains can be driven to suprathermal rotation with subsonic drift, while WIS grains are only driven to suprathermal rotation with supersonic drift. The suprathermal rotation rate is found to depend on the drift angle and the grain shape.

\item[3.] For three HIS, we find that MAT tends to align grains with low-J attractors and high-J repellors/attractors. MAT alignment appears to depend on the grain drift direction about the magnetic field.

\item[4.] We find that superparamagnetic inclusions in HIS grains can help MAT alignment to align with high-J attractor points, producing a high degree of grain alignment. For supersonic drift, we also find that irregular grains can be perfectly aligned by MATs, while Gold classical mechanism only induces imperfect alignment.

\item[5.] Our numerical results demonstrate the importance of MAT alignment and its dependence on the grain surface irregularity. Due to the uncertainty of grain shapes in the ISM, the efficiency of MAT alignment requires further theoretical and observational studies.

\end{itemize}

\acknowledgements
We thank Stefan Reissl for useful comments. TH acknowledges the support by the Basic Science Research Program through the National Research Foundation of Korea (NRF), funded by the Ministry of Education (2017R1D1A1B03035359). JC's work is supported by the National R \& D Program funded by the Ministry of Education (NRF-2016R1D1A1B02015014). AL acknowledges the financial support from NASA grant NNX11AD32G, NSF grant AST 1109295, and NASA grant NNH 08ZDA0090. 

\appendix


\section{Aligning, spin-up, and precessing torque components}\label{apd:FGH}
In the alignment coordinate system $\ehat_{1}\ehat_{2}\ehat_{3}$, the MAT components that cause spin-up, precession, and alignment of grains, are given by
\bea
F(\xi,\psi,\phi)&=&Q_{e1}(-\sin\psi \cos\xi \cos\phi-\sin \xi \cos\psi)+Q_{e2}(\cos\psi\cos\xi \cos\phi-\sin\xi\sin\psi)+Q_{e3}\cos\xi\sin\phi,\\
G(\xi,\psi,\phi)&=&Q_{e1}\sin\psi\sin\phi- Q_{e2}\cos\psi\sin\phi+ Q_{e3}\cos\phi,\label{eeq10}\\ 
H(\xi,\psi,\phi)&=&Q_{e1}(\cos\psi\cos\xi -\sin\psi\sin\xi\cos\phi) + Q_{e2}(\sin\psi \cos\xi +\cos\psi\sin\xi\cos\phi)+Q_{e3}\sin\xi\sin\phi, \label{eeq11} 
\ena
where $Q_{e1}(\xi, \psi, \phi), Q_{e2}(\xi, \psi, \phi), Q_{e3}(\xi, \psi, \phi)$, as functions of $\xi, \psi$ and $\phi$, are components of the RAT efficiency vector in the
lab coordinate system (see DW97; LH07a). 

To obtain $Q_{e{1}}(\xi, \psi, \phi),
Q_{e{2}}(\xi, \psi, \phi)$ and $Q_{e{3}}(\xi, \psi, \phi)$ from ${\bf Q}_{\Gamma}(\Theta, \beta, \Phi)$, we need to use the relations between $\xi, \psi, \phi$ and $\Theta, \beta, \Phi$ (see \citealt{2003ApJ...589..289W}; HL08).

The average of the aligning torque over the Larmor precession is given by
\bea
\langle F(\xi,\psi)\rangle = \frac{1}{2\pi}\int_{0}^{2\pi}F(\xi,\psi,\phi) d\phi.
\ena

\bibliography{ms.bbl}

\begin{thebibliography}{}
\expandafter\ifx\csname natexlab\endcsname\relax\def\natexlab#1{#1}\fi

\bibitem[{Ade {et~al.}(2015)Ade, Aghanim, Ahmed, \& et~al. {(BICEP2/Keck and
  Planck Collaborations)}}]{Ade:2015ee}
Ade, P. A.~R., Aghanim, N., Ahmed, Z., \& et~al. {(BICEP2/Keck and Planck
  Collaborations)}. 2015, PRL, 114, 101301

\bibitem[{Alves {et~al.}(2014)Alves, Frau, Girart, Franco, Santos, \&
  Wiesemeyer}]{2014A&A...569L...1A}
Alves, F.~O., Frau, P., Girart, J.~M., {et~al.} 2014, A\&A, 569, L1

\bibitem[{Andersson {et~al.}(2015)Andersson, Lazarian, \&
  Vaillancourt}]{Andersson:2015bq}
Andersson, B.-G., Lazarian, A., \& Vaillancourt, J.~E. 2015, ARA\&A, 53, 501

\bibitem[{Andersson {et~al.}(2011)Andersson, Pintado, Potter, Strai{\v z}ys, \&
  Charcos-Llorens}]{2011A&A...534A..19A}
Andersson, B.-G., Pintado, O., Potter, S.~B., Strai{\v z}ys, V., \&
  Charcos-Llorens, M. 2011, A\&A, 534, 19

\bibitem[{Andersson {et~al.}(2013)Andersson, Piirola, De~Buizer, Clemens,
  Uomoto, Charcos-Llorens, Geballe, Lazarian, Hoang, \&
  Vornanen}]{2013ApJ...775...84A}
Andersson, B.-G., Piirola, V., De~Buizer, J., {et~al.} 2013, \apj, 775, 84

\bibitem[{Chandrasekhar \& Fermi(1953)}]{1953ApJ...118..113C}
Chandrasekhar, S., \& Fermi, E. 1953, \apj, 118, 113

\bibitem[{Cho \& Lazarian(2005)}]{2005ApJ...631..361C}
Cho, J., \& Lazarian, A. 2005, \apj, 631, 361

\bibitem[{Cho \& Lazarian(2007)}]{2007ApJ...669.1085C}
Cho, J., \& Lazarian, A. 2007, \apj, 669, 1085

\bibitem[{Cortes {et~al.}(2006)Cortes, Crutcher, \&
  Matthews}]{2006ApJ...650..246C}
Cortes, P.~C., Crutcher, R.~M., \& Matthews, B.~C. 2006, \apj, 650, 246

\bibitem[{Das \& Weingartner(2016)}]{2016MNRAS.457.1958D}
Das, I., \& Weingartner, J.~C. 2016, \mnras, 457, 1958

\bibitem[{Davis(1951)}]{1951PhRv...81..890D}
Davis, L. 1951, Physical Review, 81, 890

\bibitem[{Davis \& Greenstein(1951)}]{1951ApJ...114..206D}
Davis, L.~J., \& Greenstein, J.~L. 1951, \apj, 114, 206

\bibitem[{Dolginov \& Mitrofanov(1976)}]{1976Ap&SS..43..291D}
Dolginov, A.~Z., \& Mitrofanov, I.~G. 1976, Ap\&SS, 43, 291

\bibitem[{Draine \& Weingartner(1996)}]{1996ApJ...470..551D}
Draine, B.~T., \& Weingartner, J.~C. 1996, \apj, 470, 551

\bibitem[{Draine \& Weingartner(1997)}]{1997ApJ...480..633D}
Draine, B.~T., \& Weingartner, J.~C. 1997, \apj, 480, 633

\bibitem[{Gold(1952{\natexlab{a}})}]{Gold:1952p5848}
Gold, T. 1952{\natexlab{a}}, Nature, 169, 322

\bibitem[{Gold(1952{\natexlab{b}})}]{1952MNRAS.112..215G}
Gold, T. 1952{\natexlab{b}}, \mnras, 112, 215

\bibitem[{Hall(1949)}]{Hall:1949p5890}
Hall, J.~S. 1949, Science, 109, 166

\bibitem[{Hildebrand(1988)}]{Hildebrand:1988p2566}
Hildebrand, R.~H. 1988, Royal Astronomical Society, 29, 327

\bibitem[{Hiltner(1949)}]{Hiltner:1949p5856}
Hiltner, W.~A. 1949, Science, 109, 165

\bibitem[{Hoang(2017)}]{2017ApJ...836...13H}
Hoang, T. 2017, \apj, 836, 13

\bibitem[{Hoang {et~al.}(2010)Hoang, Draine, \& Lazarian}]{Hoang:2010jy}
Hoang, T., Draine, B.~T., \& Lazarian, A. 2010, \apj, 715, 1462

\bibitem[{Hoang \& Lazarian(2008)}]{Hoang:2008gb}
Hoang, T., \& Lazarian, A. 2008, \mnras, 388, 117

\bibitem[{Hoang \& Lazarian(2009{\natexlab{a}})}]{2009ApJ...697.1316H}
Hoang, T., \& Lazarian, A. 2009{\natexlab{a}}, \apj, 697, 1316

\bibitem[{Hoang \& Lazarian(2009{\natexlab{b}})}]{2009ApJ...695.1457H}
Hoang, T., \& Lazarian, A. 2009{\natexlab{b}}, \apj, 695, 1457

\bibitem[{Hoang \& Lazarian(2014)}]{2014MNRAS.438..680H}
Hoang, T., \& Lazarian, A. 2014, \mnras, 438, 680

\bibitem[{Hoang \& Lazarian(2016)}]{2016ApJ...831..159H}
Hoang, T., \& Lazarian, A. 2016, \apj, 831, 159

\bibitem[{Hoang {et~al.}(2015)Hoang, Lazarian, \&
  Andersson}]{2015MNRAS.448.1178H}
Hoang, T., Lazarian, A., \& Andersson, B.-G. 2015, \mnras, 448, 1178

\bibitem[{Hoang {et~al.}(2012)Hoang, Lazarian, \& Schlickeiser}]{Hoang:2012cx}
Hoang, T., Lazarian, A., \& Schlickeiser, R. 2012, \apj, 747, 54

\bibitem[{Jones {et~al.}(2015)Jones, Bagley, Krejny, Andersson, \&
  Bastien}]{2015AJ....149...31J}
Jones, T.~J., Bagley, M., Krejny, M., Andersson, B.-G., \& Bastien, P. 2015,
  \apj, 149, 31

\bibitem[{Kimura {et~al.}(2017)Kimura, Tanaka, Nozawa, Takeuchi, \&
  Inatomi}]{2017SciA....3E1992K}
Kimura, Y., Tanaka, K.~K., Nozawa, T., Takeuchi, S., \& Inatomi, Y. 2017,
  Science Advances, 3, e1601992

\bibitem[{Lazarian(1994)}]{1994MNRAS.268..713L}
Lazarian, A. 1994, \mnras, 268, 713

\bibitem[{Lazarian(1995{\natexlab{a}})}]{1995MNRAS.277.1235L}
Lazarian, A. 1995{\natexlab{a}}, \mnras, 277, 1235

\bibitem[{Lazarian(1995{\natexlab{b}})}]{Lazarian:1995p3034}
Lazarian, A. 1995{\natexlab{b}}, \apj, 451, 660

\bibitem[{Lazarian(1997)}]{1997ApJ...483..296L}
Lazarian, A. 1997, \apj, 483, 296

\bibitem[{Lazarian(2007)}]{2007JQSRT.106..225L}
Lazarian, A. 2007, J. Quant. Spectrosc. Rad. Trans., 106, 225

\bibitem[{{Lazarian} {et~al.}(2015){Lazarian}, {Andersson}, \& {Hoang}}]{LAH15}
{Lazarian}, A., {Andersson}, B.-G., \& {Hoang}, T. 2015, in Polarimetry of
  stars and planetary systems, ed. L.~{Kolokolova}, J.~{Hough}, \& A.-C.
  {Levasseur-Regourd} ((New York: Cambridge Univ. Press)), 81

\bibitem[{Lazarian \& Draine(1999)}]{1999ApJ...516L..37L}
Lazarian, A., \& Draine, B.~T. 1999, \apj, 516, L37

\bibitem[{Lazarian \& Efroimsky(1996)}]{Lazarian:1996p6083}
Lazarian, A., \& Efroimsky, M. 1996, \apj, 466, 274

\bibitem[{Lazarian \& Hoang(2007{\natexlab{a}})}]{2007MNRAS.378..910L}
Lazarian, A., \& Hoang, T. 2007{\natexlab{a}}, \mnras, 378, 910

\bibitem[{Lazarian \& Hoang(2007{\natexlab{b}})}]{2007ApJ...669L..77L}
Lazarian, A., \& Hoang, T. 2007{\natexlab{b}}, \apj, 669, L77

\bibitem[{Lazarian \& Hoang(2008)}]{Lazarian:2008fw}
Lazarian, A., \& Hoang, T. 2008, \apj, 676, L25

\bibitem[{Lazarian \& Roberge(1997)}]{1997ApJ...484..230L}
Lazarian, A., \& Roberge, W.~G. 1997, \apj, 484, 230

\bibitem[{Lazarian \& Yan(2002)}]{2002ApJ...566L.105L}
Lazarian, A., \& Yan, H. 2002, \apj, 566, L105

\bibitem[{Netzer \& Elitzur(1993)}]{1993ApJ...410..701N}
Netzer, N., \& Elitzur, M. 1993, \apj, 410, 701

\bibitem[{Patat {et~al.}(2015)Patat, Taubenberger, Cox, Baade, Clocchiatti,
  H{\"o}flich, Maund, Reilly, Spyromilio, Wang, Wheeler, \&
  Zelaya}]{Patat:2015bb}
Patat, F., Taubenberger, S., Cox, N. L.~J., {et~al.} 2015, A\&A, 577, A53

\bibitem[{Purcell(1969)}]{Purcell:1969p3641}
Purcell, E.~M. 1969, Physica, 41, 100

\bibitem[{Purcell(1979)}]{1979ApJ...231..404P}
Purcell, E.~M. 1979, \apj, 231, 404

\bibitem[{Purcell \& Spitzer(1971)}]{1971ApJ...167...31P}
Purcell, E.~M., \& Spitzer, L.~J. 1971, \apj, 167, 31

\bibitem[{Rao {et~al.}(1998)Rao, Rao, Crutcher, Crutcher, Plambeck, Plambeck,
  Wright, \& Wright}]{1998ApJ...502L..75R}
Rao, R., Rao, R., Crutcher, R.~M., {et~al.} 1998, \apjl, 502, L75

\bibitem[{Roberge \& Hanany(1990)}]{1990BAAS...22..862R}
Roberge, W.~G., \& Hanany, S. 1990, Bulletin of the American Astronomical
  Society, 22, 862

\bibitem[{Roberge {et~al.}(1995)Roberge, Hanany, \&
  Messinger}]{1995ApJ...453..238R}
Roberge, W.~G., Hanany, S., \& Messinger, D.~W. 1995, \apj, 453, 238

\bibitem[{Tang {et~al.}(2009)Tang, Tang, Ho, Ho, Girart, Girart, Rao, Rao,
  Koch, Koch, Lai, \& Lai}]{2009ApJ...695.1399T}
Tang, Y.-W., Tang, Y.-W., Ho, P. T.~P., {et~al.} 2009, \apj, 695, 1399

\bibitem[{Tazaki {et~al.}(2017)Tazaki, Lazarian, \&
  Nomura}]{2017ApJ...839...56T}
Tazaki, R., Lazarian, A., \& Nomura, H. 2017, \apj, 839, 56

\bibitem[{Weingartner \& Draine(2001)}]{Weingartner:2001p3480}
Weingartner, J.~C., \& Draine, B.~T. 2001, \apj, 553, 581

\bibitem[{Weingartner \& Draine(2003)}]{2003ApJ...589..289W}
Weingartner, J.~C., \& Draine, B. 2003, \apj, 589, 289

\bibitem[{Whittet {et~al.}(2008)Whittet, Hough, Lazarian, \&
  Hoang}]{2008ApJ...674..304W}
Whittet, D. C.~B., Hough, J.~H., Lazarian, A., \& Hoang, T. 2008, \apj, 674,
  304

\bibitem[{Yan \& Lazarian(2003)}]{2003ApJ...592L..33Y}
Yan, H., \& Lazarian, A. 2003, \apj, 592, L33

\bibitem[{Yan {et~al.}(2004)Yan, Lazarian, \& Draine}]{Yan:2004ko}
Yan, H., Lazarian, A., \& Draine, B.~T. 2004, \apj, 616, 895

\end{thebibliography}

\end{document}